\documentclass[11pt,a4paper]{article}                           

\usepackage{amsmath}
\usepackage{amssymb}
\usepackage[usenames]{color} 
\usepackage{multirow}
\usepackage{graphicx}
\usepackage{epstopdf}
\usepackage{array}
\usepackage{hhline}
\usepackage{cite}
\usepackage{microtype,colortbl}
\usepackage[bf]{caption}
\usepackage{color}
\usepackage[usenames,dvipsnames]{xcolor}
\usepackage{pstool}
\usepackage{tikz}
\usepackage{ytableau}
\usepackage{latexsym,stmaryrd}
\usepackage[bookmarks=false]{hyperref}
 \usepackage{lscape}
 \usepackage{subfig}
 \usepackage{tensor}
 \usepackage{dsfont}
\usepackage{ifpdf}
\def\hybrid{\topmargin -20pt    \oddsidemargin 0pt
        \headheight 0pt \headsep 0pt
        \textwidth 6.25in       % A4 paper
        \textheight 9 in       % A4 paper
        \marginparwidth .875in
        \parskip 5pt plus 1pt 
          \jot = 1.5ex
   }
\hybrid
\numberwithin{equation}{section}
\numberwithin{table}{section}

\newcommand{\beq}{\begin{equation}}  \newcommand{\eeq}{\end{equation}}
\newcommand{\bal}{\begin{aligned}}   \newcommand{\eal}{\end{aligned}}
\newcommand{\bea}{\begin{eqnarray}}  \newcommand{\eea}{\end{eqnarray}}

\newcommand{\bmat}{\left(\begin{array}}
\newcommand{\emat}{\end{array}\right)}

\newcommand{\bbC}{\mathbb{C}}
\newcommand{\bbR}{\mathbb{R}}

\newcommand{\nn}{\nonumber}

\newcommand{\cO}{\mathcal{O}}

\newcommand{\cK}{\mathcal{K}}
\newcommand{\cN}{\mathcal{N}}

\newcommand{\cH}{\mathcal{H}}

\newcommand{\cM}{\mathcal M}

\usepackage{cancel}

\newcommand{\be}{\begin{equation}}
\newcommand{\ee}{\end{equation}}

\newcommand{\half}{\frac{1}{2}}

\DeclareMathOperator{\ad}{ad}

\newcommand{\slt}{\mathfrak{sl}(2)}
\newcommand{\SLt}{\mathrm{Sl}(2)}

\newcommand{\bbZ}{\mathbb{Z}}

\newcommand*{\Scale}[2][4]{\scalebox{#1}{$#2$}}

\newcommand{\dd}{\mathrm{d}}

\definecolor{Gray}{gray}{0.95}

\begin{document}
\baselineskip=14pt
\parskip 5pt plus 1pt 

\vspace*{2cm}
\begin{center}
	{\LARGE\bfseries Bulk Reconstruction in Moduli Space Holography}\\[.3cm]
	
	\vspace{1cm}
	{\bf Thomas W.~Grimm}\footnote{t.w.grimm@uu.nl},
	{\bf Jeroen Monnee}\footnote{j.monnee@uu.nl},
	{\bf Damian van de Heisteeg}\footnote{d.t.e.vandeheisteeg@uu.nl},
	
	{\small
		\vspace*{.5cm}
		Institute for Theoretical Physics, Utrecht University\\ Princetonplein 5, 3584 CC Utrecht, The Netherlands\\[3mm]
	}
\end{center}
\vspace{1cm}
\begin{abstract}\noindent
	It was recently suggested that certain UV-completable supersymmetric actions 
	can be characterized by the solutions to an auxiliary non-linear sigma-model with special asymptotic boundary conditions.  
	The space-time of this sigma-model is the scalar field space of these effective theories while the target space is a coset space. 
	We study this sigma-model without any reference to a potentially underlying geometric
description. 
Using a holographic approach reminiscent of the bulk reconstruction in the AdS/CFT correspondence, we then derive its near-boundary solutions
	for a 	two-dimensional space-time.   
	Specifying a set of $ Sl(2,\mathbb{R})$ boundary data we show that the near-boundary solutions are 
	uniquely fixed after imposing a single bulk-boundary matching condition. The reconstruction exploits an elaborate set 
	of recursion relations introduced by Cattani, Kaplan, and Schmid in the proof of the $Sl(2)$-orbit theorem.
	We explicitly solve these recursion 
	relations for three sets of simple boundary data and show that they model asymptotic periods 
	of a Calabi--Yau threefold near the conifold point, the large complex structure point, and the Tyurin degeneration.

\end{abstract}

\newpage

\tableofcontents
\setcounter{footnote}{0}

\newpage
%%%%%%%%%%%%%%%%%%%%%%%%%%%%%%%%%%%%%%%%%%%%%%%%%%%%%%

\section{Introduction}

In recent years the study of general constraints on effective theories that can be coupled to quantized gravity has received 
much attention, culminating in a vast network of `swampland conjectures' \cite{Palti:2019pca,vanBeest:2021lhn}. These conjectures
have been mostly tested in string theory and highlight the observation that the known effective theories 
arising from string theory are not generic. This should be contrasted with the fact that there is an immense number of 
possibilities for connecting string theory with a four-dimensional theory, for example, by choosing different 
compactification geometries. One approach to uncover the special properties of effective theories arising 
from string compactifications is to consider the behaviour of their effective theories when taking limits in scalar field spaces. 
This lead to a prominent set of swampland conjectures and emergence proposals \cite{Ooguri2007,Klaewer:2016kiy,GPV,Heidenreich:2018kpg,Lee:2019wij,Lanza:2020qmt,Calderon-Infante:2020dhm} 
that describe the behaviour and validity of the effective theories when 
moving to an asymptotic boundary after traversing an infinite geodesic distance. Furthermore, it is long known \cite{Ashok:2003gk,Acharya:2006zw} that 
also finiteness statements about the number of flux vacua are dependent on the asymptotic behaviour of the effective theory 
when approaching the boundary of field space. Taken together this provides motivation to develop new strategies to determine 
and characterize the structure of field spaces and effective theories with a focus on their asymptotic properties. 

Recently, it was proposed in  \cite{Grimm:2020cda} to develop a holographic description of the near-boundary effective theories. 
More concretely, it was suggested that the properties of certain supersymmetric near-boundary effective theories 
are determined holographically from a constrained set of boundary data and the solutions to an auxiliary bulk action principle \cite{Cecotti:2020rjq,Cecotti:2020uek,Grimm:2020cda}. This action is a non-linear sigma-model with a space-time that is 
the field space, or rather moduli space, of the effective theory while the target space is a certain coset space built out of the 
dualities and symmetries of the theory.  
The most direct applications of these ideas lie in constraining the form of the vector sector in general 
four-dimensional $\cN=2$ supergravity theories \cite{GPV,Cecotti:2020rjq,Cecotti:2020uek,Grimm:2020cda}. 
More indirectly the holographic approach can also be applied for certain scalar potentials in $\cN=1$ supergravity actions that naturally 
appear in flux compactifications \cite{Grimm:2019ixq,Grimm:2020cda}. There are two perspectives that one can take on this `moduli space 
holography'. First, one can use it as an alternative way to study moduli spaces appearing in geometric compactifications of 
string theory. Most notably, it applies rather directly to the moduli spaces of Calabi--Yau manifolds in any dimension and
has been used to investigate various swampland conjectures, see e.g.  \cite{GPV,Grimm:2018cpv,Corvilain:2018lgw,Font:2019cxq, Grimm:2019wtx,Grimm:2019ixq,Lanza:2020qmt,Gendler:2020dfp,BastianGrimmHeisteeg,Grimm:2020ouv}. Second, it 
can be viewed abstractly with no reference to any geometric realization and studied as a sigma-model whose solutions 
with special boundary conditions admit some remarkable features. We will mostly take the latter perspective in this work, extend 
the ideas of \cite{Grimm:2020cda}, and show that they are applicable in concrete examples.

To give a concrete description of the model we constrain ourselves to an auxiliary matter bulk theory  
with a space-time being a real two-dimensional manifold. The matter field is valued in a non-compact group $G_\bbR$, which preserves 
a bilinear form on a finite-dimensional Hilbert space. Focusing on the non-linear sigma-model action for the matter fields, we find that it enjoys both global and local symmetries turning 
the resulting target space into a coset construction reminiscent of a WZW model. 
Being group-valued, the matter field acts as an operator on a finite-dimensional Hilbert space. One of the crucial aspects of the construction is that the Hilbert space enjoys a decomposition into a finite number of complex vector spaces, which can be thought of as assigning a background charge $Q$ to the various spaces. 
In the geometric settings this corresponds to the Hodge decomposition of the cohomology groups of a K\"ahler manifold. 
Much of the intricacy of our model comes from the interplay between the near-boundary behaviour of the bulk matter fields and the boundary charge decomposition of said Hilbert space. 
Within this setting our aim is to find near-boundary solutions for the matter fields in a two-dimensional background metric that is flat up to a general overall Weyl factor. Furthermore, we require that the solutions admit a global continuous 
symmetry, obey a special $Q$-constraint, and match a set of $\mathrm{Sl}(2,\mathbb{R})$ boundary conditions. The appearance of an $\SLt$-symmetry is consistent with the
asymptotic behaviour of a natural metric on two-dimensional moduli spaces known as the Hodge metric.\footnote{Although we will not discuss the coupling to gravity of the model, we note that two different proposal for including gravity have appeared in \cite{Cecotti:2020uek} and \cite{Grimm:2020cda}, in which the Hodge metric plays an important role.} The latter can be constructed from the conserved currents associated to the global symmetry of the action, when evaluated on a solution to the equations of motion, and asymptotes to the Poincar\'e metric, which is invariant under an $\SLt$ isometry group. We will find that these 
solutions admit remarkable properties and provide, in special geometric situations, asymptotic models for the periods of Calabi--Yau  
manifolds.

A crucial restriction on the considered solutions is provided by the matching with a set of $\mathrm{Sl}(2,\mathbb{R})$ boundary data.  
This boundary information consists of five special operators $Q_\infty, N^+, N^0, N^-$ and $\delta$, which act on, and define, the boundary Hilbert space. The charge operator $Q_\infty$ decomposes this  Hilbert space into various charge eigenspaces and 
induces a well-defined inner product. 
The operators $N^+, N^0$ and $N^-$ form a $\mathfrak{sl}(2,\mathbb{R})$-triple, which can be thought of as a completion 
of the asymptotic global symmetry enjoyed by the considered set of solutions. By complexifying the algebra the $\slt$-triple can be used to further refine the split of the Hilbert space through a highest-weight decomposition that is compatible with the charge-decomposition. Finally, the phase operator  $\delta$ is crucial in order to uniquely fix the bulk solution. It controls the subleading corrections in the $1/y$ expansion of the 
solutions with the boundary located at $y=\infty$. Note that the boundary data $Q_\infty,N^+, N^0, N^-$ together with the 
sigma-model groups $G_\bbR$ can be classified 
using $\SLt$-representation theory \cite{robles_2016,Kerr2017}. Within this vast set of possible boundary configurations we introduce a 
special subclass that we call being of Calabi--Yau type. It will be these special types of boundary data that we later connect with period integrals 
for a set of Calabi--Yau threefold examples. 

With a detailed understanding of the bulk theory and boundary data, we set out to solve the equations of motion by reproducing large 
parts of the proof of the $\SLt$-orbit theorem by Schmid \cite{Schmid} and Cattani, Kaplan, and Schmid (CKS) \cite{CKS}, albeit in a slightly different formulation. 
We first note that the equations of motion of the bulk theory reduce to Nahm's equations when evaluated for the solutions satisfying 
the specified asymptotic constraints. The strategy is to then decompose the bulk fields according to the charge associated to $Q_\infty$, and the weight and highest-weight associated to the $\mathfrak{sl}(2,\mathbb{C})$ acting on the Hilbert space at the boundary. This leads to a system of coupled recursion relations for the various components of the bulk field, which we term the CKS recursion relations \cite{CKS}. We will review 
the $\SLt$-representation theory that ensures the recursive properties of these relations and recast them in a useful basis for explicit computations.  
We then find that the recursion is so restrictive that the leading behaviour determined by the $\slt$-boundary condition, together with a single matching condition involving the phase operator, fixes all terms subleading in the near-boundary expansion uniquely. The construction of the near-boundary solutions starting 
from the boundary data is reminiscent of the bulk reconstruction methods employed in the AdS/CFT duality \cite{Harlow:2018fse,Hamilton:2006az,Nakayama:2015mva}. We are considering massless fields, which simplifies the analysis, but note that the highly non-trivial structure arises due to the non-canonically normalized kinetic terms 
in the action. This issue arises similarly in the holographic study of WZW models, see e.g.~\cite{Beccaria:2020qtk}. Let us stress that our approach differs both conceptually and technically from applying the AdS/CFT correspondence directly to the effective theories. This latter strategy has been successfully applied to testing swampland conditions in \cite{Nakayama:2015hga,Harlow:2015lma,Benjamin:2016fhe,Montero:2016tif,Montero:2017mdq,Harlow:2018tng,Harlow:2018jwu,Bae:2018qym,Conlon:2018vov,Montero:2018fns,Lin:2019kpn,Conlon:2020wmc,Ooguri:2020sua,Perlmutter:2020buo}.

To elucidate these general consideration we will present a set of three simple examples with Calabi--Yau type boundary conditions. 
We explicitly solve the CKS recursion relations and determine the bulk matter fields, containing all polynomially sub-leading corrections in powers of $1/y$ near the boundary $y=\infty$. We then explain how the bulk field solutions determine the asymptotic form of period integrals of general Calabi--Yau threefolds with a one-dimensional complex structure moduli space. In fact, we find that the three considered examples exactly capture the 
three possible degenerations of such Calabi--Yau threefolds with unipotent monodromy, namely the conifold point, 
the Tyurin degeneration, and the large complex structure point. This exemplifies the fact that a rather simple set 
of algebraic boundary data fixes an infinite series of subleading corrections in the period vectors.

The paper is structured as follows. In section \ref{bulk-theory} we introduce the bulk theory defined on a real two-dimensional space-time. In particular, we write down an action for the bulk matter fields, discuss its symmetries and derive the equations of motion. We also describe some aspects of the geometry of the moduli space to motivate the particular class of solutions we will be considering. In section \ref{sec:boundary} we introduce the boundary data in detail and highlight the key properties that are required for the bulk reconstruction. Furthermore, the induced decomposition of the Hilbert space and its operators is discussed, together with a classification of the data. In table \ref{table:data} one may find the complete set of data that is used in this work. The main body of our work is comprised of sections \ref{Bulk-reconstruction} and \ref{reconstructing_examples}. In section \ref{Bulk-reconstruction} we give a detailed account of how the equations of motions are rewritten in terms of the CKS recursions and how the boundary data enters as an initial conditions in these recursions. Then in section \ref{reconstructing_examples} we solve the CKS recursion explicitly for the three examples. Finally, in section \ref{period_section} we describe the relation between the bulk fields and the period vector, by translating our description back to the Hodge-theoretical setting. There are two appendices. In appendix \ref{app:input} we have collected some computations on the $Q$-constraint and what it imposes on the input data of the CKS recursion. In appendix \ref{app:LCS_basis} one may find some explicit expressions that are used for the bulk reconstruction for $\mathrm{IV}_1$ boundaries.

\section{The bulk theory}\label{bulk-theory}

In this section we introduce the bulk matter theory and describe the set of solutions to the classical field 
equations that will be studied in the remainder of this work. The bulk action is a two-dimensional non-linear sigma-model 
of a group-valued matter field acting on a finite-dimensional Hilbert space and will be introduced in section \ref{bulk_matter_action}. 
We discuss the symmetries of this theory and note that gauge equivalent matter field configurations live in a coset space $G_{\bbR} / M$. 
In section \ref{local-solutions} we will then specify the considered asymptotic solutions that admit a global symmetry parametrized 
by a nilpotent matrix $N^-$. These solutions turn out to be uniquely fixed by a set of boundary data consisting of an 
sl$(2)$-algebra and two special operators, which are introduced later in section \ref{sec:boundary}. 

Note that the constructions presented here are motivated by the behaviour of period maps from a moduli space 
into the arithmetic quotients $\Gamma \backslash G_{\bbR} / M$. The considered action can be viewed as the two-dimensional 
generalization of the action introduced in \cite{donaldson1984} that encodes Nahm's equations. It can also be derived as an 
action encoding the $tt^*$ equations \cite{Cecotti:2020rjq}.   

\subsection{The bulk matter action} \label{bulk_matter_action}

In this section we introduce the bulk theory for a group-valued matter field $h$. It is given by a non-linear sigma-model 
on a two-dimensional space $\cM$, the moduli space, and a target space being a real group $G_{\bbR}$. This real group 
is defined by its action on some finite dimensional complex vector space $\cH$ and required to 
preserve a bilinear product $\langle v , w \rangle$ on $\cH$, i.e.~we 
require 
\beq
g\in G_{\bbR}:\quad \langle g v ,g w \rangle = \langle v , w \rangle\ . 
\eeq
Here we note that $G_{\bbR}$ need not comprise the full symmetry group of the bilinear product. To define the sigma-model, we also introduce a special  inner product $\langle v|w \rangle$, which renders 
$\cH$ into a Hilbert space. This inner product is induced by an operator $Q \in i\mathfrak{g}_{\bbR}$ that defines 
a split of $\cH$ into $D+1$ vector spaces
\beq
\cH_q\ , \quad q=-\frac{D}{2}, -\frac{D}{2}+1,..., \frac{D}{2}\ , 
\eeq
where $D$ will be called the \textit{weight} associated to the Hilbert space $\cH$. 
We can think of $q$ as being the charge 
of a state in $\cH_q$, i.e.~we have 
\beq \label{ref-Q}
|w\rangle \in \cH_q:   \quad  Q |w\rangle =q |w\rangle \ ,
\eeq
where we have used bra-ket notation to denote states in $\cH$. 
Note that by definition $\bar Q = - Q$, which implies that all eigenstates with non-vanishing real charge are complex.
Given such a charge operator $Q$ the inner product  $ \langle v|w \rangle$  is defined as 
\beq \label{def-innerproduct}
\langle v|w \rangle :=  \langle \bar v, e^{\pi i Q} w \rangle\ .
\eeq
This inner product allows us to associate to each operator $\cO$ acting on $|w \rangle$ an 
adjoint operator $\cO^\dagger$. 

We are now in the position to introduce the sigma-model action for the fields $h$. In order to do this 
we denote by $\sigma^\alpha$,  $\alpha =1,2$ real local coordinates on $\cM$ and define $g_{\alpha \beta}$ to be 
a background metric on $\cM$. The action that we consider for the matter field takes the form  
\beq \label{Smat_1}
S_{\rm mat}(h) = \frac{1}{4}\int_{\cM} d^2 \sigma \sqrt{g} g^{\alpha \beta} \mathrm{Tr}\left(h^{-1} \partial_\alpha h+\left(h^{-1} \partial_\alpha h\right)^\dagger \right)\left(h^{-1} \partial_\beta h+\left(h^{-1} \partial_\beta h\right)^\dagger \right)\ ,
\eeq
where the trace is evaluated by using the inner product $ \langle v|w \rangle$ defined in \eqref{def-innerproduct}. It will 
later be useful to employ form notation on $\cM$, writing e.g.~$\dd h = \partial_\alpha h \, \dd \sigma^\alpha$. The action \eqref{Smat_1}
can then be written more compactly as 
\begin{equation}\label{Smat_2}
S_{\rm mat}(h) = \frac{1}{4}\int \mathrm{Tr}\left| h^{-1} \text{d} h+\left(h^{-1} \text{d} h\right)^\dagger \right|^2\ ,
\end{equation} 
where $|A|^2 = A \wedge * A$ with $*$ being the Hodge star on $\cM$ in the background metric $g_{\alpha \beta}$.  Let us stress that $h$ is a group-valued field and thus, in general, combines the degrees of freedom of multiple massless scalar fields. Due to the non-linearity of the action \eqref{Smat_2} in $h$ any simple parametrization of $h$ in terms of scalar fields will result in non-canonically normalized kinetic terms, which highlights the complexity of the model.

Let us point out that this action has two sets of symmetries, which we will discuss in turn. Firstly, we see that \eqref{Smat_2}
has a global invariance under left-multiplication 
$h(\sigma) \rightarrow g_L h(\sigma) $ with $g_L \in G_{\bbR}$.
This global symmetry yields  a conserved current of the form
\begin{equation}
\label{eq:current}
J_L = \frac{1}{2} *\big[(\text{d}h)h^{-1}+h\left(h^{-1}\text{d}h\right)^\dagger h^{-1}\big]\ .
\end{equation}
Secondly, one checks that \eqref{Smat_2} has a local invariance under right-multiplication 
$h(\sigma) \rightarrow h(\sigma) g_R(\sigma)$, with $g_R^\dagger(\sigma) = g^{-1}_{R}(\sigma)$. 
The presence of this gauge symmetry shows that the matter action \eqref{Smat_2} actually 
describes fields in a coset $G/M$, where $M$ consists of the group elements $g_R$ that satisfy the unitarity condition
$g_R^\dagger = g^{-1}_{R}$ with respect to the inner product \eqref{def-innerproduct}. Put differently, $M$ can be realized as the stabilizer of the group element $e^{\pi i Q}$.\footnote{For this one uses the fact that $g^\dagger = e^{-\pi iQ}g^{-1} e^{\pi i Q}$ for a group element $g\in G$, cf. \eqref{def-innerproduct}} It is a general result that this always yields a maximal compact subgroup of $G$. In fact, the converse is also true: to each maximal compact subgroup of $G$ one can associate an inner product which is stabilized by it \cite{Borel1973}.

\subsection{Near-boundary solutions at the center of the Poincar\'e disc} \label{local-solutions}

We now turn to the study of the solutions to the classical equations of motion 
of the matter action \eqref{Smat_2}. Concretely, we will consider a background 
metric with the line element 
\beq \label{conformally-flat}
g_{\alpha \beta} d\sigma^\alpha d\sigma^\beta = f^2(\sigma) \big[(d\sigma^1)^2 + (d\sigma^2)^2 \big]\ , 
\eeq
where $f(\sigma)$ is a (non-vanishing) function of $\sigma$. It is easy to check that the Weyl factor 
$f^2$ drops from the action and the equations of motion. As long as we do not couple this theory 
to two-dimensional gravity, we are thus free to work with any choice of $f(\sigma)$. 

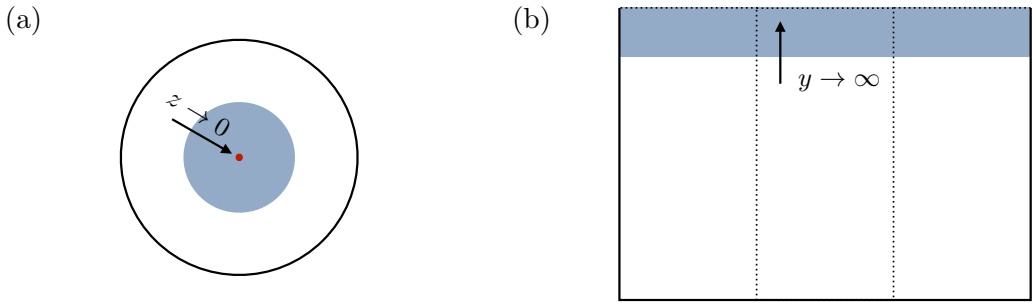
\begin{figure}[h!]
	\begin{center}
		\includegraphics[width=0.8\textwidth]{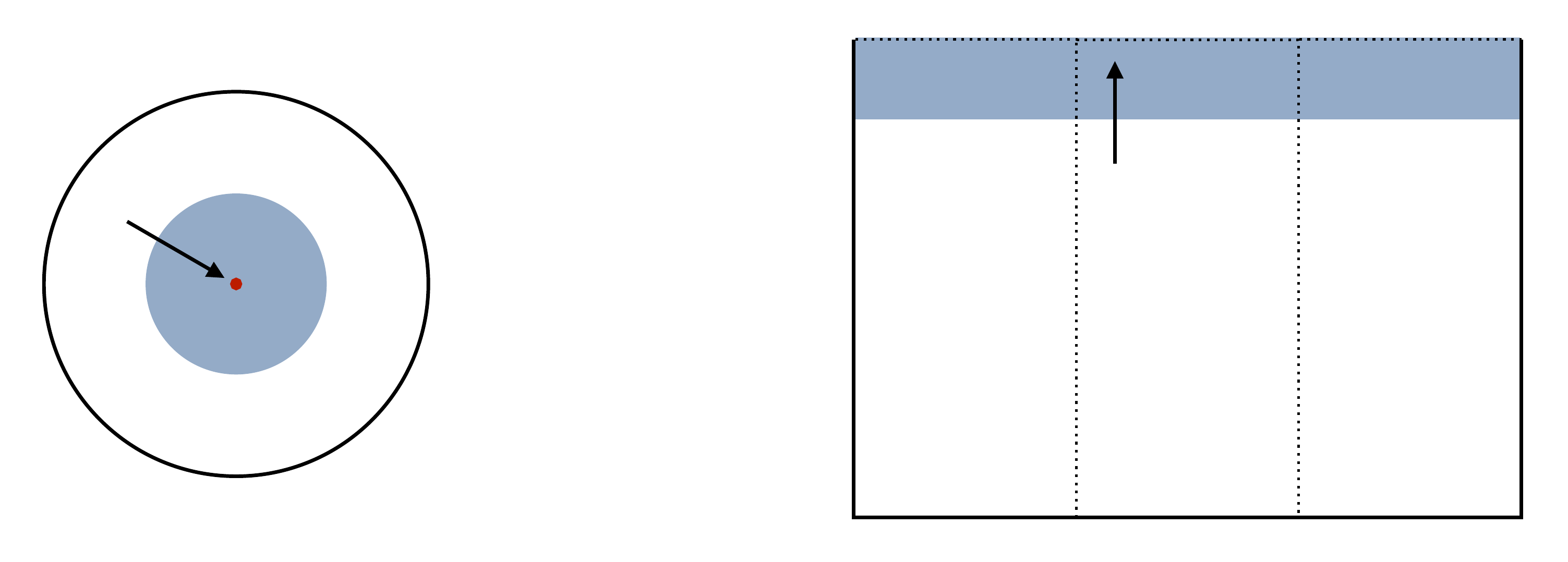}
		\vspace*{-1cm}
	\end{center}
	\begin{picture}(0,0)
	\put(70,82){\rotatebox{-31}{\small $z \rightarrow 0$}}
	\put(307,88){\small $y \rightarrow \infty$}
	\put(10,110){(a)}
	\put(200,110){(b)}
	\end{picture}
	
	\caption{Local patch in the space $\cM$ represented as Poincar\'e disc, Figure (a), and upper half-plane, Figure (b). 
		The asymptotic solutions  are expanded near the boundary at $z=0$ in the disc, which is mapped to 
		$t = i \infty$ on the upper half plane. \label{fig:disc}}
	
\end{figure}

In the following we will consider a local patch in $\cM$ that can be represented by 
the punctured Poincar\'e unit disc or its universal cover, the upper half plane. We will 
be interested in determining solutions near the puncture of the disc, which is mapped 
to imaginary infinity in the upper half plane, as depicted in Figure \ref{fig:disc}. 
Working on the disc we pick complex coordinates $z,\bar z$, such that $0<|z|\leq 1$, with the puncture 
located at $z=0$. The upper half plane covering the disc can be parametrized 
by introducing coordinates $t=\frac{1}{2\pi i} \log z = x + i y$. 
The metric on the upper half plane can be taken 
to be the standard Poincar\'e metric given by 
\beq  \label{Poincare-metric}
g^{\text{\tiny Poincar\'e}}_{\alpha \beta}\,  d\sigma^\alpha d\sigma^\beta = \frac{1}{y^2} \big[(d x)^2 + (dy)^2 \big]\ , \qquad \sigma^1=x, \ \sigma^2 = y\ . 
\eeq
We note that the precise form of this metric will be relevant when attempting to couple the matter 
theory \eqref{Smat_2} to a two-dimensional gravity theory that is not Weyl invariant. 

Two proposals of coupling of the matter action \eqref{Smat_2} to a gravity theory have been studied in \cite{Cecotti:2020uek} and \cite{Grimm:2020cda}. In both constructions a 
special metric on the moduli space, the so-called Hodge metric, played a central role. To see how this metric 
arises from our general perspective, let us note that one can construct a natural metric $g_{\alpha\beta}(J_L)$ on $\cM$ using the conserved currents \eqref{eq:current} as 
\begin{equation}\label{eq:hodgemetric}
g_{\alpha\beta}(J_L) d\sigma^\alpha d\sigma^\beta= \mathrm{Tr}\left[ (J_L)_\alpha (J_L)_\beta \right]d\sigma^\alpha d\sigma^\beta \ ,
\eeq
where $(J_L)_\alpha$ denotes the $\alpha$-component of the 1-form $J_L$.
We will show in section \ref{ssec:reconstructmetric} that one can evaluate this metric on a solution of the bulk matter theory and find the general asymptotic form 
\beq \label{eq:metricexpand}
g_{\alpha\beta}(J_L^{\rm sol}) d\sigma^\alpha d\sigma^\beta=  \frac{1}{y^2}\left( c^{(0)}+ \frac{c^{(1)}}{y}+\frac{c^{(2)}}{y^2}+\cdots\right) \big[(d x)^2 + (dy)^2 \big]\ .
\end{equation}
We find that for all our solutions $ c^{(0)} \neq 0$, such that the metric has the leading asymptotic 
behaviour of the Poincar\'e metric \eqref{Poincare-metric}. In fact, we will see that concrete expressions for the coefficients $c^{(i)}$ can be derived in terms of the boundary data in \eqref{eq:ci}. It was argued in \cite{Cecotti:2020uek,Grimm:2020cda} that in geometric settings the metric $g_{\alpha\beta}(J_L^{\rm sol}) $ coincides with the Hodge metric. 

Let us now return to analyzing the near-boundary solutions for the bulk theory we have introduced. Evaluating the equations of motion of the action \eqref{Smat_2} in a metric of the form \eqref{conformally-flat}
yields the equation
\beq \label{eom_in_flat_gauge}
\sum_\alpha \partial_{\sigma^\alpha}  \big(  h^{-1}  \partial_{\sigma^\alpha}   h + ( h^{-1}  \partial_{\sigma^\alpha}   h)^\dagger \big) - \big[(  h^{-1} \partial_{\sigma^\alpha}   h)^\dagger,   h^{-1} \partial_{\sigma^\alpha}   h \big] = 0\ .
\eeq
Our aim is to determine solutions to this equation near the boundary $y=\infty$. 
The class of solutions that we determine is constrained by a number of conditions. Firstly, we 
note that $h(z,\bar z)$ can transform under a monodromy transformation when encircling the center 
of the disc. More precisely, such a transformation would rotate
\beq
h(z,\bar z) \ \rightarrow \ T h(z,\bar z)\ , 
\eeq
when sending $z \rightarrow e^{2\pi i }z$. 
Here $T$ is an element of the monodromy group $\Gamma \subset G_{\mathbb{R}}$ of $\cM$.  
The fields $h(z,\bar z)$ are thus single-valued if we work on the quotient $G_{\bbR}/\Gamma$.
In the following we will assume that $T$ is unipotent, i.e.~that it 
can be written as 
\beq \label{nil_mod}
T = e^{N^-}\ , 
\eeq
where $N^-$ is a nilpotent matrix $(N^-)^k = 0$, $k>D$. When formulated in terms of the coordinates 
$t = x + i y$ on the covering space of $\cM$, the matter fields $h$ thus behave as 
\beq \label{discrete_sym}
h(x+ 1,y) = e^{N^-} h(x, y)\ .
\eeq
Clearly, we can obtain this transformation behaviour by setting $h(x,y) = e^{xN^-} \tilde h(x,y)$ with $\tilde h(x,y)$ being a periodic function invariant under the shift $x \rightarrow x+1$. In the following we will require that sufficiently close to the puncture the 
discrete symmetry \eqref{discrete_sym} becomes continuous, i.e.~that  $x \rightarrow x + c$ for any constant $c$ 
is a symmetry of the action \eqref{Smat_2}. Together with \eqref{discrete_sym} this implies that $h$ is of the form 
\beq \label{h-split-form}
h(x,y) = e^{x N^-} \tilde h(y)\ .
\eeq
Solving the equations of motions \eqref{eom_in_flat_gauge} we will thus have to solely determine the $G_\bbR$-valued function $\tilde h(y)$.

Finding solutions of \eqref{eom_in_flat_gauge} that take the form \eqref{h-split-form} is still a notoriously difficult task. One additional constraint 
that we will impose is a compatibility condition of such solutions with $Q$. Concretely, we will demand that 
\beq \label{Q-constraint}
-2\big[Q,  h^{-1} \partial_{y}   h\big]= i \Big( (  h^{-1} \partial_{x}   h)^\dagger +    h^{-1} \partial_{x}   h \Big) \ , \qquad     \big[Q,  h^{-1} \partial_{x}   h \big] =i  h^{-1} \partial_{y}   h \ . 
\eeq 
These conditions imply that $ (  h^{-1} \partial_{y}   h)^\dagger =   h^{-1} \partial_{y}  h $.

We are now in the position to summarize the equations that we will solve recursively in the remainder of this paper. First, we introduce the following shorthand notation
\beq  \label{def-cN}
\cN^0(y) := - 2   h^{-1}  \partial_y   h\ ,  \qquad \cN^-(y)  :=    h^{-1}  \partial_x   h \ ,\qquad
\cN^+(y)  := (\cN^-(y))^\dagger\ .
\eeq
It is then straightforward to check that \eqref{eom_in_flat_gauge} together with \eqref{Q-constraint} imply the 
differential equations 
\beq \label{Nahm_eq}
\partial_{y} \cN^{\pm} = \pm \tfrac{1}{2} [\cN^{\pm},\cN^0] \ , \qquad  
\partial_{y} \cN^0 =- [\cN^{+} ,\cN^{-}]\ ,
\eeq
and the algebraic conditions 
\beq \label{Q-constr_2}
\big[Q,\cN^{0}\big]= i  (\cN^+ + \cN^-) \ , \qquad     \big[Q,\cN^{\pm}\big] =-\tfrac{i}{2} \cN^0 \ . 
\eeq
Note that the differential equations \eqref{Nahm_eq} are also known as Nahm's equations. The two equations on the left are automatically solved for $\cN^0,\cN^\pm$ of the form \eqref{def-cN}, while the equation on the right 
corresponds to the equations of motion \eqref{eom_in_flat_gauge}.

\section{The boundary data}
\label{sec:boundary}

In this work we will consider solutions to the equations \eqref{Nahm_eq}, \eqref{Q-constr_2} and determine bulk fields that are of the 
general form \eqref{h-split-form}. We note, however, that these conditions do not fix the bulk solutions entirely. In section \ref{sec:boundary_theory} a set of boundary data is introduced, which fixes the solution uniquely upon imposing a single matching condition \eqref{zetadelta-eq} as is explained in section \ref{Bulk-reconstruction}. We then provide a classification of this boundary data  in section \ref{boundary_class} for boundaries of Calabi--Yau type with $D=3$. This classification applies to moduli spaces of any dimension, and when restricting to two-dimensional moduli spaces there will remain only three non-trivial possibilities, which we call type $\mathrm{I}_1$, $\mathrm{II}_0$ and $\mathrm{IV}_1$. We construct the most general set of boundary data for these three types of boundaries in section \ref{ssec:data}, which will be the starting point for the examples we study later. 

\subsection{The sl(2) boundary data}
\label{sec:boundary_theory}
We start by discussing the general aspects of the boundary data. The following structures can be extracted from the Hodge theory constructions of \cite{Schmid,CKS}. We follow here the presentation of \cite{Grimm:2020cda}, which states 
the relevant information without reference to Hodge structures. From the mathematical perspective the following presentation uses ideas 
put forward in \cite{robles_2016}. To begin with we require that the vector space $\cH$ also admits a splitting into vector spaces $\cH^\infty_q$ that is attached to the boundary at $y=\infty$. We encode this splitting by a boundary charge operator $Q_\infty\in i\mathfrak{g}_{\mathbb{R}}$
\begin{equation}\label{eq:Qdecomp}
|w\rangle \in \cH^\infty_q:\quad Q_\infty |w\rangle = q |w\rangle\ .
\end{equation}
In order to find well-behaved solutions $h$ that match this boundary splitting, we require that $Q_\infty$ is 
at most a rotated version of the charge operator $Q$ introduced in \eqref{ref-Q}. Concretely, we require that there exists a 
$\zeta \in \mathfrak{g}_{\mathbb{R}}$ such that $Q = e^{-\zeta} Q_\infty e^{\zeta}$.
From $Q_\infty$ one may again construct an inner product via
\begin{equation}
\langle v | w \rangle_\infty := \langle \bar{v}, e^{i\pi Q_\infty} w\rangle\ .
\end{equation}
which is simply a rotated variant of the inner product introduced in \eqref{def-innerproduct}. 

The second structure that we require to exist on the boundary is an $\slt$-algebra of operators acting on $\cH$. Recall that we associated a nilpotent matrix $N^-$ to the puncture in \eqref{nil_mod}. Also note that there are many ways to pick two other generators $N^+$, $N^0$ such that 
the triple $N^\pm, N^0\in\mathfrak{g}_{\mathbb{R}}$ commute as $\slt$ generators. The crucial point here is that we consider a choice that 
is compatible with the splitting induced by $Q_\infty$
We will formulate the properties of this algebra both in the real and complex setting. The real algebra $\mathfrak{sl}(2,\mathbb{R})$ is generated by $N^\pm, N^0\in\mathfrak{g}_{\mathbb{R}}$ and the complex algebra is generated by $L_{\pm 1}, L_0\in\mathfrak{g}_{\mathbb{C}}$, satisfying the commutation relations
\begin{equation} \label{real-sl2alg}
[N^0, N^\pm] = \pm 2N^\pm, \quad [N^+, N^-]=N^0,
\end{equation}
and
\begin{equation}
[L^0, L_{\pm 1}] =\pm  2L_{\pm 1}\, , \quad [L_{1}, L_{-1}]=L_0\, .
\end{equation}
The two sets of generators are related by
\begin{equation}\label{eq:Ldef}
L_{\pm 1} =\frac{1}{2}\left(N^+ + N^- \mp i N^0 \right),\quad L_0 = i\left(N^- - N^+\right).
\end{equation}
It turns out that there is another way to describe the relation between $N^0,N^\pm$ and $L_0,L_{\pm 1}$, namely by 
a transformation with a group element $\rho \in G_\bbC$ defined as
\begin{equation} \label{def-rho}
\rho=\mathrm{exp}\Big[\frac{i\pi}{4}\big(N^+ + N^-\big)\Big]=\mathrm{exp}\Big[\frac{i\pi}{4}\left(L_{1} + L_{-1}\right)\Big]\ .
\end{equation}
It is not hard to check\footnote{Here one can use the identity for the adjoint actions $\text{Ad}_{\text{exp} \cO} =  \text{exp}(\text{ad}_\cO)  $.} that $\rho$ allows us to move between the real and complex algebra via
\begin{equation}
\label{eq:rho_switch}
L_{\pm 1} = \rho N^\pm \rho^{-1}\, ,\quad L_{0} = \rho N^0 \rho^{-1}\, .
\end{equation}
The compatibility with $Q_\infty$ can be formulated as
\begin{equation}
[Q_\infty, N^0] = i\left(N^++N^-\right), \quad [Q_\infty, N^\pm] = -\frac{i}{2}N^0\, ,
\end{equation}
or
\begin{equation}
[Q_\infty, L_\alpha] =\alpha L_\alpha \,,\quad \alpha = \pm 1,  0 \, .
\end{equation}
A crucial observation is that $L_0$ commutes with $Q_\infty$, which allows us to find a common eigenbasis for the two operators. For this reason we will mostly work with the complex algebra. There is, of course, one more operator which commutes with both $L^0$ and $Q_\infty$, namely the Casimir operator $L^2$. It is given by
\begin{equation} \label{def-Casimir}
L^2=2L_{1}L_{-1} + 2L_{-1} L_{1} +(L_0)^2 \, .
\end{equation}
Let us note that there is another way of interpreting the operators $L^\alpha, Q_\infty$. In fact, we see that 
$\hat Q \equiv  Q_\infty - \frac{1}{2} L_0$ commutes with all $L_{\alpha}$ and hence we have the algebra 
\beq
\mathfrak{sl}(2,\bbR) \oplus \mathfrak{u}(1): \ L^\alpha, \hat Q\ . 
\eeq
in this work we prefer to work with the charge operator $Q_\infty$ instead of $\hat{Q}$, but note that $\hat{Q}$ does appear naturally in the bulk reconstruction in e.g.~\eqref{charges_cL}.

Much of our work revolves around solving operator equations, hence it will be extremely useful to split the space of operators using the eigenvalues of $L^2, L^0$ and $Q_\infty$. Concretely, given any operator $\cO $ we may decompose it as
\beq \label{weight-charge-expansion-operator1}
\cO = \sum_{0 \leq d \leq D} \sum_{-d\leq s\leq d} \sum_{-D\leq q \leq D} \cO^{(d,s)}_q \ , 
\eeq
with 
\bea \label{weight-charge-expansion-operator2}
(\ad L)^2 \cO^{(d,s)}_q &=& d(d+2)\, \cO^{(d,s)}_q  \ , \nn \\
\big[L_0,\cO^{(d,s)}_q\big] &=&  s\, \cO^{(d,s)}_q\ ,  \\
\big[Q_\infty ,\cO^{(d,s)}_q \big] &=& q\, \cO^{(d,s)}_q\ ,  \nn 
\eea
where we have used the shorthand notation $(\ad L)^2$ to denote replacing each left-multiplication in \eqref{def-Casimir} with an 
adjoint action, i.e.~we have set
\beq
(\ad L)^2 \cO :=2\big[L_{1},\big[L_{-1},\cO \big]\big]+2\big[L_{-1},\big[L_{1},\cO \big]\big]+\big[L_0,\big[L_0,\cO \big]\big]\ .
\eeq
We call $d, s$ and $q$ the \textit{highest weight}, \textit{weight} and \textit{charge} of the operator, respectively. 
In the following, it is sometimes not necessary to perform all three decompositions \eqref{weight-charge-expansion-operator2}.
We will then employ the notation
\beq \label{weightonly}
\cO^{(l)}_q \equiv  \sum_{d \in \bbZ} \cO^{(d,l)}_q\ ,\qquad \cO^{(d,l)} \equiv  \sum_{q \in \bbZ} \cO^{(d,l)}_q \ , \qquad \cO^{[d]} \equiv  \sum_{q,l \in \bbZ} \cO^{(d,l)}_q \ ,
\eeq
when we do not perform the highest weight decomposition, the charge decomposition, or only perform the highest weight decomposition, respectively.

Note that in the case we do not perform a charge decomposition, one could also have chosen to perform the highest weight and weight decomposition with respect to the real $\slt$-algebra \eqref{real-sl2alg}. 
These two decompositions are related precisely by $\rho$ introduced in \eqref{def-rho}. In other words, if $\cO$ is an operator with weight $s$ under $N^0$, then $\hat{\cO}=\rho\cO\rho^{-1}$ is an operator with weight $s$ under $L^0$, and similarly for the highest weight. In the following we will add a hat to an operator if it is obtained via the transformation with $\rho$. This is particularly relevant if $\cO$ is a real operator. Such operators are naturally decomposed with respect to the real $\slt$-algebra \eqref{real-sl2alg} and cannot be an eigenoperator under $Q_\infty$, unless they are uncharged.  

The last boundary operator that we need to introduce is a real operator $\delta \in \mathfrak{g}_{\mathbb{R}}$. It is an essential part of the boundary data and determines crucially the complexity of the associated bulk solutions. We will call this operator the \textit{phase operator} as in ref.~\cite{Grimm:2020cda}. In accordance with the notation introduced above, we use $\delta$ to define
\begin{equation}
\hat{\delta} = \rho\delta\rho^{-1} \ ,
\end{equation} 
which is thus an element of $ \mathfrak{g}_{\bbC}$.
The phase operator $\delta$ or $\hat{\delta}$ is required to have a number of special properties such that it can be part of a consistent set of boundary data. 
Firstly, it has to commute with $L_{-1}$, i.e.~we require
\begin{equation}
\label{eq:delta_Lmin}
[L_{-1},\hat{\delta}]=0\ .
\end{equation}
Secondly, its components all have weight less than or equal to $-2$, and charge less than or equal to $-1$. In other words, we require that 
it admits an expansion
\begin{equation}\label{eq:deltaexpansion}
\hat{\delta} = \sum_{s\leq -2} \sum_{q \leq -1} \hat{\delta}^{(s)}_q\ .
\end{equation}
We will show in section \ref{Bulk-reconstruction} that these properties of the phase operator $\delta$ suffice to fix the associated bulk solution uniquely 
after imposing one matching condition. 

\subsection{Classifying boundary data} \label{boundary_class}
The data we have just introduced can now be used to classify the boundary data. Recall that a given boundary is characterized by an $\mathfrak{sl}(2,\mathbb{C})$-triple $(L_1,L_0,L_{-1})$, a charge operator $Q_\infty$ and a phase operator $\delta$. Since the boundary operators lie in the algebra of $G_{\bbR}$, it is important to stress the dependency of $G_{\bbR}$ on the choice of the bilinear $\langle \cdot, \cdot \rangle$. Moreover, since $G_{\bbR}$ need not comprise the full symmetry group of the bilinear, there are many choices that can be made, which greatly complicates the classification. Nevertheless, a full classification of allowed groups $G_{\bbR}$ and boundary data has been given by Robles in \cite{robles_2016}. As part of this classification one has to classify the charge operator and $\slt$-triple based on the eigenvalues of the weight operator $L_0$, the Casimir $L^2$ and the charge operator $Q_\infty$. This classification then captures what spectra are allowed for the operators $L^2,L^0$ and $Q_\infty$ in order to obtain a consistent boundary theory. In other words, it tells us which splittings of the vector space $\cH$ into eigenspaces of these operators have to be considered, which allows us to systematically go through all possible cases.

Let us begin with the splitting of the vector space $\cH$
induced by the charge operator $Q_\infty$ into $D+1$ eigenspaces as in \eqref{eq:Qdecomp}, and gather the dimensions of these eigenspaces $\cH^\infty_q$ together as
\begin{equation} \label{H-dimq}
\dim \cH^\infty_q = (h_{D/2},\, h_{D/2-1}, \, \ldots,\, h_{-D/2+1},\, h_{-D/2} )\, ,
\end{equation}
These dimensions satisfy $h_{q}=h_{-q}$, since under complex conjugation the eigenspaces are related by $\overline{\cH}_{q}=\cH_{-q}$. 
The splitting of the vector space $\cH$ into eigenspaces $\cH^{\infty}_{q}$ is then refined by considering the $\mathfrak{sl}(2,\mathbb{C})$-triple $(L_1,L_0,L_{-1})$. The weight and Casimir operators $L_0,\, L^2 $ commute with the charge operator $Q_\infty$, so the vector space $\cH$ admits a common eigenbasis. We therefore decompose each eigenspace $\cH^\infty_q$ as
\begin{equation} 
\cH^\infty_q = \sum_{0\leq d \leq D} \sum_{-d \leq s \leq d} \left(\cH^\infty \right)_q^{(d,s)}\, , 
\end{equation}
where for an element $|d,s;q\rangle  \in \left(\cH^\infty \right)_q^{(d,s)}$ one has
\begin{equation}\begin{aligned}
L_0 |d,s;q\rangle  &= s |d,s;q\rangle\, , \\
L^2 |d,s;q\rangle &= d(d+2) |d,s;q \rangle\, , \\
Q_\infty |d,s;q\rangle  &= q |d,s;q \rangle \, . \\
\end{aligned}\end{equation}
One can now use this refined splitting to classify the boundary data. While we will present in section \ref{Bulk-reconstruction} the general formalism to solve the bulk theory for any boundary data, our explicit examples in section \ref{reconstructing_examples} will consist of data of a distinguished type, which we call being of Calabi--Yau type:\footnote{In this setting the $h_q$ are related to the Hodge numbers $h^{p,q}=\mathrm{dim}_\mathbb{C}\; H^{p,q}(Y_D,\mathbb{C})$ by $h^{p,q}=h_{(p-q)/2}$.}
\beq \label{CYtype-cond}
\text{Calabi--Yau type:} \quad h_{D/2}=h_{-D/2}=1\, , \qquad G_{\bbC}=
\begin{cases}
	\mathrm{Sp}(\dim\cH, \bbC)\, ,& D \,\mathrm{ odd }\, ,\\
	\mathrm{SO}(\dim\cH, \bbC)\, , & D\, \mathrm{ even }\, .
\end{cases}
\eeq
From a geometric perspective the first condition is related to the fact that there exists a unique holomorphic $(D,0)$-form on a Calabi--Yau manifold
of complex dimension $D$. The second condition implies that $G_{\bbR}$ comprises the full symmetry group of the bilinear form $\langle\cdot,\cdot\rangle$, which is (skew-)symmetric for (odd) even $D$. In particular, we will consider the case $D=3$. For this reason it will be sufficient to review only part of the classification, restricting to the case where we have dimensions $(1,h_{1/2},h_{1/2},1)$ and group $G_{\bbR}=\mathrm{Sp}(2+2h_{1/2},\bbR)$ as in \cite{Kerr2017,Grimm:2018cpv}. 

Based on the eigenvalue of $\cH^\infty_{3/2}$ under $L_0$ we separate into four principal types by
\begin{equation}\label{eq:principaltypes}
\left| \hat{d}, \hat{d} ; \tfrac{3}{2} \right\rangle \in  \cH^\infty_{3/2}\ , \qquad \begin{cases}
\mathrm{I}&: \qquad \hat{d}=0\, ,\\
\mathrm{II}&: \qquad \hat{d}=1\, ,\\
\mathrm{III}&: \qquad \hat{d}=2\, , \\
\mathrm{IV}&: \qquad \hat{d}=3\, .
\end{cases}
\end{equation}
Alternatively this condition can be phrased as that the lowering operator $L_{-1}$ (or $N^-$) can act at most $\hat{d}$ times on an element in the $q=3/2$ eigenspace before vanishing, since it is part of a $(\hat{d}+1)$-dimensional irreducible representation. Note that $\hat{d}$ is at most equal to three, since applying $L_{-1}$ also lowers the charge of the state, and $q\geq -\frac{3}{2}$ when $D=3$. 

In the same spirit we can characterize how the $q=1/2$ eigenspace decomposes under the operators $L^2$ and $L_0$. Note that the decomposition of the $q=-1/2$ and $q=-3/2$ eigenspaces then follows by complex conjugation. We can encode the splitting of $\cH^\infty_{1/2}$ by a single integer $n$ defined as
\begin{equation}
\label{eq:secondary_type}
n = \sum_{0\leq d \leq 3} \dim \left(\cH^\infty \right)_{1/2}^{(d, 1)}.
\end{equation}
In other words, $n$ denotes the number of linearly independent states with weight $s=1$ and charge $q=1/2$. It turns out that the principal type and the integer $n$ together classify all types of boundary splittings of $\cH$, which we denote by\footnote{This classification naturally applies to limits in the complex structure moduli space of Calabi--Yau threefolds, and by using mirror symmetry it can be used to characterize limits in K\"ahler moduli spaces \cite{Grimm:2018cpv,Corvilain:2018lgw}. In turn, it was suggested in \cite{Grimm:2019bey} that the limit types on the K\"ahler cone can be combined into graphs to classify the Calabi--Yau threefolds themselves.} 
\begin{equation}
\mathrm{I}_n\, , \quad \mathrm{II}_n\, , \quad \mathrm{III}_n \, ,  \quad \mathrm{IV}_n\, .
\end{equation} 
Here the range for the index $n$ changes per principal type and depends on the value of $h_{1/2}$. For example, for type $\mathrm{IV}_n$ there is at least one vector with weight $s=1$ and charge $q=1/2$, namely $L_{-1}|3, 3;3/2\rangle$,  hence $n\geq 1$. On the other hand, $n$ cannot exceed the dimension of $\cH^\infty_{1/2}$, so $n\leq h_{1/2}$. One can consider the other principal types in a similar fashion to arrive at the following ranges
\begin{align} \label{general-types_3}
\mathrm{I}_n&:\quad 0\leq\; n \leq h_{1/2}\, , &
\mathrm{II}_n&:\quad 0\leq \;n \leq h_{1/2}-1 \, ,& \\
\mathrm{III}_n&:\quad 0\leq \;n \leq h_{1/2}-2 \, ,&
\mathrm{IV}_n&:\quad 1\leq \;n\leq h_{1/2}\ . &
\end{align}
To close our discussion, let us point out that the phase operator $\delta$ is not captured by the classification of the boundary data that we have just reviewed. 
While it is a crucial element of the set of boundary data, it is not involved in the splitting of the vector space $\cH$ into eigenspaces under the operators $Q_\infty, L^2, L_0$. Given this data, the phase operators $\delta$ then has to be an operator acting on this splitting that satisfies \eqref{eq:delta_Lmin} and \eqref{eq:deltaexpansion}. In the next subsection we discuss the simplest situation with $h_{1/2} = 1$ in which the phase operators can 
be parameterized easily. 

\subsection{Classifying phase operators in simple Calabi--Yau type settings}
\label{ssec:data}

Having briefly reviewed the classification of the boundary data associated to $Q_\infty, L^2, L_0$, we now want to have a closer look 
at a certain set of special cases. More precisely, we will consider boundary data of Calabi--Yau type and weight $D=3$, which 
also satisfies the condition $h_{1/2} = 1$. This latter assumption will be particularly useful when identifying the allowed phase operators 
$\delta$ compatible with the algebra obtained from $Q_\infty$, $L_\alpha$. 
It follows from \eqref{general-types_3} that there are only three possible non-trivial types, namely
\begin{equation} \label{simple-types}
\quad h_{3/2}=h_{1/2}=1:\quad \mathrm{I}_1\, ,\quad \mathrm{II}_0\, ,\quad \mathrm{IV}_1 \ . 
\end{equation}
In a geometric realization these boundary data can arise in the complex one-dimensional moduli space of a Calabi--Yau threefold. 
The boundary types respectively cover the conifold point, the Tyurin degeneration, and the large complex structure point as we will 
see in detail in section \ref{reconstructing_examples}. We exclude type $\mathrm{I}_0$ because it corresponds to the trivial representation where $L_{-1}=0$.

In the following we explicitly determine the boundary data associated to the types \eqref{simple-types}. The reader can find the result in table \ref{table:data}. To begin with, recall that for the three boundary types under consideration the group $G_{\mathbb{R}}$ is given by the symplectic group $\mathrm{Sp}(4,\mathbb{R})$. Its Lie algebra $\mathfrak{sp}(4,\mathbb{R})$ has ten generators, which satisfy
\begin{equation}
X\in\mathfrak{sp}(4,\mathbb{R}):\quad X^TS+SX = 0\, ,
\end{equation}
where $S$ represents the symplectic product $\langle \cdot, \cdot \rangle $ on $\cH$ by $\langle v,w\rangle = v^T Sw$.
In particular, all boundary operators, such as the $\slt$-triple, will be represented by $4\times 4$ matrices in $\mathfrak{sp}(4,\mathbb{R})$ or its complexification $\mathfrak{sp}(4,\mathbb{C})$. Note that in table \ref{table:data} we have chosen a different realization of $S$ for each type, simply for convenience.

\begin{table}
	\centering
	\begin{tabular}{ | c | c | c | c |} \hline
		operator &$\mathrm{I}_{1}$ & $\mathrm{II}_{0}$ & $\mathrm{IV}_{1}$ \\ \hline \hline
		$N^+$ &  \rule[-1.05cm]{.0cm}{2.3cm} $\begin{pmatrix}
		0 & 0 & 0 & 0 \\
		0 & 0 & 1 & 0 \\
		0 & 0 & 0 & 0 \\
		0 & 0 & 0 & 0 \\
		\end{pmatrix}$  & $\begin{pmatrix}
		0 & 0 & 1 & 0 \\
		0 & 0 & 0 & 1 \\
		0 & 0 & 0 & 0 \\
		0 & 0 & 0 & 0 \\
		\end{pmatrix}$ & $\begin{pmatrix}
		0 & 3 & 0 & 0 \\
		0 & 0 & 2 & 0 \\
		0 & 0 & 0 & 1 \\
		0 & 0 & 0 & 0 \\
		\end{pmatrix}$ \\ \hline
		$N^0$ & \rule[-1.05cm]{.0cm}{2.3cm} $\begin{pmatrix}
		0 & 0 & 0 & 0 \\
		0 & 1 & 0 & 0 \\
		0 & 0 & -1 & 0 \\
		0 & 0 & 0 & 0 \\
		\end{pmatrix}$ & $\begin{pmatrix}
		1 & 0 & 0 & 0 \\
		0 & 1 & 0 & 0 \\
		0 & 0 & -1 & 0 \\
		0 & 0 & 0 & -1 \\
		\end{pmatrix}$ &  $\begin{pmatrix}
		3 & 0 & 0 & 0 \\
		0 & 1 & 0 & 0 \\
		0 & 0 & -1 & 0 \\
		0 & 0 & 0 & 3 \\
		\end{pmatrix}$ \\ \hline
		$N^-$ & \rule[-1.05cm]{.0cm}{2.3cm} $\begin{pmatrix}
		0 & 0 & 0 & 0 \\
		0 & 0 & 0 & 0 \\
		0 & 1 & 0 & 0 \\
		0 & 0 & 0 & 0 
		\end{pmatrix}$  & $\begin{pmatrix}
		0 & 0 & 0 & 0 \\
		0 & 0 & 0 & 0 \\
		1 & 0 & 0 & 0 \\
		0 & 1 & 0 & 0 \\
		\end{pmatrix}$ &   $\begin{pmatrix}
		0 & 0 & 0 & 0 \\
		1 & 0 & 0 & 0 \\
		0 & 2 & 0 & 0 \\
		0 & 0 & 3 & 0 
		\end{pmatrix}$  \\ \hline
		$Q_{\infty}$ &  \rule[-1.05cm]{.0cm}{2.3cm} $\begin{pmatrix}
		0 & 0 & 0 & -\frac{3 i}{2} \\
		0 & 0 & -\frac{i}{2} & 0 \\
		0 & \frac{i}{2}  & 0 & 0 \\
		\frac{3 i}{2} & 0 & 0 & 0 \\
		\end{pmatrix}$ & $\begin{pmatrix}
		0 & -i & -\frac{i}{2} & 0 \\
		i & 0 & 0 & -\frac{i}{2} \\
		\frac{i}{2} & 0 & 0 & -i \\
		0 & \frac{i}{2} & i & 0 \\
		\end{pmatrix}$ & $\begin{pmatrix}
		0 & -\frac{3 i}{2} & 0 & 0 \\
		\frac{i}{2} & 0 & -i & 0 \\
		0 & i & 0 & -\frac{i}{2} \\
		0 & 0 & \frac{3 i}{2} & 0 \\
		\end{pmatrix}$ \\ \hline
		$\delta$ & \rule[-1.05cm]{.0cm}{2.3cm}  $\begin{pmatrix}
		0 & 0 & 0 & 0 \\
		0 & 0 & 0 & 0 \\
		0 & - c & 0 & 0 \\
		0 & 0 & 0 & 0 
		\end{pmatrix}$ & $\begin{pmatrix}
		0 & 0 & 0 & 0 \\
		0 & 0 & 0 & 0 \\
		-c & 0 & 0 & 0 \\
		0 & -c & 0 & 0 \\
		\end{pmatrix}$ & $\begin{pmatrix}
		0 & 0 & 0 & 0 \\
		0 & 0 & 0 & 0 \\
		0 & 0 & 0 & 0 \\
		\chi & 0 & 0 & 0 
		\end{pmatrix}$\\ \hline
		$S$ & \rule[-1.05cm]{.0cm}{2.3cm}  $\begin{pmatrix} 0 & 0 & 0 & 1\\
		0 & 0 & -1 & 0\\
		0 & 1 & 0 & 0\\
		-1 & 0 & 0 & 0 \end{pmatrix} $& $\begin{pmatrix}
		0 & 0 & 1 & 0 \\
		0 & 0 & 0 & 1 \\
		-1 & 0 & 0 & 0 \\
		0 & -1 & 0 & 0
		\end{pmatrix}$ & $\begin{pmatrix}
		0 & 0 & 0 & -1 \\
		0 & 0 & 3 & 0 \\
		0 & -3 & 0 & 0 \\
		1 & 0 & 0 & 0 \\
		\end{pmatrix}$ \\ \hline
	\end{tabular}
	\caption{\label{table:data} Characteristic data for the three possible kinds of boundaries of Calabi--Yau type with weight $3$ in two-dimensional (complex one-dimensional) moduli spaces. }
\end{table}

Let us now outline how the $\slt$-triples are constructed. The idea is to determine the irreducible $\slt$-representations fixed by the boundary type. Given this set of states, we can then simply write down generators that produce the corresponding spectrum. For the one-modulus boundaries we find
\begin{equation}\label{eq:spectra}
\begin{aligned}
\mathrm{I}_{1} &: \qquad \left|0, 0 ; \tfrac{3}{2}\right\rangle\, , \ \  \left|1,1;\tfrac{1}{2}\right\rangle\, , \ \ \left|1,-1;-\tfrac{1}{2}\right\rangle\, , \ \ 
\left|0, 0 ; -\tfrac{3}{2}\right\rangle \, , \\
\mathrm{II}_{0} &: \qquad \left|1, 1 ; \tfrac{3}{2}\right\rangle\, , \ \  \left|1,-1;\tfrac{1}{2}\right\rangle\, , \ \  \left|1,1;-\tfrac{1}{2}\right\rangle\, , \ \ 
\left|1, -1 ; -\tfrac{3}{2}\right\rangle \, , \\
\mathrm{IV}_{1} &: \qquad \left|3, 3 ; \tfrac{3}{2}\right\rangle\, , \  \ \left|3,1;\tfrac{1}{2}\right\rangle\, , \  \ \left|3,-1;-\tfrac{1}{2}\right\rangle\, , \ \ 
\left|3,-3 ; -\tfrac{3}{2}\right\rangle \, . \\
\end{aligned}
\end{equation}
These spectra are obtained as follows. The presence of the states $|\hat{d},\hat{d};3/2\rangle$ and their descendants is given by \eqref{eq:principaltypes}. Under complex conjugation states $|d,s;q\rangle$ are mapped to $|d,-s;-q\rangle$, resulting in an additional set of states for $\mathrm{I}_{1}$ and $\mathrm{II}_{0}$ boundaries related to $|\hat{d},-\hat{d};-3/2\rangle$. For $\mathrm{II}_{0}$ and $\mathrm{IV}_{1}$ boundaries this information is already sufficient to fix their spectra. For $\mathrm{I}_{1}$ boundaries we infer from \eqref{eq:secondary_type} the presence of the states $|1,1;1/2\rangle$ and $|1,1;-1/2\rangle$ by the fact that $n=1$, which completes the spectrum. One can then take a simple form for the $\mathfrak{sl}(2,\mathbb{R})$-triple that has the right set of irreducible $\slt$-representations.  
By switching to the $\mathfrak{sl}(2,\mathbb{C})$-triple via \eqref{eq:rho_switch} and matching the charges according to \eqref{eq:spectra} the generators $(L_{1},L_{0},L_{-1})$ and the charge operator $Q_{\infty}$ can then be constructed as well. 

Finally we have to construct the phase operator $\delta$ for the boundaries, so let us briefly recall its properties. The operator $\delta$ needs to be a real map, an infinitesimal isometry of the symplectic pairing $S$ and commute with $N^{-}$. Furthermore, when we rotate to the complex basis by the transformation $\rho$ and expand into eigenoperators of $Q_{\infty}$ and $L_{0}$ it needs to have charge $1 \leq -q \leq 3$ and weight $1+q \leq -s \leq 4+q$. For $\mathrm{I}_{1}$ and $\mathrm{II}_{0}$ boundaries the only map that satisfies these criteria is the lowering operator $N^{-}$ itself, so the phase operator $\delta$ is fixed up to an overall proportionality constant $c$. For $\mathrm{IV}_{1}$ boundaries there are two independent maps, either $N^{-}$ or the map given for $\delta$ in table \ref{table:data}. For convenience we have set the component along $N^{-}$ to zero. This component simply amounts to a coordinate shift $y \to y-c$ for the bulk fields, as will be demonstrated by the bulk reconstruction for the $\mathrm{I}_{1}$ and $\mathrm{II}_{0}$ boundaries.

%%%%%%%%%%%%%%%%%%%%%%%%%%%%%%%%%%%%%%%%%%%%%%%%%%%%%%
\section{Bulk reconstruction and the CKS recursion} \label{Bulk-reconstruction}

As announced in section \ref{local-solutions}, we are interested  in determining the bulk matter fields $h$ that satisfy the bulk equations of motion \eqref{eom_in_flat_gauge}, 
admit the near-boundary split form \eqref{h-split-form}, and are restricted by the $Q$-constraint \eqref{Q-constraint}.  
In order to determine such solutions we recall from equation \eqref{def-cN} that they can 
be used to define three real operators $\cN^0(y), \cN^\pm(y)$ that satisfy 
Nahm's equations \eqref{Nahm_eq} and the $Q$-constraint \eqref{Q-constr_2}.
In this section we will describe a procedure developed by Cattani, Kaplan and Schmid (CKS) to iteratively solve these equations
and then recover $h(x,y)$ \cite{CKS}. 
More precisely, we first show in section \ref{CKS-recursion} that after combining $\cN^0, \cN^\pm$ in a clever way one can translate 
Nahm's equations into an intricate set of recursion relations. We explain how the boundary data, namely the $\slt$-triple and the phase operator $\delta$, provide the initial conditions of the recursion. This recursion relation can then be solved, which yields a unique solution for $\cN^0(y), \cN^\pm(y)$. Having found a solution for $\cN^0(y), \cN^\pm(y)$ we will then 
describe in section \ref{subsec:h_reconstruction} how this leads to a solution for the bulk matter fields $h(x,y)$.
In this section 
we will show that such solutions take the form 
\beq\label{eq:hexpansion}
h(x,y) = e^{x N^-}  e^{\zeta} \Big( 1 + \frac{g_1}{y} + \frac{g_2}{y^2} + \ldots \Big)   y^{-\frac{1}{2}\tilde N^0} \ , 
\eeq
and explain how the matrices $\zeta$, $g_k$ and $\tilde N^0$ are determined in terms of the boundary data. 

\subsection{The CKS Recursion} \label{CKS-recursion}

In order to determine solutions $h(x,y)$ to the bulk theory, we first recall from \eqref{def-cN} that each such $h(x,y)$ allows us to 
define an $\cN^0(y)$ and $\cN^\pm(y)$, which obey the conditions 
\bea
\partial_{y} \cN^{\pm} &=& \pm \tfrac{1}{2} [\cN^{\pm},\cN^0] \ , \qquad  
\partial_{y} \cN^0 =- [\cN^{+} ,\cN^{-}]\ , \label{Nahm3}\\ 
\big[Q,\cN^{0}\big]&=& i  (\cN^+ + \cN^-) \ , \qquad     \big[Q,\cN^{\pm}\big] =-\tfrac{i}{2} \cN^0 \ , \label{Q_constr3}
\eea
as already given in \eqref{Nahm_eq} and \eqref{Q-constr_2}. 
Note that these equations are formulated in the real algebra $\mathfrak{g}_{\mathbb{R}}$ and adapted to
the reference charge operator $Q$ of the bulk theory introduced in section~\ref{bulk_matter_action}.  
In the following we will solve these equations for $\cN^0,\cN^\pm$ and identify solutions that match the boundary 
data. 

In order to match the bulk theory to a set of boundary data we now pick a $Q$ which is 
equivalent to $Q_\infty$ up to an adjoint transformation with an element of $G_\bbR$. More precisely, we 
introduce a $\zeta \in \mathfrak{g}_\bbR$ such that 
\beq \label{matchzeta}
Q_\infty = e^{\zeta} Q e^{-\zeta}\ . 
\eeq
Two comments are in order here. Firstly, we realize that this condition gives a 
consistency requirement on the $Q$ used in the bulk theory. Secondly, we will see 
later on that $\zeta$ is fixed via the boundary data for a given $\hat \delta$. 
As advocated in \cite{Grimm:2020cda} it will be convenient to also transform 
the $\cN^0,\cN^\pm$ with $e^\zeta$ and $\rho$ defined in~\eqref{def-rho}. The 
latter transformation allows us to work in the complex algebra, which is necessary 
to discuss eigenoperators under $Q_\infty$ as alluded to in section \ref{sec:boundary_theory}. 
Concretely, we define operators 
\begin{equation} \label{def-bfLbullet}
\mathbf{L}^\bullet = \rho e^\zeta \cN^\bullet e^{-\zeta}\rho^{-1}\ , 
\end{equation}
where $ \bullet $ stands for either $0$, $+$, or $-$. In terms of these operators the bulk equations of motion take the form
\begin{equation}
\label{eq:Nahm_L}
\partial_y \mathbf{L}^\pm = \pm \frac{1}{2}[\mathbf{L}^\pm, \mathbf{L}^0],\quad \partial_y\mathbf{L}^0 = -[\mathbf{L}^+,\mathbf{L}^-]\, .
\end{equation}
Moreover, the $Q$-constraint \eqref{Q_constr3} becomes 
\bea \label{charges_cL}
\big[2Q_\infty -  L_0, \mathbf{L}^{0}\big]&=& 2 i  ( \mathbf{L}^+ +  \mathbf{L}^-) +i \big[  L_1, \mathbf{L}^0\big]- i[L_{-1},  \mathbf{L}^{0}\big] \ ,  \nn \\
\big[2Q_\infty  -  L_0 , \mathbf{L}^{\pm}\big]   &=& -i  \mathbf{L}^0 + i \big[ L_1, \mathbf{L}^\pm\big] - i[L_{-1},  \mathbf{L}^{\pm}\big] \ .
\eea
Note that this $Q$-constraint does not appear in this form in \cite{CKS}. However, as was shown in \cite{Grimm:2020cda}, and will be recalled in appendix \ref{app:input}, this 
approach allows us to more easily impose it on the solution.

We now discuss the general procedure to solve \eqref{eq:Nahm_L} under the constraint \eqref{charges_cL}. Let us recall that the $\mathbf{L}^\bullet$ are operators which act on the finite-dimensional Hilbert space $\cH$ and may therefore be represented by matrices. Alternatively, one may pick a basis of $\mathfrak{g}_\bbC$ and represent each $\mathbf{L}^\bullet$ as a vector with respect to this basis. The main strategy of CKS is to solve Nahm's equations by combining the $\mathbf{L}^\bullet$ into a single vector $\Phi$ as
\begin{equation} \label{def-Phi_L}
\Phi = \begin{pmatrix}
\mathbf{L}^+\\ \mathbf{L}^0 \\ \mathbf{L}^-\\
\end{pmatrix}.
\end{equation}
We note that $\Phi$ can either be viewed as a 3-vector with matrices as entries, or as a $(3\times\mathrm{dim}\;\mathfrak{g})$-component vector. The former interpretation will be most useful for formal manipulations, whereas the latter will be more practical to use in concrete examples, as we will see in section \ref{reconstructing_examples}. The reason for introducing $\Phi$ is that one can construct another $\slt$-triple that acts on it, which allows one to perform further decompositions besides the ones for the separate $\mathbf{L}^\bullet$. We will introduce this triple shortly. First, in order to write down Nahm's equations in terms of $\Phi$, we introduce a bilinear $B$ acting on two vectors $\Phi$ and $\tilde{\Phi}$ as\footnote{In \cite{CKS} the notation $Q$ is used for the bilinear $B$, whose expressions differ by a choice of basis. Our basis is the same as in \cite{Pearlstein2006}.}
\begin{equation}
\label{eq:bilinear_B}
B(\Phi,\tilde{\Phi}) =\frac{1}{4} \begin{pmatrix}
[\mathbf{L}^0,\tilde{\mathbf{L}}^+] - [\mathbf{L}^+,\tilde{\mathbf{L}}^0]\\
2 [\mathbf{L}^+,\tilde{\mathbf{L}}^-] - 2 [\mathbf{L}^-,\tilde{\mathbf{L}}^+]  \\
[\mathbf{L}^-,\tilde{\mathbf{L}}^0] - [\mathbf{L}^0,\tilde{\mathbf{L}}^-]
\end{pmatrix}.
\end{equation}
Note that $B(\Phi,\tilde{\Phi})$ is symmetric under exchanging $\Phi$ and $\tilde{\Phi}$, hence for $\Phi=\tilde{\Phi}$ this takes the simple form
\begin{equation}
B(\Phi,\Phi) = \half\begin{pmatrix}
[\mathbf{L}^0,\mathbf{L}^+] \\ 2[\mathbf{L}^+, \mathbf{L}^-] \\ [\mathbf{L}^-,\mathbf{L}^0]
\end{pmatrix}.
\end{equation}
We readily see that \eqref{eq:Nahm_L} can then be written as
\begin{equation}
\label{eq:eom_B}
\frac{d\Phi}{dy}=-B\left(\Phi,\Phi\right).
\end{equation}
To turn the differential equation \eqref{eq:eom_B} into an algebraic recursion relation we now perform a series expansion of $\Phi$ around $y\rightarrow\infty$. In order to match the solution to the boundary data, we impose that the leading behaviour of $\Phi$ is given by
\begin{equation} \label{leading-Phi}
\Phi= y^{-1}\begin{pmatrix}
L_1 \\ L_0 \\ L_{-1}
\end{pmatrix}+\cO(y^{-3/2})\, .
\end{equation}
In other words, the leading behaviour of $\Phi$ is given precisely by the $\slt$-triple $(L_{0},L_{\pm 1})$ of the boundary data. 
Indeed, one checks that this ansatz satisfies \eqref{eq:eom_B} to leading order in $y^{-1}$. To parametrize possible sub-leading terms in $\Phi$, we make the ansatz
\begin{equation} \label{Phi-expand}
\Phi=\sum_{n\geq 0} \Phi_n y^{-1-n/2}= \sum_{n\geq 0} \begin{pmatrix}
L_n^+\\ L_n^0 \\ L_n^-\\
\end{pmatrix}y^{-1-n/2},\qquad \Phi_0:=\begin{pmatrix}
L_{1} \\ L_0 \\ L_{-1}
\end{pmatrix}.
\end{equation}
In terms of the $\Phi_n$, \eqref{eq:eom_B} reduces to a recursion relation
\begin{equation}
\label{eq:recursion_CKS}
(n+2)\Phi_n-4B(\Phi_0,\Phi_n)=2\sum_{0<k<n}B(\Phi_k,\Phi_{n-k})\, .
\end{equation}
In principle, this equation allows one to determine the $\Phi_n$ in terms of the previous $\Phi_k$, $k<n$. However, the expression for $B(\Phi_0,\Phi_n)$ will generically be very complex. In order to simplify this, we proceed in two steps. First, we make use of the highest weight decomposition \eqref{weight-charge-expansion-operator1} to decompose the operators $L_n^\bullet$ as 
\begin{equation}
L_n^\bullet = \sum_{d\geq 0} (L_n^\bullet)^{[d]}\ ,
\end{equation}
where each $(L_n^\bullet)^{[d]}$ is an operator of highest weight $d$ in the notation introduced in \eqref{weightonly}. Note that in the discussion of the recursion relations it will not be necessary to perform the weight or charge decomposition. Using this decomposition we have split the various components of $\Phi_n$ into different pieces, which can be collected into
\begin{equation}
\Phi_n^d = \begin{pmatrix}
(L_n^+)^{[d]} \\ (L_n^0)^{[d]} \\ (L_n^-)^{[d]}
\end{pmatrix}.
\end{equation}
We can, however, perform a further decomposition of the full $\Phi_n^d$ by diagonalizing the operator $B(\Phi_0,\cdot)$ which appears in \eqref{eq:recursion_CKS}. This will lead to a great simplification of the recursion. For convenience, let us abbreviate $B_0:=B(\Phi_0,\cdot)$. To be explicit, using \eqref{eq:bilinear_B} this operator can be written as
\begin{equation}
\label{eq:B_identity}
B_0= \frac{1}{4}\begin{pmatrix}
\ad L_0 & -\ad L_{1} & 0\\
-2\ad L_{-1} & 0 & 2 \ad L_{1}\\
0 & \ad L_{-1} & -\ad  L_0
\end{pmatrix}.
\end{equation}
To evaluate the action of $B_0$ on $\Phi$ we use the interpretation of $\Phi$ as a 3-vector consisting of matrices $\{\mathbf{L}^+,\mathbf{L}^0,\mathbf{L}^-\}$ on which $(L_0,L_{\pm 1})$ can act via the adjoint action. In other words
\begin{equation} \label{B0Phi}
B_0(\Phi) = \frac{1}{4}\begin{pmatrix}
[L_0,\mathbf{L}^+] - [L_1,\mathbf{L}^0]\\
2 [L_{1},\mathbf{L}^-]- 2 [L_{-1},\mathbf{L}^+]\\
[L_{-1},\mathbf{L}^0] - [L_0,\mathbf{L}^-]
\end{pmatrix}\ .
\end{equation}
Our next goal will be to find a split of the $\Phi$, which diagonalizes $B_0$ and hence lets us 
evaluate the second term in \eqref{eq:recursion_CKS}. The remarkable idea of CKS is to introduce yet another $\slt$-decomposition, which now acts on the 3-vectors $\Phi^d_n$. This new decomposition 
allows us to split 
\begin{equation} \label{Phi-epsilon-split}
\Phi_n^d= \sum_{\epsilon=-1,0,1} \Phi_n^{d,\epsilon}\ .
\end{equation}
We stress that this is \textit{not} the $(d,s)$-decomposition introduced in \eqref{weightonly} for which the indices were written with 
brackets. 
The second eigenvalue $\epsilon$ arises from an $\slt$-triple $(\Lambda^0,\Lambda^\pm)$ which acts on $\Phi$ by also 
mixing the 3-vector components $\mathbf{L}^\bullet$. This $\slt$-triple is given by 
\begin{equation}
\label{eq:sl2_tensor}
\begin{aligned}
\Lambda^+ &= \begin{pmatrix}
\ad L^+ & 0 & 0 \\
2 & \ad L^+& 0 \\
0 & -1 & \ad L^+ \\
\end{pmatrix} , \\
\Lambda^0 &= \begin{pmatrix}
\ad L^0-2 & 0 & 0 \\
0 & \ad L^0 & 0 \\
0 & 0 & \ad L^0+2 \\
\end{pmatrix} ,\\
\Lambda^- &= \begin{pmatrix}
\ad L^- & 1 & 0 \\
0 & \ad L^- & -2 \\
0 & 0 & \ad L^- \\
\end{pmatrix}.
\end{aligned}
\end{equation}
By slight abuse of notation the integer entries are proportional to identity matrices.\footnote{To elaborate, the $3\times 3$ matrices
	\begin{equation}
	\begin{pmatrix}
	0 & 0 & 0\\
	2 & 0 & 0\\
	0 & -1 & 0
	\end{pmatrix},\qquad \begin{pmatrix}
	-2 & 0 & 0\\
	0 & 0 & 0\\
	0 & 0 & 2
	\end{pmatrix}, \qquad \begin{pmatrix}
	0 & 1 & 0\\
	0 & 0 & -2\\
	0 & 0 & 0
	\end{pmatrix}\, .
	\end{equation}
	also form an $\slt$-triple (more precisely, they correspond to the co-adjoint representation). The $\slt$-triple $(\Lambda^+, \Lambda^0, \Lambda^-)$ is then obtained by taking the tensor product between the above generators and the $\slt$-triple $(L_{1} ,  L_0 , L_{-1})$.} The label $\epsilon$ appearing in \eqref{Phi-epsilon-split} is then related to the eigenvalue under the Casimir $\Lambda^{2}$ via
\begin{equation}
\Lambda^{2}= 2 \Lambda^+  \Lambda^- +2\Lambda^-  \Lambda^+ +(\Lambda^0)^2: \qquad \Lambda^{2} \Phi^{d,\epsilon} = (d+2\epsilon)(d+2\epsilon+2) \Phi^{d,\epsilon}\, .
\end{equation}
In other words, for a given $d$, each $\Phi^d$ splits into three components $\Phi^{d,\epsilon}$, which have highest weight $d+2\epsilon$ with respect to the Casimir $\Lambda^2$.
Using \eqref{eq:sl2_tensor} it is straightforward to compute $\Lambda^2$ explicitly as
\begin{equation}
\Lambda^2 = \begin{pmatrix}
(\ad L)^2+8-4\ad L^0 & 4\ad L^+ & 0\\
8\ad L^- & (\ad L)^2+8& -8 \ad L^+\\
0 & -4\ad L^- & (\ad L)^2+8 +4\ad L^0 
\end{pmatrix}\ . 
\end{equation}
We can now compare this expression with the expression \eqref{eq:B_identity} for $B_0$ and 
observe that the Casimir can also be written as
\beq
\Lambda^2   = \big( (\ad L)^2+8 \big)\mathbb{I}_{3\times 3} - 16 B_0\ .
\eeq
In other words, we see that the components $\Phi_n^{d,\epsilon}$ are also eigenvectors of $B_0$. In fact,
we evaluate 
\beq
\label{eq:B_eigenvalue}
-4B_0(\Phi_n^{d,\epsilon})=\left\{\epsilon(1+d+\epsilon)-2\right\}\Phi_n^{d,\epsilon}\ . 
\end{equation}
Returning to the recursion relation \eqref{eq:recursion_CKS}, we see that for such eigenvectors it simplifies to
\begin{equation}
\label{eq:CKS_recursion_epsilon}
\boxed{\rule[-.4cm]{.0cm}{1cm} \quad \big(n+\epsilon(1+d+\epsilon) \big) \Phi_n^{d,\epsilon}=2\sum_{0<k<n} B(\Phi_k,\Phi_{n-k})^{d,\epsilon} \, . \quad}
\end{equation}
The recursion \eqref{eq:CKS_recursion_epsilon} is the master equation that encodes the constraints on the 
coefficients $\Phi_n^{d,\epsilon}$ for any solution $\Phi$ of~\eqref{eq:eom_B}. 

Let us make some further remarks regarding the structure of the recursion. To begin with, we note that the 
representation theory of $\slt$ implies that the number of linearly independent operators with a given $d$ 
and $\epsilon$ is equal to $d+2\epsilon+1$. This implies, in particular, that $\Phi^{1,-1}_1 = 0$. 
Furthermore, for $n=1$ the right-hand side of \eqref{eq:CKS_recursion_epsilon} vanishes and we conclude that also 
$\Phi_1^{d,\epsilon}=0$ for $(d,\epsilon)\neq (1,-1)$. Taken together, we thus find that 
\begin{equation}
\Phi_1=0\ .
\end{equation}
Applying this result to the expansion \eqref{Phi-expand} of $\Phi$ this means that the term proportional to $y^{-3/2}$ vanishes and the first sub-leading term is of order $y^{-2}$. Inspecting the recursion \eqref{eq:CKS_recursion_epsilon} we see that for $n>1$ the $\Phi_n^{d,\epsilon}$ can be obtained recursively by computing the action of $B$ on $\Phi_k$ and $\Phi_{n-k}$, $k<n$, and projecting the result onto its $d,\epsilon$ components. The only $\Phi_n^{d,\epsilon}$ which are not determined recusively from \eqref{eq:CKS_recursion_epsilon} are those with highest weight $d=n$.
It is easy to check that $\Phi_n^{n,-1}$ actually does not appear on the left-hand side, since its pre-factor vanishes for this component. 
To obtain the constraints on $\Phi_n^{n,1}$ and $\Phi_n^{n,0}$ one has to use the properties of $B$ \footnote{To be precise, one uses the fact that for two operators $S$ and $T$, one has that $B(S^{d,\epsilon}, T^{d',\epsilon'})^{d'',\epsilon''}=0$ unless the following three conditions hold: (1) $0\leq d'' \leq d+d'$, (2) $d'' \equiv d+d' \;\mathrm{mod}\;2,$ and (3) $0\leq d''+2\epsilon'' \leq d+d'+2\epsilon+2\epsilon'$. These properties can be derived from the specific expression for $B$ and its behaviour with respect to the underlying $\slt$-structure. We refer the reader to \cite{CKS} for more details. } 
to realize that the right-hand side in \eqref{eq:CKS_recursion_epsilon} vanishes and therefore implies that  
\beq
\Phi_n^{n,1}=\Phi_n^{n,0}= 0 \ . 
\eeq
In conclusion, we find that we need to supply 
\beq \label{input-data}
\text{input data}:   \quad \Phi_n^{n,-1}\ 
\eeq  
for the recursion. Recalling that there 
are maximally  $d+2\epsilon+1$ independent $\Phi_n^{d,\epsilon}$, we conclude that there are generically $n-1$ linearly independent $\Phi_n^{n,-1}$ that need to be given. 

We will now discuss a way to encode the input data  $\Phi_n^{n,-1}$ in an efficient way, which makes the $n-1$ linearly independent degrees of freedom manifest. Furthermore, note that we have so far only discussed the differential constraint \eqref{eq:Nahm_L} and it remains to impose \eqref{charges_cL} on any 
solution. In the following we will evaluate the conditions \eqref{charges_cL} imposes on the input data $\Phi_n^{n,-1}$.
From \eqref{eq:B_eigenvalue} we know that $4B_0(\Phi_n^{n,-1})=(n+2)\Phi_n^{n,-1}$. As we will recall in appendix~\ref{app:input} this equation can be solved by the following ansatz \cite{CKS}
\begin{equation}
\label{eq:input_data}
\boxed{ \rule[-.8cm]{.0cm}{1.8cm}
\quad \Phi_n^{n,-1}= \sum_{1\leq s,q \leq n-1} a^{n,s}_q \begin{pmatrix}
-\frac{1}{n-s} \left(\mathrm{ad}\;L_1\right)^{s+1}\\
2 \left(\mathrm{ad}\;L_1\right)^{s} \\
(n-s+1) \left(\mathrm{ad}\;L_1\right)^{s-1}
\end{pmatrix}\hat{\eta}^{(-n)}_{-q} \ ,\quad }  
\end{equation}
where $\hat \eta \in \mathfrak{g}_{\bbC}$ has to obey
\beq \label{eta-properties}
\hat \eta= \sum_{s\leq -2} \sum_{q \leq -1} \hat{\eta}^{(s)}_q \ , \qquad [L_{-1},\hat \eta]= 0 \ .
\eeq
Note that $\Phi$ is a 3-vector made out of operators $\mathbf{L}^\bullet$, which themselves stem 
from real operators $\cN^\bullet \in \mathfrak{g}_\bbR$ by transformation with $\rho$ as given in \eqref{def-bfLbullet}. 
This implies that also $\hat \eta$ can 
be obtained from a real operator $\eta = \rho^{-1} \hat \eta \rho \in \mathfrak{g}_\bbR $. We also see in \eqref{eq:input_data} that there are indeed 
$n-1$ linearly independent contributions to $\Phi_n^{n,-1}$, which correspond to the various charge components of the operator $\hat{\eta}$. 
The operator $\hat \eta$ now 
encodes the input data of the recursion relation. 
It is important to stress, however, that the condition 
$4B_0(\Phi_n^{n,-1})=(n+2)\Phi_n^{n,-1}$ does not yet fix the complex coefficients $a^{n,s}_q$. In order to 
fix these coefficients we now impose the $Q$-constraint on the entries of $\Phi_n^{n,-1}$. A direct computation, which can be found 
in appendix~\ref{app:input}, reveals that 
\begin{equation}
a^{n,s}_q=i^{s-1} \frac{(n-s)!}{n!} b^{s-1}_{q-1,n-q-1}\, ,
\end{equation}
where the coefficients $b^{s}_{p,q}$ are defined by
\begin{equation}
(1-x)^p(1+x)^q = \sum_{s=0}^{p+q} b^s_{p,q} x^s \, ,\quad p,q\geq 0\, .
\end{equation}
To summarize, we have reduced Nahm's equations to a set of recursion relations \eqref{eq:CKS_recursion_epsilon} for the components $\Phi_n^{d,\epsilon}$ which encode the full $\Phi$ defined in \eqref{def-Phi_L}. The initial conditions are determined by the boundary data $\hat \eta$ and 
the $\slt$-triple via \eqref{eq:input_data}. In order to show this we have  imposed the $Q$-constraint \eqref{charges_cL} on $\Phi^{n,-1}_n$. Using 
the recursion relations \eqref{eq:CKS_recursion_epsilon} this ensures that the full solution obeys this constraint.  
We next discuss how one can relate a solution $\Phi$ back to the matter fields $h$ and how $\hat \eta$ is determined in terms of the 
phase operator $\hat \eta$ which was part of the boundary data.

\subsection{Bulk matter field reconstruction and a matching condition}
\label{subsec:h_reconstruction}

Recall that the bulk theory introduced in section \ref{bulk-theory} has $h$ as a dynamical set of fields, while in the preceding section we have 
discussed the solutions $\Phi$ to Nahm's equation and the Q-constraint. In this section we will show how a solution for $\Phi$ can be used to 
obtain a matter field solution $h$. Furthermore, we will see how a single matching condition \eqref{zetadelta-eq} allows us to fix the input data 
$\hat \eta$ for the CKS recursion as well as the transformation $\zeta$ in \eqref{matchzeta} in terms of the boundary data. 

Recall that the vector $\Phi$ contains $\mathbf{L}^0$ as one its components. Furthermore, we can recall from~\eqref{def-cN}, \eqref{def-bfLbullet} the relations
\begin{equation}
\label{eq:h_cN}
-2h^{-1}\partial_y h = \cN^0\, ,\qquad \cN^0 = e^{-\zeta}\rho^{-1} \mathbf{L}^0\rho e^{\zeta}\, .
\end{equation}
In essence, we need to solve this equation to fix the $y$-dependence in $h(x,y) = e^{xN^-} \tilde h(y)$ for a given $\mathbf{L}^0$. Note that if $h$ were simply a number, one could write this relation as $-2\partial_y \log(h)=\cN^0$ and solve it relatively straightforwardly. However, because $h$ is matrix-valued one needs to do a bit more work. First, it is useful to introduce a new function
\begin{equation}
\label{eq:def_g}
g(y) = e^{-xN^-} h(x,y) y^{\frac{1}{2} \tilde N^0}\ , \qquad \tilde N^0 = e^\zeta N^0 e^{-\zeta}\ . 
\end{equation}
Clearly, the resulting function is $x$-independent. Furthermore, it is convenient to introduce the factor $y^{\frac{1}{2}N^0}$, which turns out
to remove the overall scaling of $h$ as we will see in the following. Firstly, note that \eqref{leading-Phi} with \eqref{eq:h_cN} implies that
\begin{equation}
\cN^0(y) = \frac{N^0}{y}+\cO(y^{-2})\, .
\end{equation}
The factor  $y^{\frac{1}{2} \tilde N^0}$ in the definition of $g(y)$ ensures that the leading $N^0$-term in $\cN^0$ drops when
computing $g^{-1}\partial_y g$.  In other words, we find that $g^{-1}\partial_y g=\cO(y^{-2})$. Secondly, since $\Phi$ is described in terms of the complex algebra, it also convenient to introduce a rotated version of $g(y)$ using $\rho$
\begin{equation}
\label{eq:ghat}
\hat{g}(y) =\rho e^{-\zeta}g(y) e^{\zeta}\rho^{-1}\, .
\end{equation}
By combining \eqref{eq:h_cN}, \eqref{eq:def_g} and \eqref{eq:ghat}, together with the fact that $\mathrm{Ad}\, {y^{\frac{1}{2} \tilde N^0}}=y^{\frac{1}{2}\ad \tilde N^0}$ one can obtain the following relation between $\hat{g}$ and $\mathbf{L}^0$
\begin{equation}
\label{eq:invg_dg}
\hat{g}^{-1}\partial_y\hat{g} = \sum_{n\geq 2} B_n y^{-n}\, ,
\end{equation}
where the $B_n\in\mathfrak{g}_{\mathbb{C}}$ are comprised of particular $(d,s)$-components of the $L^0_n$ as follows
\begin{equation}
\label{eq:Bn}
B_n = -\frac{1}{2}\sum_{s\leq n-2} \sum_{d\leq 2n-2-s}\left(L_{2n-2-s}^0\right)^{(d,s)}.
\end{equation}
Note that we did not yet solve the differential equation \eqref{eq:invg_dg}, but have merely identified how the solution for $\Phi$ contributes to it through the $B_n$. We are now in a position to solve it, by again performing a series expansion of $\hat{g}$ around $y=\infty$ by writing 
\begin{equation}
\hat{g}(y) = \sum_{k\geq 0} \frac{\hat{g}_k}{y^{k}}\ ,\qquad  \hat{g}_0=1\ .
\end{equation}
Inserting this ansatz into \eqref{eq:invg_dg} we find
\begin{equation}
\partial_y \hat{g} = \hat{g}(y)\sum_{n\geq 2} B_n y^{-n}=\sum_{m\geq 0} \sum_{n\geq 2} \hat{g}_m B_n y^{-m-n}=\sum_{k\geq 1} \Big[\sum_{j=1}^k \hat{g}_{k-j} B_{j+1}\Big] y^{-k-1},
\end{equation}
where in the last line we have changed summation variables. Comparing this result with the series expansion of $\partial_y \hat{g}$ we  find
the general solution 
\beq \label{eq:recursion_g_1}
\hat g_k = P_k (B_2,...,B_{k+1})\, ,
\eeq
where the $P_k$ are iteratively defined non-commutative polynomials
\begin{equation}
\label{eq:recursion_g_2}
P_0=1\ ,\qquad P_k=-\frac{1}{k} \sum_{j=1}^k P_{k-j} B_{j+1} \ .
\end{equation}
We see that each $\hat{g}_k$ is recursively given in terms of the $\hat{g}_{k-j}$ and $B_{j+1}$. Of course, solving this recursion may still be very complicated, but we will show that it can be done in our examples. It is also interesting to note that while the $B_n$ are elements of the algebra, the $\hat{g}_k$ are, generically, not. Indeed, the algebra is closed under the commutator, but \eqref{eq:recursion_g_1}, \eqref{eq:recursion_g_2} contains only the product of matrices. As a result the full function $\hat{g}(y)$ is also not an algebra element. 
Taking these findings together, we thus arrive at bulk solutions of the form  
\beq \label{solution-expansion}
h(x,y) = e^{x N^-}  e^{\zeta} \Big( 1 + \frac{g_1}{y} + \frac{g_2}{y^2} + \ldots \Big)   y^{-\frac{1}{2}\tilde N^0}
\eeq
with $\tilde N^0 = e^\zeta N^0 e^{-\zeta}$. The coefficients in this expansion satisfy 
\beq \label{g-conditions}
(\ad {L_{-1}})^{n+1} \hat g_n = 0 \ , \qquad (\hat g_n)^{(l)}_q = 0\, ,\quad l\geq n\, .
\eeq
Note that these conditions arise a as a very non-trivial consequence of this iterative process 
and the fact that the $B_i$ are determined by the CKS recursion and are part of the famous sl(2)-orbit theorem \cite{CKS}. 

It remains to show how the $\hat{g}_k$ and $\hat \zeta = \rho \zeta \rho^{-1}$ in a bulk solution are actually fixed in terms of the boundary data. This can 
be done by imposing a single constraint 
\begin{equation} \label{zetadelta-eq}
\boxed{
\quad e^{i\hat{\delta}} = e^{\hat{\zeta}} \Big(1+\sum_{k>0} \frac{(-i)^k}{k!} (\ad L_{-1})^k \hat{g}_k \Big) \ .  \quad }
\end{equation}
Clearly, this condition can be equivalently formulated in the real algebra by removing the hats and replacing $L_{-1}$ by $N^-$.
To see that this condition fixes $\hat{g}_k, \hat \zeta$ uniquely, we first note that 
\beq
C_{k+1}(\eta) \equiv \frac{(-i)^k}{k!} (\ad L_{-1})^k B_{k+1}  =  i  \sum_{l\geq k+1} \sum_{q\geq 1}\ b^{k-1}_{q-1,l-q-1}  \  \hat \eta^{(-l)}_{-q}\ ,
\eeq
as can be shown by using \eqref{eq:Bn}, \eqref{eq:input_data}, \eqref{g-conditions} and \eqref{eta-properties}. Applying 
the properties \eqref{g-conditions} of $\hat g_k$ in the recursive solution \eqref{eq:recursion_g_1} with \eqref{eq:recursion_g_2} we  find that \eqref{zetadelta-eq} can be written as
\beq \label{delta-zeta-eta}
e^{i \hat \delta}   = e^{\hat{\zeta}} \Big( 1+\sum_{k\geq 1 } P_k(C_2,...,C_{k+1})  \Big)\ .
\eeq
Note that the right-hand side only depends on $\hat \zeta$ and $\hat \eta$, while the left-hand side depends on $\hat \delta$. Recalling 
that $\hat \delta, \hat \eta,\hat \zeta$ stem from the real counterparts $\delta,\eta,\zeta$ we realize that \eqref{delta-zeta-eta} gives a complex matrix equation determining two real unknowns $\eta$ and $\zeta$. Let us stress that the sums appearing in \eqref{zetadelta-eq} and \eqref{delta-zeta-eta} have maximally 
$2D$ terms, due to the presence of the lowering operators $ (\ad L_{-1})^k$ in \eqref{zetadelta-eq}. However, in general there can be 
infinitely many non-trivial $g_k$ in matter field solution \eqref{solution-expansion}. We will see this explicitly for our examples 
with $D=3$ in the remainder of this work.

In order to prepare for these examples, let us record here how the relation \eqref{delta-zeta-eta} yields a matching of
the components $\hat \delta^{(p)}_q$ of $\hat \delta$ with the components of  $\hat \eta,\hat \zeta$.
For $\zeta$ we find the relations 
\begin{align}\label{eq:zeta}
&&\hat \zeta_{-1}^{(-2)}=\hat \zeta_{-2}^{(-4)}=0\ ,
\qquad \hat \zeta_{-1}^{(-3)} =- \frac{i}{2} \hat \delta_{-1}^{(-3)} \ , \qquad
\hat \zeta_{-1}^{(-4)} =  -\frac{3i}{4} \hat \delta_{-1}^{(-4)}\ ,& &\nn \\
&&  \hat \zeta_{-2}^{(-5)} = -\frac{3i}{8}\hat \delta_{-2}^{(-5)} -\frac{1}{8}\Big[\hat\delta_{-1}^{(-2)},\hat\delta_{-1}^{(-3)} \Big]\ , 
\qquad \hat \zeta_{-3}^{(-6)} = - \frac{1}{8} \Big[\hat\delta_{-1}^{(-2)},\hat\delta_{-2}^{(-4)} \Big]\ . &&
\end{align}
The components of $\eta$ are fixed by 
\begin{align}\label{eq:eta_delta}
&& \hat \eta_{-1}^{(-2)}=-\hat \delta_{-1}^{(-2)}  \, ,
\qquad \hat \eta_{-1}^{(-3)} =-  \hat \delta_{-1}^{(-3)} \, , \qquad
\hat \eta_{-1}^{(-4)} =  -\frac{3}{4} \hat \delta_{-1}^{(-4)}\, , \qquad \hat \eta_{-2}^{(-4)} =  -\frac{3}{2} \hat \delta_{-2}^{(-4)}\, ,& &\nn \\
&&   \hat \eta_{-2}^{(-5)} =  -\frac{3}{2} \hat \delta_{-2}^{(-5)}+\frac{i}{2} \Big[\hat \delta_{-1}^{(-2)},  \hat \delta_{-1}^{(-3)} \Big] \, , \qquad  \hat \eta_{-3}^{(-6)} =  -\frac{15}{8} \hat \delta_{-3}^{(-6)}-\frac{5i}{4} \Big[\hat \delta_{-1}^{(-3)},  \hat \delta_{-2}^{(-3)} \Big]\, . &&
\end{align}
In conclusion, we have seen that the full bulk matter field solution $h(x,y)$ can be recovered using $\hat \delta$, and the operators $Q_\infty,L_\alpha$. 
The strategy is to fix the leading coefficients through \eqref{zetadelta-eq} and then run the CKS recursion to determine the complete 
series for $\hat g(y)$. This requires us to find the appropriate components of each $L^0_n$ which combine to the $B_n$ in \eqref{eq:Bn}, which yield 
$\hat{g}$ via \eqref{eq:recursion_g_1}, \eqref{eq:recursion_g_2}.

\subsection{Reconstructing the Hodge metric}\label{ssec:reconstructmetric}
In this section we discuss how we can reconstruct the Hodge metric from the bulk solution. We know that the Hodge metric can be expressed in terms of the conserved current \eqref{eq:current} via \eqref{eq:hodgemetric}. Let us therefore first rewrite this conserved current into a form familiar from the CKS recursion. By using \eqref{def-cN} we can write the conserved current componentwise as
\begin{equation}
(J_L)_x = \frac{1}{2} h \big( \cN^- + \cN^+ \big)h^{-1}\, , \qquad (J_L)_y =-\frac{1}{2}h \,  \cN^0 \, h^{-1} \, .
\end{equation}
This allows us to express the Hodge metric \eqref{eq:hodgemetric} as
\begin{equation}\label{eq:metric-cN}
g_{\alpha \beta}(J_L) \dd\sigma^\alpha \dd\sigma^\beta = \frac{1}{4 } \text{Tr}\big[ (\cN^0 )^2 \big] \, ( \dd x^2 + \dd y^2 ) \, ,
\end{equation}
where we used cyclicity of the trace, $\text{Tr}\big[  \cN^0 (\cN^- + \cN^+) \big] = 0$ and $\text{Tr}\big[  (\cN^- + \cN^+)^2 \big]= \text{Tr}\big[  (\cN^0)^2 \big]$.\footnote{This follows by using the commutation relations \eqref{Q-constr_2} to show that $\text{Tr}\big[ \cN^0 (\cN^- + \cN^+) \big] = -i\text{Tr}\big[ \cN^0 [Q, \cN^0] \big] = 0$ and $\text{Tr}\big[ (\cN^- + \cN^+)^2 \big] = -i \text{Tr}\big[ (\cN^- + \cN^+) [Q, \cN^0] \big] =i \text{Tr}\big[  [Q, \cN^+ + \cN^-]  \cN^0 \big] = \text{Tr}\big[ (\cN^0)^2 \big] $.} The idea is now to expand $\cN^0$ in inverse powers of $y$ in order to determine the coefficients $c^{(n)}$ in the expansion for the Hodge metric \eqref{eq:metricexpand}. To gain some intuition for the problem at hand, we can first study only the leading terms in this near-boundary expansion. We then find that the current behaves asymptotically as
\begin{equation}
(J_L)_x =\frac{  y N^-+N^+/y}{2y}  + \ldots\, , \qquad (J_L)_y = -\frac{N^0}{2y} +\ldots\, .
\end{equation}
In turn we compute the leading coefficient of the Hodge metric to be
\begin{equation}\label{eq:leadingcoef}
c^{(0)} = \frac{1}{4} \text{Tr}[ (N^0)^2 ] = \frac{1}{12}\sum_i d_i(d_i+1)(d_i+2)\, .
\end{equation}
where the sum over $i$ runs over the irreducible $\slt$-representations, with $d_i$ their highest weights. In order to determine the subleading coefficients $c^{(n)}$ as well, we need to write out $\cN^0$ further. We can write it in terms of the matter field $h(x,y)$ via \eqref{def-cN} as
\begin{equation}
\begin{aligned}
\frac{1}{4}\text{Tr}[ (\cN^0)^2 ] &= \text{Tr}\big[ \big( h^{-1} \partial_y h \big)^2 \big]  =  \text{Tr}\big[ \big( g^{-1} \partial_y g - \frac{1}{2y} N^0  \big)^2 \big]  \, ,
\end{aligned}
\end{equation}
where we used \eqref{eq:def_g} in the second equality to express $h(x,y)$ in terms of $g(y)$. We can now switch to the complex basis under the trace, which allows us to expand $g^{-1} \partial_y g$ in the coefficients $B_n$ by using \eqref{eq:invg_dg}. Writing $B_1=-L_0/2$, we can collect the terms at order $y^{-n}$ together as
\begin{equation}\label{eq:ci}
c^{(n)} = \frac{1}{4} \sum_{1 \leq k \leq n+1} \text{Tr}[B_k B_{n-k+2}]\, . 
\end{equation}
Note that for $n=0$ this result matches indeed with the leading coefficient given in \eqref{eq:leadingcoef}.

%%%%%%%%%%%%%%%%%%%%%%%%%%%%%%%%%%%%%%%%%%%%%%%%%%%%%%
\section{Bulk reconstruction of weight 3 examples of Calabi--Yau type} \label{reconstructing_examples}

Having discussed the general procedure to reconstruct the bulk solution, we now turn our attention to explicit examples where we demonstrate how this works in practice. In section \ref{boundary_class} we classified the possible boundary data with a real two-dimensional moduli spaces that are of 
weight $D=3$ and of Calabi--Yau type. These settings are characterized by 
$(h_{3/2},h_{1/2} ,h_{-1/2}, h_{-3/2})= (1,h_{1/2},h_{1/2},1)$ and $G_{\mathbb{C}}=\text{Sp}(2h_{1/2}+2, \mathbb{C})$ as noted in \eqref{H-dimq}, \eqref{CYtype-cond}. We have found that there are $4 h_{1/2}$ different types $\text{I}_n$, 
$\text{II}_n$, $\text{III}_n$, and $\text{IV}_{n}$. In principle, we can now construct the most general form of the boundary data $N^0,N^\pm$, $Q_\infty$, and $\delta$. In this generality, however, it will be very involved to reconstruct the associated bulk solution. In particular, this can 
be traced back to the fact that $\delta$ can depend on many free parameters for large $h_{1/2}$. To illustrate how the reconstruction 
works in practice we therefore restrict to the case $h_{1/2}=1$. For this situation, we have constructed the boundary data in section \ref{ssec:data} and listed 
the result in table \ref{table:data}. This simple set of matrices will now be used in the CKS recursion introduced in section \ref{CKS-recursion}
and will then allow us to determine the bulk fields $h(x,y)$ using the steps described in section \ref{subsec:h_reconstruction}. 

Let us already note that we will show later, in section \ref{period_section}, that the reconstructed bulk solutions can be used to derive period vectors familiar from the study of geometric moduli spaces of Calabi--Yau manifolds. 
Concretely, we find
\bea
\text{I}_1 &\quad& \Rightarrow \quad \text{conifold point} \nn \\[-.1cm]
\text{II}_0 &\quad& \Rightarrow \quad \text{Tyurin degeneration, K-point}\\[-.1cm]
\text{IV}_1 &\quad& \Rightarrow \quad \text{large complex structure point}\nn 
\eea 
In order to make such an identification we need to determine the near-boundary solution $h(x,y)$. In this section we will 
do this for all three cases in turn.

%%%%%%%%%%%%%%%%%%%%%%%%%%%%%%%%%%%%%%%%%%%%%%%%%%%%%%
\subsection{Bulk reconstruction for $\mathrm{I}_{1}$ boundary data}
\label{sec:conifold}

Recall from table \ref{table:data} the defining data for $\mathrm{I}_1$ boundaries. As noted in our general discussion of the 
bulk reconstruction it is preferable to work in a complex basis obtained by the transformation matrix \eqref{def-rho}. 
For the $\slt$-triple this means that we redefine our generators via \eqref{eq:Ldef} as
\begin{equation}
L_{1} = \begin{pmatrix}
0 & 0 & 0 & 0 \\
0 & -i & 1 & 0 \\
0 & 1 & i & 0 \\
0 & 0 & 0 & 0 \\
\end{pmatrix},\quad
L_0 = \begin{pmatrix}
0 & 0 & 0 & 0 \\
0 & 0 & -i & 0 \\
0 & i & 0 & 0 \\
0 & 0 & 0 & 0 \\
\end{pmatrix},\quad L_{-1} = \begin{pmatrix}
0 & 0 & 0 & 0 \\
0 & i & 1 & 0 \\
0 & 1 & -i & 0 \\
0 & 0 & 0 & 0 
\end{pmatrix}.
\end{equation}
The phase operator $\delta$ plays an important role in the bulk reconstruction, so we also want to express it in terms of the complex basis. Noting that it is proportional to $N^-$ in the real basis, we find that it simply transforms to $\hat{\delta}=\hat{\delta}_{-1}^{(-2)} = -c L_{-1}$ in the complex basis.

\subsubsection*{Solving the CKS recursion}
\label{ssec:CKSI1}
Let us now solve the CKS recursion for the type $\mathrm{I}_1$ boundary data. We begin by determining the initial data of our recursion, which is fixed by the operator  $\hat{\eta}$. This operator can be expressed in terms of the phase operator $\hat{\delta}$ by using \eqref{eq:eta_delta} as
\begin{equation}
\hat{\eta} = \hat{\eta}_{-1}^{(-2)} = - \delta_{-1}^{(-2)} = c L_{-1}\, .
\end{equation} 
The initial data can then be computed from $\hat{\eta}$ by using \eqref{eq:input_data}. This gives us just $\Phi_2^{2,-1}$ as input for the recursion, since $\hat{\eta}$ only has a component with weight $n=2$. Furthermore the sums over the weights $s$ and charges $q$ only run over a single term, hence
\begin{equation}
\Phi_2^{2,-1}= \begin{pmatrix}
(L^{+}_{2})^{[2]} \\
(L^{0}_{2})^{[2]} \\
(L^{-}_{2})^{[2]}
\end{pmatrix}= c\begin{pmatrix}
L_{1}\\ 
L_{0} \\
L_{-1}
\end{pmatrix} = c\, \Phi_0\, .
\end{equation}
where in the last equality we noticed that the initial data of the recursion is simply proportional to the leading term $\Phi_0=(L_{1},L_{0},L_{-1})$ of the expansion \eqref{Phi-expand}. In order to perform the CKS recursion we then need to evaluate the bilinear $B$ defined in \eqref{eq:bilinear_B} for this initial data. For our initial data it is interesting to point out that in general
\begin{equation}
B(\Phi_{0} \, , \, \Phi_{0}) = \Phi_{0}\ .
\end{equation}
This implies that each subsequent term $\Phi_{n}$ in the CKS recursion will be proportional to $\Phi_{0}$. Furthermore, from the structure of the recursion relation \eqref{eq:CKS_recursion_epsilon} we find at odd orders that $\Phi_{2n+1}=0$. Let us therefore make the ansatz
\begin{equation}
\Phi_{2n}^{2,-1}=c_n \Phi_0\, ,
\end{equation}
for some coefficients $c_{n}$ with $c_0=1$ and $c_{1}=c$. By plugging our ansatz into \eqref{eq:CKS_recursion_epsilon} we then obtain the recursion relation
\begin{equation}
(2n-2)c_{n} = 2 \sum_{0<k<n}c_k c_{n-k},\quad n>1\ .
\end{equation}
One can easily verify that under the specified initial conditions this recursion is solved by
\begin{equation}
c_{n} =c^n\ .
\end{equation}
Hence the full solution to the recursion relation is given by
\begin{equation}\label{eq:phiI1}
\Phi_{2n} =c^n\Phi_0,\quad \mathrm{or}\quad \begin{pmatrix}
L_{2n}^+ \\
L_{2n}^0\\ 
L_{2n}^-
\end{pmatrix} = c^n\begin{pmatrix}
L_{1}\\L_0\\L_{-1}
\end{pmatrix}\ .
\end{equation}
We see that the result for $\mathrm{I}_{1}$ boundaries is remarkably simple. From the observation that the phase operator can be expressed in terms of the lowering operator as $\hat{\delta}=-cL_{-1}$ we have been able to solve the recursion completely algebraically. In particular we did not need to perform the highest-weight decompositions with respect to the Casimir operators $L^{2}$ and $\Lambda^{2}$ explicitly. In anticipation of the other examples, let us already note that for boundaries of type $\mathrm{II}_{0}$ we find a similar recursion, whereas for $\mathrm{IV}_{1}$ boundaries we do have to make use of this $\mathfrak{sl}(2)$-machinery.

\subsubsection*{Reconstructing the bulk solution}

Despite the apparent simplicity of the $\Phi_{2n}$ given in \eqref{eq:phiI1}, there are still some non-trivial steps to perform in order to complete the bulk reconstruction of $h(y)$, to which we now turn our attention. Following the procedure laid out in section \ref{subsec:h_reconstruction} we first write down the middle component $L^0_{2n}$ of $\Phi_{2n}$ as
\begin{equation}
(L^0_{2n})^{(2,0)}= c^n L_0\ .
\end{equation}
By using \eqref{eq:Bn} to compute the coefficients $B_{n}$ we then find that the sum only runs over a single term, yielding
\begin{equation}
B_n = -\frac{1}{2} L^0_{2n-2}=-\frac{1}{2}c^{n-1} L_{0}\ .
\end{equation}
In turn these coefficients can be used to recursively determine the $\hat{g}_k$ by using \eqref{eq:recursion_g_2}. Plugging our expression for the $B_{n}$ into this recursion relation we obtain
\begin{equation}\label{eq:gkrecursionI1}
-k \hat{g}_k =-\frac{1}{2} \sum_{j=1}^k \hat{g}_{k-j} c^j L_{0}\, .
\end{equation}
with the initial condition $\hat{g}_0=1$. In order to solve this recursion it is convenient to rotate back to the real basis via the transformation matrix \eqref{def-rho} given by
\begin{equation}
\rho = \begin{pmatrix}
1 & 0 & 0 & 0 \\
0 & \frac{1}{\sqrt{2}} & \frac{i}{\sqrt{2}} & 0 \\
0 & \frac{i}{\sqrt{2}} & \frac{1}{\sqrt{2}} & 0 \\
0 & 0 & 0 & 1 \\
\end{pmatrix}\, .
\end{equation}
Namely, in this basis we simply need to work with a diagonal matrix $N^{0}$ instead of $L_{0}$, and consequently the terms $g_{k}$ are diagonal as well. Therefore let us take as ansatz
\begin{equation}
g_k=c^k \, \mathrm{diag}\left(0 ,\, g^+_k , \, g_k^- , \, 0\right)\, ,
\end{equation}
where $g^\pm_k$ are arbitrary coefficients with $g^\pm_1=1$ for which we want to solve the recursion. By using this ansatz we find that \eqref{eq:gkrecursionI1} reduces to two decoupled recursion relations
\begin{equation}
2k \, g^\pm_k = \pm \sum_{j=1}^k g^\pm_{k-j}\, .
\end{equation}
Combining the equations at levels $k$ and $k+1$ one can rewrite these equations as
\begin{equation}
g^\pm_{k+1} = \frac{(2k\pm 1)}{2(k+1)} g^\pm_k\, ,
\end{equation}
which are solved by
\begin{equation}
g^+_k = \frac{1}{2}\frac{\Gamma(k+1/2)}{\Gamma(3/2)\Gamma(k+1)}\, ,  \qquad g^-_k = -\frac{1}{2}\frac{\Gamma(k-1/2)}{\Gamma(1/2)\Gamma(k+1)}\, ,
\end{equation}
where $\Gamma(n+1)=n!$ denotes the gamma function. Putting everything together, one finds
\begin{equation}\label{eq:gkI1}
g_k = -\frac{1}{2}\frac{c^k}{k!}\begin{pmatrix}
0 & 0 & 0 & 0\\
0 &  -\frac{\Gamma(k+1/2)}{\Gamma(3/2)}  & 0 & 0\\
0 & 0 & \frac{\Gamma(k-1/2)}{\Gamma(1/2)} & 0\\
0 & 0 & 0 & 0
\end{pmatrix},\quad k\geq 1\, .
\end{equation}
One can verify that this solution to the CKS recursion indeed satisfies the matching condition given in \eqref{zetadelta-eq}. Resumming the series in $y^{-k}$ for the matrix-valued function $g(y)$ we find
\begin{equation}
g(y)=\sum_{k\geq 0} g_k y^{-k}=\begin{pmatrix}
1 & 0 & 0 & 0\\
0 & \sqrt{\frac{y}{y-c}} & 0 & 0\\
0 & 0 &  \sqrt{\frac{y-c}{y}}& 0\\
0 & 0 & 0 & 1
\end{pmatrix}.
\end{equation}
By multiplying with $e^{xN^{-}}$ from the left and $y^{-N^{0}/2}$ from the right we obtain the group-valued bulk field
\begin{equation}
\label{eq:I1_h}
\boxed{\rule[-1.1cm]{.0cm}{2.4cm} \quad
h(x,y) = \begin{pmatrix}
1 & 0 & 0 & 0\\
0 & \frac{1}{\sqrt{y-c}} & 0 & 0\\
0 & \frac{x}{\sqrt{y-c}} &  \sqrt{y-c} & 0\\
0 & 0 & 0 & 1
\end{pmatrix}\ . \quad }
\end{equation}
Note that the appearance of $c$ in $h(x,y)$ can be interpreted simply as a coordinate shift $y \to y-c$. In other words, by absorbing $c$ via a coordinate redefinition we can set the component of the phase operator $\delta$ along $N^{-}$ to zero.

%%%%%%%%%%%%%%%%%%%%%%%%%%%%%%%%%%%%%%%%%%%%%%%%%%%%%%
\subsection{Bulk reconstruction for $\mathrm{II}_{0}$ boundary data}
\label{sec:II0}
In our next example we study bulk reconstruction for boundary data of type $\mathrm{II}_{0}$.
Geometrically we find that we can interpret these boundaries as a so-called Tyurin degeneration of a Calabi--Yau threefold \cite{Tyurin:2003}. Recently these boundaries have also been studied in \cite{Joshi:2019nzi} as K-points in complex structure moduli space. Let us again begin by reformulating the boundary data given in table \ref{table:data} for the CKS recursion in terms of the complex basis. By using \eqref{eq:Ldef} we find that the $\mathfrak{sl}(2,\mathbb{C})$-triple takes the form
\begin{equation}
L_{1} = \begin{pmatrix}
-\frac{i}{2} & 0 & \frac{1}{2} & 0 \\
0 & -\frac{i}{2} & 0 & \frac{1}{2} \\
\frac{1}{2} & 0 & \frac{i}{2} & 0 \\
0 & \frac{1}{2} & 0 & \frac{i}{2} \\
\end{pmatrix}, \quad L_{0} = \begin{pmatrix}
0 & 0 & -i & 0 \\
0 & 0 & 0 & -i \\
i & 0 & 0 & 0 \\
0 & i & 0 & 0 \\
\end{pmatrix}, \quad 
L_{-1} = \begin{pmatrix}
\frac{i}{2} & 0 & \frac{1}{2} & 0 \\
0 & \frac{i}{2} & 0 & \frac{1}{2} \\
\frac{1}{2} & 0 & -\frac{i}{2} & 0 \\
0 & \frac{1}{2} & 0 & -\frac{i}{2} \\
\end{pmatrix}.
\end{equation}
We can also express the phase operator $\delta$ in this complex basis. Similar to the $\mathrm{I}_1$ boundary the phase operator $\delta$ is proportional to the lowering operator $N^-$, so we find $\hat{\delta}=\hat{\delta}^{(-1)}_{-2} = -c L_{-1}$.

\subsubsection*{Solving the CKS recursion}
We now show how to solve the CKS recursion for boundary data of type $\mathrm{II}_{0}$. This recursion takes a similar form as we found for the $\mathrm{I}_{1}$ boundaries studied before. The underlying reason is that the phase operator $\hat{\delta}$ is again proportional to the lowering operator $L_{-1}$. The main difference between these two types of boundaries is that $L_{-1}$ now acts non-trivially on two copies of a highest weight $d=1$ state instead of one. Due to the similarities in the actual recursions that will have to be solved for the components of the various matrices, we refer to section \ref{ssec:CKSI1} for the detailed derivations. 

We begin by determining the initial data of the recursion, which is fixed in terms of the operator $\hat{\eta}$. This operator can be expressed in terms of the phase operator $\hat{\delta}$ via \eqref{eq:eta_delta}, from which we find
\begin{equation}
\hat{\eta} = \hat{\eta}^{(-2)}_{-1} = -\hat{\delta}^{(-2)}_{-1} = c L_{-1} \,  .
\end{equation}
The initial data of the recursion can then be computed from $\hat{\eta}$ by using \ref{eq:input_data}. This gives us just $\Phi_2^{2,-1}$ as input for the recursion, since $\hat{\eta}$ only has a component with weight $n=2$. Furthermore, the sums over the weights $s$ and charges $q$ only run over a single term, hence
\begin{equation}
\Phi_2^{2,-1}=\half \begin{pmatrix}
-\left(\mathrm{ad}\;L^+\right)^{2}\\
2\;\mathrm{ad}\;L^+ \\
2
\end{pmatrix} \hat{\eta}_{-1}^{(-2)} = c\begin{pmatrix}
L_1\\ 
L_0 \\
L_{-1}
\end{pmatrix} = c\, \Phi_0\, .
\end{equation}
We see that $\Phi_2^{2,-1}$ is simply proportional to the leading term $\Phi_0$ of our near-boundary expansion \eqref{Phi-expand}, similar to our result for the $\mathrm{I}_{1}$ boundaries. When we plug this initial data into the recursion relation \eqref{eq:CKS_recursion_epsilon}, we can use that the components $L_{\bullet}$ satisfy the same commutation relations when we evaluate the bilinear $B$. Thus we find again that the recursion relation is solved by
\begin{equation}
\Phi_{2n} = \begin{pmatrix}
L_{2n}^+ \\
L_{2n}^0\\ 
L_{2n}^-
\end{pmatrix} = c^n\begin{pmatrix}
L_{1}\\L_0\\L_{-1}
\end{pmatrix} = c^n\Phi_0 \, ,
\end{equation}
the difference being that the matrices $L_{\bullet}$ take a different form for $\mathrm{II}_{0}$ boundaries compared to $\mathrm{I}_{1}$ boundaries.

\subsubsection*{Reconstructing the bulk solution}

Let us next perform the relevant steps in order to complete the bulk reconstruction and determine the bulk fields $h(x,y)$. 
This amounts to recursively determining the coefficients $\hat{g}_{k}$. We first compute the coefficients $B_{n}$ from \eqref{eq:Bn} to be
\begin{equation}
B_{n} = - \half L^{0}_{2n-2} = - \half c^{n-1} L_{0}\, .
\end{equation}
The recursive formula for the $\hat{g}_{k}$ given in \eqref{eq:recursion_g_2} can then be written out as
\begin{equation}
-k \hat{g}_{k} = \sum_{j=1}^{k} \hat{g}_{k-j}B_{j+1} = - \half \left[ \sum \hat{g}_{k-j} c^{j} \right] L_{0}\, ,
\end{equation}
with initial condition $\hat{g}_{0}=1$. This recursion for $\hat{g}_k$ takes a similar form as \eqref{eq:gkrecursionI1} found for $\mathrm{I}_1$ boundaries, but recall that $L_0$ now acts on two copies of a highest weight $d=1$ instead of one. Nevertheless we can solve the recursion relation for $\hat{g}_{k}$ by taking similar steps, and we find
\begin{equation}
g_{k} =-\half \frac{c^{k}}{k!} \begin{pmatrix}
\frac{\Gamma(k+1/2)}{\Gamma(3/2)} & 0 & 0 & 0 \\
0 & \frac{\Gamma(k+1/2)}{\Gamma(3/2)}  & 0 & 0 \\
0 & 0 & -\frac{\Gamma(k-1/2)}{\Gamma(1/2)}  & 0 \\
0 & 0 & 0 & -\frac{\Gamma(k-1/2)}{\Gamma(1/2)} 
\end{pmatrix},
\end{equation}
where we chose to rotate back to the real basis by the transformation matrix \eqref{def-rho}
\begin{equation}
\rho = \frac{1}{\sqrt{2}} \begin{pmatrix} 1& 0 & i& 0 \\
0 & 1& 0 & i\\
i & 0 & 1 & 0 \\
0 & i& 0 & 1\end{pmatrix}\, .
\end{equation}
Note indeed that compared to $\mathrm{I}_{1}$ boundaries the only difference is that the the components in the coefficients $g_{k}$ given in \eqref{eq:gkI1} now appear twice. Next we can resum the expansion in $y^{-k}$ for $g(y)$ as
\begin{equation}
g(y) = \sqrt{\frac{y}{y-c} }\begin{pmatrix}
1 & 0 & 0 & 0 \\
0 &1 & 0 & 0 \\
0 & 0 & \frac{y-c}{y} & 0 \\
0 & 0 & 0 & \frac{y-c}{y}
\end{pmatrix}.
\end{equation}
By multiplying with $e^{xN^{-}}$ from the left and $y^{-N^{0}/2}$ from the right we then obtain the group-valued bulk field
\begin{equation}
\label{eq:II0_h}
\boxed{\quad \rule[-.9cm]{.0cm}{2.0cm}  h(x,y) =\sqrt{\frac{1}{y-c}}   \begin{pmatrix}
1 & 0 & 0 & 0 \\
0 & 1& 0 & 0 \\
x& 0 & y-c & 0 \\
0 & x & 0 & y-c \\
\end{pmatrix}\ . \quad }
\end{equation}
Note that again the parameter $c$ can in principle be removed by a coordinate redefinition $y \to y-c$, similar to the $\mathrm{I}_1$ boundary.

%%%%%%%%%%%%%%%%%%%%%%%%%%%%%%%%%%%%%%%%%%%%%%%%%%%%%%
\subsection{Bulk reconstruction for $\mathrm{IV}_{1}$ boundary data}
\label{ssec:IV1}
In this section we study our final example, the type $\mathrm{IV}_1$ boundary. From a geometrical perspective these boundaries arise as large complex structure points in the complex structure moduli space of Calabi--Yau threefolds, see e.g. \cite{Candelas:1990rm}. This case will be considerably more involved, owing to the fact that the phase operator $\delta$ is no longer proportional to $N^-$. Nevertheless we will obtain exact results for both $\Phi=(\mathbf{L}^+,\mathbf{L}^0,\mathbf{L}^-)$ and the bulk matter fields. This demonstrates the power of the CKS recursion, which provides us with a general formalism to perform this bulk reconstruction. 

We remind the reader that the relevant boundary data is collected in section \ref{ssec:data}. As before, let us also present the complex $\slt$-generators for the type $\mathrm{IV}_1$ boundary
\begin{equation}
L_{-1}= \begin{pmatrix}
\frac{3 i}{2} & \frac{3}{2} & 0 & 0 \\
\frac{1}{2} & \frac{i}{2} & 1 & 0 \\
0 & 1 & -\frac{i}{2} & \frac{1}{2} \\
0 & 0 & \frac{3}{2} & -\frac{3 i}{2} \\
\end{pmatrix},\quad 
L_0 = \begin{pmatrix}
0 & -3 i & 0 & 0 \\
i & 0 & -2 i & 0 \\
0 & 2 i & 0 & -i \\
0 & 0 & 3 i & 0 \\
\end{pmatrix},\quad L_{1} = \begin{pmatrix}
-\frac{3 i}{2} & \frac{3}{2} & 0 & 0 \\
\frac{1}{2} & -\frac{i}{2} & 1 & 0 \\
0 & 1 & \frac{i}{2} & \frac{1}{2} \\
0 & 0 & \frac{3}{2} & \frac{3 i}{2} \\
\end{pmatrix}.
\end{equation}
Furthermore, one may compute the complex operator $\hat{\delta}$ using the transformation $\rho$ to find
\begin{equation}
\hat{\delta} = \hat{\delta}^{(-6)}_{-3} =\frac{\chi}{8} \begin{pmatrix}
-i& -3 & 3i  & 1 \\
-1 & 3i & 3 & -i \\
i & 3 & -3i & -1 \\
1 & -3i& -3 & i \\
\end{pmatrix}\, ,
\end{equation}
where $\chi \in \mathbb{R}$.\footnote{\label{chi} It will turn out that the parameter $\chi$ is related to the Euler characteristic $\chi(Y_3)$ via $\chi=\frac{\zeta(3)\chi(Y_3)}{48 \pi^3 \cK_{111}}$, where $\cK_{111}$ denotes the triple intersection number.} Here we have indicated the weight $s=-6$ and charge $q=-3$ that follow from computing its commutators with $L_0$ and $Q_\infty$ respectively. Indeed, note that $\hat{\delta}$ is not proportional to $L_{-1}$, as was the case for the type $\mathrm{I}_1$ and $\mathrm{II}_0$ boundaries. As a result the CKS recursion will become much more complicated. 

\subsubsection*{Solving the CKS recursion}

Given this boundary data, we now turn to the CKS recursion in order to reconstruct the bulk solution. Again we start by determining the initial data of this recursion, for which we need to compute $\hat{\eta}$ from the phase operator $\hat{\delta}$. Since $\hat{\delta}=\hat{\delta}_{-3}^{(-6)}$ with respect to the gradings induced by $L_0$ and $Q_\infty$ we find using \eqref{eq:eta_delta} that
\begin{equation}
\hat{\eta} = \hat{\eta}^{(-6)}_{-3}   = -\frac{15}{8}\,\hat{\delta}^{(-6)}_{-3}.
\end{equation}
The input data for the recursion relation can now be computed from $\hat{\eta}$. Since $\hat{\eta} = \hat{\eta}^{(-6)}_{-3}$ we see that the only terms in \eqref{eq:input_data} that contribute have $n=6$ and $q=3$, and therefore the only input is $\Phi_6^{6,-1}$. Note however, that in contrast to the type $\mathrm{I}_1$ boundary, the sum over weights runs over $1\leq s \leq 5$, hence we expect $\Phi_6^{6,-1}= ( (L_6^+)^{[6]}, (L_6^0)^{[6]}, (L_6^-)^{[6]})$ to be significantly more complex. Explicitly, we find
\bea\label{eq:initialdatamatrix}
\left(L^+_6\right)^{[6]} =\frac{\chi}{8} 
\left(
\begin{array}{cccc}
-3 i & 6 & 0 & 0 \\
2 & 9 i & -6 & 0 \\
0 & -6 & -9 i & 2 \\
0 & 0 & 6 & 3 i \\
\end{array}\right),\qquad 
\left(L^0_6\right)^{[6]} = \frac{\chi}{2}\left(
\begin{array}{cccc}
0 & -3 i & 0 & 0 \\
i & 0 & 3 i & 0 \\
0 & -3 i & 0 & -i \\
0 & 0 & 3 i & 0 \\
\end{array}
\right)\ , \\
\left(L^-_6\right)^{[6]} = \frac{\chi}{8} \left(
\begin{array}{cccc}
3 i & 6 & 0 & 0 \\
2 & -9 i & -6 & 0 \\
0 & -6 & 9 i & 2 \\
0 & 0 & 6 & -3 i \\
\end{array}
\right). \hspace*{3.5cm}\nn 
\eea
As the reader may verify, computing commutators for the action of the bilinear $B$ defined in \eqref{eq:bilinear_B} on $\Phi_6^{6,-1}$, given the above expressions, is not very enlightening. At this point, it will be more convenient to leverage the interpretation of $\mathfrak{g}$ as a vector space and choose a convenient basis in terms of which the action of $B$ is easier to handle. Recall that $\mathfrak{g}$ is 10-dimensional, with three of the generators given by the $\slt$-triple, which form an irreducible representation of highest weight $d=2$. The remaining seven generators form an irreducible representation of highest weight $d=6$ and can also be constructed explicitly using the $\slt$-triple. The basis we take for the irreducible representation of highest weight $d=6$ is constructed out of its highest weight state $(L_{1})^{3}$ via
\begin{equation}
\label{eq:basis_LCS}
\begin{aligned}
T_{2k}  &= \frac{i n_k}{6}  \bigg( \frac{(\ad L_{-1})^k}{k!} + \frac{(\ad L_{-1})^{6-k}}{(6-k)!} \bigg) (L_{1})^3 \qquad (k=0,1,2,3) \, ,\\
T_{2k+1}  &= \frac{n_k}{6}    \bigg( \frac{(\ad L_{-1})^k}{k!} - \frac{(\ad L_{-1})^{6-k}}{(6-k)!} \bigg) (L_{1})^3 \qquad (k=0,1,2)\, ,\\
\end{aligned}
\end{equation}
where for briefness we defined the coefficients
\begin{equation}
n_k = i^k\sqrt{\frac{1}{2} \frac{6!}{k!(6-k)!}} \times \begin{cases}
1 \qquad &\text{ for $k=0,1,2$,}\\
1/\sqrt{2} & \text{ for $k=3$.}
\end{cases} \, 
\end{equation}
Our basis is completed by fixing a basis for the irreducible representation of highest weight $d=2$, which we span by
\begin{equation}
T_{7}  =\frac{1}{2\sqrt{5}}\big( L_{1} + L_{-1} \big)\, , \qquad
T_{8}  = \frac{i}{2\sqrt{5}}\big( L_{1} - L_{-1} \big)\, , \qquad
T_{9}  = \frac{1}{2\sqrt{5}}L_0\, .
\end{equation}
We included these normalization factors because the $T_a$ are now orthonormal in the sense that
\begin{equation}
\mathrm{Tr}(T_a T_b)=\delta_{ab}\, .
\end{equation}
This allows one to easily switch between expressing elements of $\mathfrak{g}$ as $4\times 4$ matrices or as 10-component vectors. Furthermore, the adjoint action of the $\mathfrak{sl}(2,\mathbb{C})$-triple $(\ad L_{1}, \ad L_0, \ad L_{-1})$ can now be realized by $10\times 10$ matrix multiplication, and these expressions are included in \eqref{eq:adjoints} for completeness. In the $T_a$ basis we can write the initial data of the CKS recursion \eqref{eq:initialdatamatrix} as
\begin{equation}
\begin{aligned}
\left(L_6^+\right)^{[6]}  &=\frac{\chi}{8}\sqrt{\frac{15}{2}}\, \big(i\sqrt{15}T_1 -\sqrt{15} T_2-3i T_5
+ T_6\big)\ ,\\
\left(L_6^0\right)^{[6]} &= \frac{\chi}{4} \sqrt{15}\,\big( -\sqrt{5} T_3 + \sqrt{3} T_7 \big),\\
\left(L_6^-\right)^{[6]} &=\frac{\chi}{8}\sqrt{\frac{15}{2}}\,\big(-i\sqrt{15}T_1-\sqrt{15}T_2+ 3i T_5+ T_6\big).
\end{aligned}
\end{equation} 
The next step of the CKS recursion is to combine the data $(L^+_n, L^0_n,L^-_n)$ at each step into 30-component vectors $\Phi_n$. A natural $\slt$-triple $(\Lambda^+,\Lambda^0,\Lambda^-)$ that acts on these vectors is then constructed in \eqref{eq:sl2_tensor}. We can use its Casimir $\Lambda^2$ to decompose the 30-dimensional vector space further based on the highest weight $d+2\epsilon$. By construction the highest weights of our input data $\Phi_6^{6,-1}$ under $L^2$ and $\Lambda^2$ are $d=6$ and $\epsilon=-1$, hence let us put
\begin{equation}
\Phi_6^{6,-1} = -8 i \sqrt{\frac{2}{15}}\, \chi \, e^{6,-1}\, .
\end{equation}
The crucial observation is that there exist three more vectors $e^{6,1},\;e^{2,1}$ and $e^{2,-1}$ on which the action of the bilinear $B$ closes. Together these are given by
\begin{equation}\label{eq:ebasis}
\begin{aligned}
e^{6,1} &= \begin{pmatrix}
-4\sqrt{15} \, T_{0}+3i\sqrt{15}  \, T_{1}+12  \, T_{4}-3i  \, T_{5} \\
6 i \sqrt{10}  \,T_{2} -6i\sqrt{6} \, T_{6} \\
4 \sqrt{15} \, T_{0} +3 i \sqrt{15} \, T_{1} - 12  \,T_{4} - 3i  \, T_{5}
\end{pmatrix},\quad &e^{2,1} =  \sqrt{5}  \begin{pmatrix}
2i\, T_{7} +T_{8} \\
2 \,T_{9} \\
-2i \,T_{7}+  T_{8}
\end{pmatrix}, \\
e^{6,-1} &= \begin{pmatrix}
\sqrt{15} \, T_{0}+i\sqrt{15} \, T_{1}-3  \, T_{4}-i  \, T_{5} \\
2i \sqrt{10}  \, T_{2}-2i\sqrt{6}  \, T_{6} \\
-\sqrt{15}  \, T_{0}+i \sqrt{15}  \, T_{1}+3  \,T_{4}-i  \, T_{5}
\end{pmatrix}, \quad &e^{2,-1} =  \sqrt{5} \begin{pmatrix}
-i  \, T_{7}+ T_{8}\\
2\,T_{9}\\
i \,T_{7}+T_{8}
\end{pmatrix}. \\
\end{aligned}
\end{equation}
We can then write our ansatz for  $\Phi_{6n}$ as\footnote{For convenience we chose to rotate the coefficients  appearing with the vectors $e^{6,1}$, $e^{6,-1}$, $e^{2,1}$ and $e^{2,-1}$ such that the 30-component $\Phi_{6n}$ does not have multiple $a_{n},b_{n},c_{n},d_{n}$ in its entries. This  makes it easier to find the solution to the coupled system of recursion relations later.} 
\begin{equation}
\label{eq:LCS_Phi_ansatz}
\Phi_{6n}= (a_{n}+b_{n})e^{6,1}+(-3a_{n}+4b_{n})e^{6,-1}+(c_{n}+d_{n})e^{2,1}+(2c_{n}-d_{n})e^{2,-1} \, ,
\end{equation}
with the initial data of the recursion given by
\begin{equation}
a_{1} = \frac{i\chi}{56} \sqrt{\frac{15}{2}} \, , \quad b_{1}=-\frac{i\chi}{56}\sqrt{\frac{15}{2}} \, , \quad c_{1}=0\, , \quad d_{1}=0\, .
\end{equation}
For arbitrary coefficients $a_n,b_n,c_n,d_n$ it can be verified that $ B(\Phi_{6k}, \Phi_{6(n-k)})$ is again spanned by the vectors $e^{6,1}$, $e^{6,-1}$, $e^{2,1}$ and $e^{2,-1}$, so the action of the bilinear $B$ is indeed closed on this subspace. This is a huge simplification, since there are now only four components of $\Phi$ that enter the recursion relation instead of thirty. It should be noted, however, that $B$ still acts very non-trivially on these four components. For example, in the basis $(e^{6,1},\, e^{6,-1},\, e^{2,1},\, e^{2,-1})$ one finds that $B(e^{6,-1},\cdot)$ acts as
\begin{equation}
B(e^{6,-1},\cdot)=-\frac{i}{14}\left(
\begin{array}{cccc}
-\frac{i \sqrt{30}}{7} & \frac{8}{7} i \sqrt{\frac{6}{5}} & \frac{24}{5} & 0 \\
\frac{2}{7} i \sqrt{\frac{2}{15}} & \frac{12}{7} i \sqrt{\frac{6}{5}} &
-\frac{32}{15} & -\frac{112}{15} \\
-\frac{3}{14} & \frac{8}{7} & 0 & 0 \\
0 & 2 & 0 & 0 \\
\end{array}
\right).
\end{equation}
Indeed, we see that the coefficients $a_n,b_n,c_n,d_n$ all become intertwined, and already at the next step $\Phi_{12}$ each of them is non-vanishing. The recursion relation \ref{eq:CKS_recursion_epsilon} therefore reduces to a system of four coupled recursion relations, which have been given in \eqref{eq:recursion} for completeness. For illustrative purposes we also listed the coefficients that follow from iterating the recursion for the first couple of steps of the recursion in table \ref{table:recursion}. 
\begin{table}[!ht]
\renewcommand{\arraystretch}{1.5}
\centering
\begin{tabular}{|c c c c|}
	\hline
	\rule[-.4cm]{.0cm}{1cm}
	$128i\sqrt{\frac{10}{3}}\times \left(\frac{4}{\chi}\right)^n a_n$   &  $ 28i\sqrt{\frac{10}{3}}\times \left(\frac{8}{\chi}\right)^n  b_n$ &$ 5 \left(\frac{8}{\chi}\right)^n  c_n$&$ 15 \left(\frac{4}{\chi}\right)^n  d_n$ \\ \hline
	-45 & 20 & 0 & 0 \\
	36 & -36 & -22 & 9 \\
	-126 & 224 & 48 & -9 \\
	198 & -724 & -278 & 27 \\
	-450 & 3216 & 952 & -45 \\
	846 & -12168 & -4156 & 99 \\
	-1746 & 49920 & 16000 & -189 \\
	3438 & -196788 & -65446 & 387 \\
	-6930 & 791792 & 259464 & -765 \\
	13806 & -3154456 & -1044212 & 1539 \\ \hline
\end{tabular}
\caption{Values obtained for the coefficients $a_n,b_n,c_n,d_n$ by iterating the recursion relation for the first $10$ steps, where we included overall normalization factors for convenience.}
\label{table:recursion}
\renewcommand{\arraystretch}{1.0}
\end{table}

Looking at the data listed in table \ref{table:recursion} one realizes that we are dealing with quite non-trivial series of coefficients. From the recursion relations \eqref{eq:recursion} we could have expected such complications to arise, since we need to solve a coupled system where we sum over all previous terms $0<k<n$. Nevertheless we can present an exact solution to these recursions, where the coefficients $a_n$ and $d_n$ are given by relatively simple power series, while $b_n$ and $c_n$ involve the hypergeometric function $_2F_1(\mu_1,\mu_2;\nu_1;z)$. To be precise, the coefficients are given by
\begin{align}
\label{eq:LCS_recursion_solution}
a_n &= \frac{i}{7} \sqrt{\frac{3}{10}} \left(\frac{\chi}{4}\right)^n \left(1-3 (-1)^n 2^{n-2} \right)\, ,\\
b_n &=  -\frac{i}{28}  \sqrt{\frac{3}{10}} \left(\frac{\chi}{4}\right)^n \Big(2 \sqrt{3} -3 (-1)^{n}
2^{n}-2^{n+2} \binom{\frac{1}{2}}{n+1} \, _2F_1\big(1,n+\frac{1}{2};n+2;-2\big)  \Big)\, ,\\
c_n &=  -\frac{1}{5} \left(\frac{\chi}{4}\right)^n \Big(\sqrt{3}+(-1)^n 2^n-2^{n+1} \binom{\frac{1}{2}}{n+1} \, _2F_1\big(1,n+\frac{1}{2};n+2;-2\big)\Big)\, , \\
d_n &=  \frac{1}{5}\left(\frac{\chi}{4}\right)^n \left(1+(-1)^n 2^{n-1}\right)\, .
\end{align}
By resumming the series expansion in $y^{-1-3n}$ for $(\mathbf{L}^+(y),\mathbf{L}^0(y),\mathbf{L}^-(y))$ we obtain the functions
\begin{align}
a(y) &= \frac{i}{7}\sqrt{\frac{3}{10}} \frac{\chi}{4}\times  \frac{\chi+20 y^3}{y \left(4 y^3-\chi\right) \left(\chi+2 y^3\right)}, \nonumber \\ 
b(y)&=-\frac{i}{28} \sqrt{\frac{3}{10}}\times \frac{\sqrt{F(y)}}{y \left(4 y^3-\chi \right) \left(\chi +2 y^3\right)}\, , \\
c(y) &= \frac{1}{3 y}+\frac{28i}{3}  \sqrt{\frac{2}{15}} b(y)+\frac{\sqrt{F(y)} \left(F(y)-\left(\chi +8 y^3\right) \left(\chi ^3+72 \chi ^2 y^3+96 \chi  y^6+128 y^9\right)\right)}{12 \chi  y \left(4 y^3-\chi \right) \left(\chi +2 y^3\right) \left(-\chi ^3+48 \chi ^2 y^3+96 
\chi  y^6+640 y^9\right)} \, ,\nonumber \\
d(y) &= \frac{3 \chi^2}{10 y \left(4 y^3-\chi\right) \left(\chi+2 y^3\right)} \nonumber \, .
\end{align}
where for convenience we defined
\begin{equation}
F(y) = \chi ^4+288 \chi ^2 y^6+16 \chi ^3 y^3+128 \chi  y^9-8 \sqrt{2} y^{3/2} \left(\chi -4 y^3\right)^2 \left(\chi +2 y^3\right)^{3/2}+512 y^{12}\, .
\end{equation}

\subsubsection*{Reconstructing the bulk solution}

Next, we determine the polynomials $g_k$ from the $L^0_{6n}$ components of $\Phi_{6n}$. For this recursion we first have to determine the coefficients $B_k$ from \eqref{eq:Bn}. From \eqref{eq:ebasis} we see that $L^0_{6n}$ is spanned by the generators $T_2$, $T_6$ and $T_9$, which means we only have to consider the weights $s=4,0,-4$. Therefore we can express the coefficients $B_k$ as
\begin{align}
B_{3k} &=-\half\left(L_{6k-6}^0\right)^{6,4}, \quad &k&\geq 1\, ,\\
B_{3k+1} &= -\half\left(L_{6k}^0\right)^{2,0}- \half \left( L_{6k}^0\right)^{6,0},\quad &k&\geq 1\, ,\\
B_{3k+2} & =-\half \left(L_{6k+6}^0\right)^{6,-4},\quad &k&\geq 0\, .
\end{align}
By inserting our expression \eqref{eq:LCS_Phi_ansatz} for $L^0_{6k}$ we find
\begin{equation}
\begin{aligned}
B_{3k} &=  7\sqrt{\frac{5}{2}}\,  b_{k-1} \big( i \, T_2  -   T_3 \big)\, , \\
B_{3k+1} &=  -7i\sqrt{6}\, b_{k} \, T_6  + 3\sqrt{5}\, c_{k}\, T_9 \, , \\
B_{3k+2} &=  7\sqrt{\frac{5}{2}}\,  b_{k+1} \big( i \, T_2  +   T_3 \big)\, . \\
\end{aligned}
\end{equation}
The coefficients $g_k$ are determined by the recursion relation \eqref{eq:recursion_g_2}. As an alternative approach, we want to mention that one can also compute $\mathbf{L}^0(y)$ from the coefficients $b_n,c_n$, and then solve the differential equation \eqref{eq:invg_dg} for $\hat{g}(y)$ with boundary condition $\hat{g}(0)=1$. The first approach gets rather complicated since $\hat{g}_k$ is no longer Lie algebra-valued. This means it no longer suffices to work with the basis $T_a$, but instead we have to write out the matrices explicitly. Either way we can present an exact solution for this recursion, which is given by
\begin{equation}
\begin{aligned}
g_{3k} &= \alpha_{3k} \begin{pmatrix}
1+3\beta_{k} & 0 & 0 & 0 \\
0 & 3+\gamma_{k}  & 0 & 0 \\
0 & 0 & 3+\delta_{k}& 0 \\
0 & 0 & 0 & \frac{ 9(1-2k)\varepsilon_{k}+6k+1 }{2k(3/2-k)} 
\end{pmatrix}\, , \\
g_{3k+1} &= \alpha_{3k+3}\begin{pmatrix}
0 & 0 & 0 & 0 \\
0 & 0 & 0 & 0 \\
\delta_{k+1}-1 & 0 & 0 & 0 \\
0 & \frac{9(1+2k)^2 \varepsilon_k +(6k+7)(8k^2+2k-5)}{(k+1)(4k^2-1)} & 0 & 0 
\end{pmatrix} \, , \\
g_{3k+2} &= \alpha_{3k}\begin{pmatrix}
0 & 0 & 3 \beta_{k}-3 & 0 \\
0 & 0 & 0 &  \gamma_{k}-1\\
0 & 0 & 0 & 0 \\
0 & 0 & 0 & 0
\end{pmatrix} \, , \\
\end{aligned}
\end{equation}  
where for brevity we defined
\begin{equation}\label{eq:a3k}
\alpha_{3k} = \frac{(-\chi)^k}{k!} \frac{4^{-1-k}\sqrt{\pi}}{\Gamma(1/2-k)}\, ,
\end{equation}
and denoted the generalized hypergeometric functions ${}_p F_q(\mu_1,\ldots, \mu_p; \nu_1,\ldots \nu_q;z)$ by the coefficients\footnote{Note that the coefficient $\delta_{k}$ should not be confused with the phase operator $\delta$.}
\begin{equation}
\begin{aligned}
\beta_{k} &= {}_2\mathrm{F}_1\big(\frac{1}{2},-k;\frac{1}{2}-k;-2\big)\, , \\
\gamma_{k} &= {}_2\mathrm{F}_1\big(-\frac{1}{2},-k;\frac{1}{2}-k;-2\big)\, , \\
\delta_{k} &= {}_3\mathrm{F}_2\big(-\frac{1}{2}, \frac{5}{6}, -k; -\frac{1}{6},\frac{1}{2}-k;-2 \big) \, , \\
\varepsilon_{k} &= {}_2\mathrm{F}_1\big(\frac{1}{2},1-k;\frac{1}{2}-k;-2\big) \, .
\end{aligned}
\end{equation}
Note that we have chosen to rotate back to the real basis by $g_{k}=\rho^{-1} \hat{g}_{k} \rho$.

By resumming the coefficients $g_{k}$ as a series expansion in $y^{-k}$ one recovers the matrix-valued function $g(y)$. Since the expressions for these kinds of functions are rather bulky around the large complex structure point, we choose to directly rotate to the matter fields by $\tilde h(y)=g(y)y^{-1/2 N^0}$ and find
\begin{equation}
\label{eq:LCS_h}
\boxed{\rule[-1cm]{.0cm}{2.2cm}\quad \Scale[0.90]{
	\tilde h(y) = \alpha(y) \left(
	\begin{array}{cccc}
	1 + 3 \beta(y)^{-1} & 0 &  3 \beta(y)^{-1} -3 & 0 \\
	0 & 3y +y \beta(y) & 0 & y  \beta(y) -y \\
	\frac{\gamma(y)}{\alpha(y)}-4 y^2  & 0 & \frac{\gamma(y)}{\alpha(y)}& 0 \\
	0 &  3\chi-3 y^3+3y^3 \beta(y) & 0 & y^3-\chi+3y^3  \beta(y) \\
	\end{array}
	\right)}\ ,\quad}
\end{equation}
where for convenience we defined the functions\footnote{Note that the function $\alpha(y)$ corresponds to the series $\alpha(y) = y^{-3/2}\sum_k \alpha_{3k} y^{-3k}$ for the coefficients defined in \eqref{eq:a3k}.}
\begin{equation}
\label{eq:LCS_normalisation}
\alpha(y)=\frac{1}{2\sqrt{4y^3-\chi}}\, , \qquad \beta(y)=\sqrt{1+\frac{\chi }{2 y^3}}\, , \qquad
\gamma(y) =\sqrt{2} \frac{y^3-\chi +3 y^3 \beta(y)}{2y^{5/2}  \sqrt{\frac{\chi  \left(2
		y^3-\chi \right)}{y^6}+8}}\, .
\end{equation}
In principle one can rotate the $x$-dependence of the matter fields into the solution by $h(x,y)=e^{xN^-}\tilde h(y)$, but the resulting expression would become quite cumbersome and has therefore been omitted. Since this dependence enters through the relatively simple factor $e^{xN^-}$ anyway, one can simply treat this factor separately when analyzing the bulk solution. Note that, in contrast to the $\mathrm{I}_1$ and $\mathrm{II}_0$ boundaries, the parameter $\chi$ that entered through the phase operator $\delta$ cannot be removed by some coordinate redefinition here.

\section{On reconstructing periods from the bulk solution}
\label{period_section}
In this section we describe another way to represent the information that the matter fields $h(x,y)$ carry, which is particularly elegant for Calabi--Yau type boundaries. This data can be captured by the asymptotic behaviour of the $(D,0)$-form periods of a Calabi--Yau manifold. In practice one can compute these periods directly for geometrical examples, and recently these techniques have been applied in the context of the swampland program in \cite{Blumenhagen:2018nts,Demirtas:2019sip,Demirtas:2020ffz,Blumenhagen:2020ire}. In order to match the two descriptions, we find that one has to include essential exponential corrections to these periods for most boundaries. For $D=3$ and $h_{1/2}=1$ we demonstrate how this works in practice, and show how to reinterpret the bulk solutions of section \ref{reconstructing_examples} geometrically in terms of local period expressions.

\subsection{Procedure for matching periods with the matter fields}
\label{sec:period_vector}
Let us begin with describing the periods of the $(D,0)$-form $\Omega$ of a Calabi--Yau manifold $Y_D$. By fixing an integral basis for the $D$-form cohomology $H^D(Y_D, \mathbb{R})$ one can represent $\Omega$ by its periods as a vector $\Pi$. This period vector $\Pi(t)$ varies holomorphically over the complex structure moduli space $\cM^{\rm cs}(Y_D)$. In general it depends in a highly non-trivial manner on the complex structure modulus $t$, since it requires one to solve complicated Picard-Fuchs equations. Nevertheless, when moving close to boundaries in moduli space, one can give a precise description by using the nilpotent orbit theorem \cite{Schmid}, which states that the period vector admits a near-boundary expansion
\begin{equation}
\label{eq:nilpotent_expansion}
\Pi(t) = e^{tN^-}\left(a_0 + a_1 e^{2\pi i t} + a_2 e^{4\pi i t}+\ldots\right).
\end{equation}
The factor involving the lowering operator $N^-$ captures the monodromy behavior of the period vector under circling singularities by $t \to t+1$, similar to the group-valued matter field in \eqref{discrete_sym}. One can think of the terms $a_1,a_2,\ldots$ as non-perturbative terms, which become exponentially small when the complex structure modulus is sent to the boundary $t \to i \infty$. 

Let us ignore these non-perturbative terms for the moment, and first try to recover the polynomial part of the periods in the coordinate $t=x+iy$ from the matter fields $h(x,y)$. Recall that the charge operator $Q_\infty$ induces a decomposition of the Hilbert space $\cH$ into charge eigenspaces $\cH^\infty_q$. The polynomial part of the $(D,0)$-form periods in \eqref{eq:nilpotent_expansion} can then be related to a particular charge eigenstate via
\begin{equation}\label{eq:matchperiods}
\Pi_{\rm nil} = e^{tN^-} a_0 = h(x,y) e^{-\zeta} |\tfrac{D}{2}\rangle\, .
\end{equation}
Notably this relation includes corrections to the periods due to the phase operator $\delta$, such as the $\alpha'$-correction with the Euler characteristic at the large complex structure point. Meanwhile exponential corrections to the periods are not covered in \eqref{eq:matchperiods} by the matter fields $h(x,y)$.

Nevertheless the matter fields $h(x,y)$ do capture information about essential non-perturbative terms for some boundaries, which can be seen as follows. Although $h(x,y)$ does not account for non-perturbative terms in the $(D,0)$-form periods directly via \eqref{eq:matchperiods}, it does determine the polynomial periods of every $(p,q)$-form, and not just the $(D,0)$-form. Recall that for Calabi--Yau manifolds other $(p,q)$-forms (with $p+q=3$) can be obtained as linear combinations of the $(D,0)$-form and its derivatives. For Calabi--Yau threefolds this applies for any element of the threeform cohomology $H^3(Y_3,\mathbb{C})$, while for fourfolds one has to restrict to the primitive\footnote{Here the four-forms $\omega \in H^4_p(Y_4, \mathbb{C})$ are primitive with respect to the K\"ahler form $J$ in the sense that they satisfy the primitivity condition $J \wedge \omega = 0$.}  cohomology $H^4_p(Y_4, \mathbb{C})$. The idea is now that after taking derivatives of the $(D,0)$-form periods the terms involving $a_0$ can vanish, which results in non-perturbative terms such as $a_1$ making up the other $(p,q)$-forms to leading order. To illustrate this point let us explicitly differentiate \eqref{eq:nilpotent_expansion}, which yields
\begin{equation}
\partial_t \Pi  = e^{tN^-}\left(N^- a_0 + (N^- + 2\pi i )a_1 e^{2\pi i t} + \ldots\right).
\end{equation}
When the first term vanishes, this means that the non-perturbative term $a_1$ can already enter in the periods of a $(D-1,1)$-form at polynomial order (after removing an irrelevant overall factor $e^{2\pi i t}$). This is precisely what happens at $\mathrm{I}_1$ boundaries such as the conifold point. We can extend this argument inductively to higher order derivatives. Defining $n_i$ as the lowest integer such that $N^{n_i} a_i  =  0$, one finds that the term $a_{i+1}$ has to be included whenever $n_0+\ldots + n_i < D$.  

Having discussed how other $(p,q)$-forms can capture non-perturbative data at leading order, we are ready to describe how we can determine these polynomial periods. We begin by explaining how one can compute the polynomial periods of the $(p,q)$-forms from the periods of the $(D,0)$-form. First one computes the derivatives of the period vector \eqref{eq:nilpotent_expansion}, and combines them into the increasing filtration
\begin{equation}\label{eq:Fdef}
F^p(t)= \mathrm{span}_{\mathbb{C}}\{   \partial^k_t \Pi \  | \  0 \leq k \leq D- p  \}   \, .
\end{equation}
For large $y \gg 1$ these vector spaces can then be approximated  by $F^p_{\rm pol}$ that vary, up to an overall rescaling in any direction, as a polynomial in $t=x+iy$. To be precise, we can write
\begin{equation}\label{eq:Fpol}
F^p (t) \approx F^p_{\rm pol}(t) = e^{tN^-} F^p_0\, ,
\end{equation}
where $F^p_0$ are vector spaces independent of $t$. The spaces $F^p_{\rm pol}$ are spanned by the polynomial part of the periods of the $(r,s)$-forms with $r \geq p$, and likewise the  $\overline{F}^q_{\rm pol}$ by $(r,s)$-forms with $s \geq q$. By taking the intersection between these vector spaces, we find that the $(p,q)$-forms follow from
\begin{equation}\label{eq:Fpol_to_Hpol}
H^{p,q}_{\rm pol} = F^p_{\rm pol} \cap \overline{F}^q_{\rm pol} \, .
\end{equation} 
Alternatively one can obtain the polynomial periods of the $(p,q)$-forms from the matter fields via the eigenspaces $\cH^\infty_q$ of the charge operator $Q_\infty$. These matter fields interpolate between the Hodge structure $\cH^\infty_q$ at the boundary and the Hodge structure $H^{p,q}_{\rm pol} $ in the bulk of the moduli space. To be precise, $h(x,y)$ relates the vector spaces by
\begin{equation}
\label{eq:Hpol}
H^{p,D-p}_{\mathrm{pol}}= h(x,y) e^{-\zeta}\cH^\infty_{p-D/2}\ .
\end{equation}
We can now relate the two descriptions of the bulk solution  by computing the vector spaces $H^{p,D-p}_{\mathrm{pol}}$ via these two approaches. Either we compute the polynomial periods of the $(p,q)$-forms via \eqref{eq:Fpol} and \eqref{eq:Fpol_to_Hpol} from the periods of the $(D,0)$-form, or we determine the eigenspaces of $Q_\infty$ and apply the group-valued matter field $h(x,y)$ following \eqref{eq:Hpol}. Let us mention that, in principle, one can even re-engineer the $(D,0)$-form periods including essential non-perturbative terms from the polynomial periods that follow from the matter fields by using \eqref{eq:Hpol}. However, it turns out that there is a more convenient approach to integrate the boundary data directly into periods instead of computing $h(x,y)$ first, and this direction will be explored further in \cite{Bastiantoappear}.

\subsection{Matching periods for bulk reconstruction examples}
\label{sec:period_vector_examples}
In this section we discuss how we can reformulate the bulk solutions obtained in section \ref{reconstructing_examples} for our three examples in terms of $(3,0)$-form periods. We start with $\mathrm{IV}_1$ boundaries: these do not require us to work with non-perturbative corrections, making the match of the two descriptions more straightforward. We then move on to $\mathrm{I}_1$ and $\mathrm{II}_0$ boundaries, where we will demonstrate how non-perturbative terms for the $(3,0)$-form periods are needed to match with the bulk solutions for the matter fields $h(x,y)$. For each of these examples we find that $\zeta=0$ by using \eqref{eq:zeta}, so the factor of $e^{-\zeta}$ can be ignored in \eqref{eq:matchperiods} and \eqref{eq:Hpol} in the following.

\subsubsection*{Type $\mathrm{IV}_1$ boundaries}
Let us begin by writing down the periods of the $(3,0)$-form for $\mathrm{IV}_1$ boundaries. From a geometrical perspective we identify the $\mathrm{IV}_1$ boundary as a large complex structure point, which allows one to write down the periods in terms of the topological data of the mirror Calabi--Yau manifold. In order to simplify our expressions, we have used a basis transformation and a rescaling of the period vector compared to e.g.~\cite{Grimm:2018cpv}.\footnote{To be precise, these redefinitions get rid of the intersection number $\cK_{111}$ appearing with $t^2$ and $t^3$, and the integrated second Chern class in the third entry. The parameter $\chi$ is related to the Euler characteristic as described in footnote \ref{chi}.} The period vector then takes the form
\begin{equation}\label{eq:LCSperiods}
\Pi = \begin{pmatrix}
1 \\
t \\
t^2 \\
t^3+i \chi
\end{pmatrix}\, ,
\end{equation}   
where we wrote $t=x+iy$. We now want match these periods to the bulk solution for the matter fields \eqref{eq:LCS_h} by using \eqref{eq:matchperiods}. In order to make this match we must first identify the state with charge $q=3/2$ under $Q_\infty$. Recalling the boundary data from table \ref{table:data} for the $\mathrm{IV}_1$ boundary, we find that
\begin{equation}
\left|3,3;\tfrac{3}{2} \right\rangle  = \frac{1}{\alpha(y)}\begin{pmatrix} 1 \\ 
i \\
-1 \\
-i \end{pmatrix}\, .
\end{equation}
where we normalized by a factor of $\alpha(y)$ for later convenience, which is defined in \eqref{eq:LCS_normalisation}. Following \eqref{eq:matchperiods} we then act with the group-valued matter field \eqref{eq:LCS_h} on the above charge eigenstate, which yields
\begin{equation}
\label{eq:LCS_period}
h(x,y) \left|3,3; \tfrac{3}{2} \right\rangle  =  \begin{pmatrix}
1 \\
t \\
t^2 \\
t^3+i \chi
\end{pmatrix}.
\end{equation}
Note that this indeed agrees with the periods given in \eqref{eq:LCSperiods}. Thus the bulk matter solution we found in this work \eqref{eq:LCS_h} matches with the familiar periods from the large complex structure point. In principle one can now continue and check the polynomial periods of the other $(p,q)$-forms by using \eqref{eq:Hpol}. However, these periods would follow directly from the polynomial periods in \eqref{eq:LCSperiods} which have already been matched, so these checks do not lead to any further insights for $\mathrm{IV}_1$ boundaries.

\subsubsection*{Type $\mathrm{I}_1$ boundaries}
For our next example we treat the $\mathrm{I}_1$ boundaries. From a geometric perspective these boundaries are identified with conifold points, which arise for instance in the moduli space of the mirror quintic \cite{Candelas:1990rm}. The period vector of the $(3,0)$-form for these boundaries takes the form
\begin{equation}
\label{eq:period_vector_conifold}
\Pi = \begin{pmatrix}
1 \\
b \, e^{2\pi i t} \\
b \, e^{2\pi i t}( t-ic - \frac{1}{2\pi i}) \\
i +\frac{b^2}{4\pi i}e^{4\pi it}
\end{pmatrix}\, ,
\end{equation}
where $b \in \mathbb{C}$ is some model-dependent coefficient. Note that, in contrast to the period vector of the type $\mathrm{IV}_1$ boundary, this period vector has multiple essential exponential corrections. 

We now want to match these periods to the bulk solution \eqref{eq:I1_h}. While one could match the polynomial periods of the $(3,0)$-form via \eqref{eq:matchperiods}, these are simply constants so this check is rather trivial. Let us therefore directly consider the match \eqref{eq:Hpol} for the polynomial periods of all $(p,q)$-forms. First we need to compute the filtration of polynomial periods $F^p_{\rm pol}$ as defined in \eqref{eq:Fpol}. We can represent these periods by a so-called period matrix as
\begin{equation}\label{eq:I1periodmatrix}
\Pi_{\rm pol} = e^{tN^-} \begin{pmatrix}
1 & 0 & 0 & 0\\
0 & 1 & 0 & 0\\
0 & -ic & 1 & 0 \\
i & 0 & 0 & 1\\
\end{pmatrix} \, ,
\end{equation}
where the first $k$ columns span the vector space $F^{3-k}_{\rm pol}$. One can compute these vector spaces $F^{p}_{\rm pol}$ explicitly by taking derivatives of the period vector according to \eqref{eq:Fdef} and thereafter truncating to the polynomial periods. To be precise, the second and third column follow from the periods at order $e^{2\pi i t}$ in \eqref{eq:period_vector_conifold}, while the last column follows from the periods at order $e^{4\pi i t}$. We can then compute the polynomial periods of the $(p,q)$-forms explicitly by determining the intersection of $F^{p}_{\rm pol}$ with $\overline{F}^{q}_{\rm pol}$ (for $p+q=3$) according to \eqref{eq:Fpol_to_Hpol}, which yields
\begin{equation}\label{eq:I1periods}
\begin{aligned}
H^{3,0}_{\rm pol} &: \quad\begin{pmatrix} 1 , & 0 , & 0 , & i \end{pmatrix}, \\
H^{2,1}_{\rm pol} &: \quad \begin{pmatrix} 0 , & 1 , & t-ic , & 0 \end{pmatrix}, \\
H^{1,2}_{\rm pol} &: \quad \begin{pmatrix} 0 , & 1 , & \bar{t}+ic , & 0 \end{pmatrix}, \\
H^{0,3}_{\rm pol} &: \quad \begin{pmatrix} 1 , & 0 , & 0 , & -i \end{pmatrix}, \\
\end{aligned}
\end{equation}
where we indicated a representative for each vector space. We now want to show that we can obtain the same periods from the bulk matter solution \eqref{eq:I1_h} by using \eqref{eq:Hpol}. In order to make this match we first have to determine the charge eigenstates with respect to $Q_\infty$. Recalling the boundary data from table \ref{table:data} we find the states
\begin{equation}
\begin{aligned}
\left|0,0;\tfrac{3}{2}\right\rangle &= \begin{pmatrix}
1, & 0, & 0, & i
\end{pmatrix}, \\
\left|1,1;\tfrac{1}{2}  \right\rangle &= \sqrt{y-c} \begin{pmatrix}
0, & 1, & i, & 0
\end{pmatrix},\\
\left|1,-1;-\tfrac{1}{2}  \right\rangle &= \sqrt{y-c} \begin{pmatrix}
0, & 1, & i, & 0
\end{pmatrix}, \\
\left|0,0;-\tfrac{3}{2}\right\rangle &= \begin{pmatrix}
1, & 0, & 0, & -i
\end{pmatrix}, \\
\end{aligned}
\end{equation}
where we normalized some states by a factor of $\sqrt{y-c} $ for later convenience. Following \eqref{eq:Hpol}, one can easily verify that application of the group-valued matter field \eqref{eq:I1_h} on these charge eigenstates indeed reproduces the polynomial periods \eqref{eq:I1periods}.

\subsubsection*{Type $\mathrm{II}_0$ boundaries}
Finally we discuss the periods for $\mathrm{II}_0$ boundaries. Geometrically these boundaries can arise in the complex structure moduli space of Calabi--Yau threefolds by performing Tyurin degenerations. The period vector of the $(3,0)$-form near these singularities reads
\begin{equation}\label{eq:II0periods}
\Pi(t) = \begin{pmatrix}
1 \\
i \\
-i c+t \\
c+i t
\end{pmatrix} + b \, e^{2\pi i t}\begin{pmatrix}
1 \\
-i \\
-ic+t+i/\pi  \\
-c-i t+1/\pi
\end{pmatrix},
\end{equation} 
where $ b \in \mathbb{C}$ is some model-dependent coefficient. Again, in contrast to the type $\mathrm{IV}_1$ case, this period vector contains a (single) essential non-perturbative term. 

We now want to match these periods to the bulk solution \eqref{eq:II0_h}. For brevity we directly discuss the match for the polynomial periods of all $(p,q)$-forms according to \eqref{eq:Hpol}, instead of just the match for the $(3,0)$-form given by \eqref{eq:matchperiods}. First we compute the filtration of polynomial periods $F^p_{\rm pol}$ as defined in \eqref{eq:Fpol}. We can represent this filtration by the period matrix
\begin{equation}
\Pi_{\rm pol} = e^{t N^-} \begin{pmatrix}
1 & 0 & 1 & 0\\
i & 0 & -i & 0 \\
-ic & 1 & -ic & 1\\
c & i & -c & -i
\end{pmatrix}\, ,
\end{equation}
where the first $k$ columns again span the vector space $F^{3-k}_{\rm pol}$. These vector spaces can be computed directly from the period vector \eqref{eq:II0periods} by taking derivatives and truncating to the leading terms. The first two columns correspond to the polynomial periods of the $(3,0)$-form, while the last two columns correspond to the periods at order $e^{2\pi it}$. By intersecting the vector spaces according to \eqref{eq:Fpol_to_Hpol} we find that the polynomial periods of the $(p,q)$-forms are given by
\begin{equation}\label{eq:II0periods_2}
\begin{aligned}
H^{3,0}_{\rm pol} &: \quad \begin{pmatrix} 1, & i, & t-ic, & it+c \end{pmatrix} , \\ 
H^{2,1}_{\rm pol} &: \quad \begin{pmatrix} 1, & i, & \bar{t}+ic, & i\bar{t}-c \end{pmatrix} , \\ 
H^{1,2}_{\rm pol} &: \quad \begin{pmatrix} 1, & -i, & t-ic, & -it-c \end{pmatrix} , \\ 
H^{0,3}_{\rm pol} &: \quad \begin{pmatrix} 1, & -i, & \bar{t}+ic, & -i\bar{t}+c \end{pmatrix} , \\ 
\end{aligned}
\end{equation}
where we indicated a representative for each vector space. We now want to show that we can obtain the same periods from the bulk matter solution \eqref{eq:II0_h} by using \eqref{eq:Hpol}. In order to make this match we first have to determine the charge eigenstates of $Q_\infty$. Recalling the boundary data from table \ref{table:data}, we find the states
\begin{equation}
\begin{aligned}
\left|1,1; \tfrac32\right\rangle &= \sqrt{y-c} \begin{pmatrix}1 , & i, & i, & -1 \end{pmatrix}, \\
\left|1,-1; \tfrac12\right\rangle &=\sqrt{y-c}  \begin{pmatrix}1, &  i, &  -i, & 1 \end{pmatrix}, \\
\left|1,1; - \tfrac12\right\rangle &=\sqrt{y-c}  \begin{pmatrix}1, &  -i, &  i, & 1 \end{pmatrix}, \\
\left|1,-1; -\tfrac32\right\rangle &= \sqrt{y-c} \begin{pmatrix}1 , & -i, & -i, & -1 \end{pmatrix}, \\
\end{aligned}
\end{equation}
where we normalized the states by a factor of $\sqrt{y-c}$. Following \eqref{eq:Hpol}, one can easily verify that application of the group-valued matter field \eqref{eq:II0_h} on these charge eigenstates indeed reproduces the polynomial periods \eqref{eq:II0periods_2}. 

\section{Conclusions}

In this work we have studied the solutions of a special non-linear sigma-model depending on group-valued matter fields $h$
varying over a two-dimensional space-time. This model was introduced to encode couplings 
of certain supersymmetric effective field theories that are compatible with a UV completion in string theory. 
In this application it is crucial to identify the two-dimensional space-time of the sigma-model
with the moduli space of the effective theory.  
Studying this sigma-model abstractly, i.e.~without any reference to the possibly underlying geometric string theory realization, we have 
introduced a recursive way to reconstruct the matter field solutions near the boundaries of the two-dimensional space-time. 
The reconstructed near-boundary solutions were uniquely fixed by a set of Sl(2)-boundary conditions 
and a single matching condition. In applications to effective actions such solutions are of particular interest, since they encode couplings that are constrained by multiple swampland conjectures. Investigating these near-boundary solutions therefore provide us 
with a general way to probe the string landscape in an abstract manner.
Furthermore, the fact that the the solutions can be reconstructed starting from the boundary supports the holographic perspective put forward in 
\cite{Grimm:2020cda} and gives an interesting realization of a bulk-boundary correspondence. 
Notably, the proposed correspondence can be formulated independently of any geometric realization and is therefore intriguing in its own right. In this work we elucidated the abstract constructions that enter in this correspondence and successfully applied it to a particular set of examples.

In the reconstruction of the matter field solutions $h$ we restricted ourselves to solving 
the equations of motion in a two-dimensional background metric that is flat up to a general Weyl factor.
Furthermore, we imposed three conditions to control their asymptotic behaviour. First, we required the existence of a continuous symmetry 
$h(x+c,y) = e^{cN^-} h(x,y)$ that arose as a continuous version of the discrete monodromy transformations. The universal emergence of such a continuous global symmetry at infinite distance boundaries in the associated effective theories is consistent with the recent discussions in  \cite{GPV,Lee:2019wij,Lanza:2020qmt}. The second restriction we imposed is the $Q$-constraint. We stress that this constraint does not follow from the action principle and should be regarded as an additional requirement to obtain `physical' solutions \cite{Cecotti:2020rjq,Grimm:2020cda}. It roughly states that 
the solutions are compatible with the charge decomposition of the Hilbert space at every point near the boundary. 
While mathematically well understood within asymptotic Hodge theory \cite{Schmid, CKS}, it would be interesting to give a 
precise physical interpretation of this constraint from the point of view of the sigma-model. 
In particular, it is desirable to check if it can be understood a gauge constraint in an extended version of the model. Lastly, we imposed specific boundary conditions on the matter fields. In particular, we demanded that the leading behaviour of the operators $\cN^\bullet(y)$, as $y\rightarrow\infty$, is governed by an $\slt$-triple. This triple plays a central role in the bulk reconstruction procedure through the boundary decomposition it induces on both the Hilbert space and the operators. Still, these three conditions do not fix the solution to the equations of motion completely. This is achieved by the matching condition with the phase operator $\delta$, which comprises the final piece of the boundary data. Indeed, while the $\slt$-triple specifies the leading behaviour of the matter fields, the CKS recursion together with the matching condition fixes all sub-leading terms $g_k$. 
The striking result is that in this way a rather simple set of boundary conditions and the CKS recursion uniquely determines an infinite series of matrix-valued coefficients in the $1/y$-expansion. 

In order to illustrate how the bulk reconstruction is performed in practice we have solved the CKS recursion for a concrete set of boundary data. From the classification of possible boundary data we have considered all examples that are of Calabi--Yau type in $D=3$ that 
have $h_{1/2}=1$. The three possible cases were denoted by $\mathrm{I}_1$, $\mathrm{II}_0$ and $\mathrm{IV}_1$. 
While it is important to stress that we did not require a geometric realization in order to arrive at this data, we have found that the corresponding 
solutions  can be associated to three well-known degenerations realized in geometric Calabi--Yau threefold moduli spaces known as the conifold point, the 
Tyurin degeneration, and the large complex structure point.  
In order to make this identification we explained the relationship between the matter field solutions matching the sl(2) boundary data of Calabi--Yau type and the asymptotic period vector of the $(D,0)$-form on a Calabi--Yau $D$-fold. Here it is interesting to 
note that the CKS recursions for the cases $\mathrm{I}_1$ and $\mathrm{II}_0$ admit rather simple solutions. Nevertheless we argued 
that matching the bulk solutions to geometric periods one is able to determine that certain instanton corrections are 
required in the asymptotic periods.
In contrast, for the example $\mathrm{IV}_1$ the CKS recursion was much more involved and required solving a set of four coupled recursion relations, while its relation to the period vector is straightforward. The intricate dependence of the complexity of the procedure on the boundary data raises the question of whether it is possible to write down a general solution to the CKS recursion, especially when considering higher-dimensional moduli spaces, and poses an interesting challenge for future investigations.

There are numerous further avenues that can be taken to extend the proposed bulk-boundary correspondence and
elucidate the physical nature of the model as a whole.
As a first step we have already studied symmetries of the bulk action in this work. We discovered a global symmetry for left multiplication with an associated conserved current, while right multiplication gave rise to a gauge symmetry. In order to extend these models, one immediate question is to find a generalization to higher-dimensional moduli spaces. While the matter action admits a straightforward higher-dimensional generalisation  \cite{Cecotti:2020rjq}, it is very non-trivial to extend the holographic perspective and the CKS recursion.  To complement this, it is also interesting to couple the matter action to gravity. For the two dimensional settings considered here, we recall 
that Einstein gravity is of topological nature. This fact, together with the desire to fix the background metric to be the so-called Weil-Petersson metric,  lead to the proposal of \cite{Grimm:2020cda} to couple the matter action to an alternative gravitational theory in two dimensions known as JT-gravity \cite{Teitelboim:1983ux,Jackiw:1984je}. In this proposal the Hodge metric, which is constructed from the on-shell conserved currents, is the crucial link between the matter action and the gravity action. While this proposal might be intriguing, it is also conceivable that the Hodge metric appears as the space-time metric itself. It is thus desirable to study the gravity coupled action in two dimensions and lay the groundwork for the generalization to higher dimensions. 

Let us also briefly comment on the reconstruction of the period vector. Our discussion focussed on the derivation of the period vectors from 
the bulk field solutions via the construction of a nilpotent orbit. We have then compared these results with the known leading 
expressions for the period vector for Calabi--Yau threefold examples. In an upcoming project, it is shown how bulk solutions with sl(2) boundary data can always be integrated to an asymptotic period vector \cite{Bastiantoappear}. This yields the immediate question of whether or not one can reconstruct the full period vector on the entire moduli space including all instanton corrections when taking into account the data supplied for all boundaries. This would imply the exciting possibility that in fact the full bulk geometry is entirely encoded in the boundary information and would give further support for the existence 
of a fully holographic description of the moduli space.

In developing further a holographic perspective one also has to face several open questions regarding the boundary information. The first is whether there exists a proper `boundary theory' that may give rise to the various components of the boundary data in a physical manner. Note that both the Hodge metric and the Weil-Petersson metric, the latter at infinite distance boundaries, asymptote to an $\mathrm{AdS}_2$ metric. This  indicates that 
the encountered bulk-boundary correspondence is some manifestation of the $\mathrm{AdS}_2$/$\mathrm{CFT}_1$ duality. 
Especially interesting in this regard is the role that the phase operator would play in such a boundary theory. Indeed, the origin of the phase operator, as well as the matching condition are crucial in order to understand the bulk-boundary correspondence further, and it is tempting to speculate that the proposed presence of emergent and axionic strings \cite{Lee:2018urn,Lee:2018spm,Marchesano:2019ifh,Lee:2019xtm,Lee:2019wij, Baume:2019sry, Lanza:2020qmt,Klaewer:2020lfg, Lanzatoappear} could serve 
as a guide to clarify their appearance. It is important to emphasize that while we have restricted our discussion of explicit examples to boundaries of Calabi--Yau type this was purely a choice made to connect with the existing swampland literature. In elucidating the bulk-boundary correspondence it would be illustrative to reconstruct the bulk solutions for boundaries that are not of Calabi--Yau type. For example it is interesting to consider the moduli spaces of Riemann surfaces, since these are already well-studied from the Hodge theory perspective \cite{Hain:2003}. It is an exciting prospect that already by changing the boundary conditions the holographic approach may also be extended to certain non-supersymmetric effective actions. 

\subsubsection*{Acknowledgments}

It is a great pleasure to thank Tarek Anous, Brice Bastian, Mike Douglas, Arno Hoefnagels, Chongchuo Li, Eran Palti, Erik Plauschinn, Christian Schnell, and Irene Valenzuela 
for very useful discussions and correspondence. 
This research is partly supported by the Dutch Research Council (NWO) via a Start-Up grant and a VICI grant.

\appendix
%%%%%%%%%%%%%%%%%%%%%%%%%%%%%%%
%%%%%%%%%%%%%%%%%%%%%%%%%%%%%%%
%%%%%%%%%%%%%%%%%%%%%%%%%%%%%%%

\section{Computations on the $Q$-constraint and the CKS input data}
\label{app:input}

In this section some properties of the input data $\Phi_n^{n,-1}$ are discussed in terms of the boundary data. We first explain how the $Q$-constraint can be written in terms of the complex algebra and the boundary charge operator $Q_\infty$. The resulting constraint is then used to fix the coefficients of the input data. We also show that the expression \eqref{eq:input_data} for the input data indeed has the desired properties, i.e.~it has $(d,\epsilon)=(n,-1)$.

\subsection{Rewriting the $Q$-constraint}

We recall the $Q$-constraint, as given in \eqref{Q-constr_2}
\begin{equation}
[Q_\infty,\cN^0]=i\left(\cN^+ + \cN^-\right)\, ,\qquad [Q,\cN^\pm] = -\frac{i}{2}\cN^0\, .
\end{equation}
Let us first write the above equations in the complex algebra by recalling \eqref{matchzeta} and \eqref{def-bfLbullet}
\begin{equation}
\label{eq:rho_Q_rhoinv}
[\rho Q_\infty \rho^{-1},\mathbf{L}^0]=i\left(\mathbf{L}^+ + \mathbf{L}^-\right)\, ,\qquad [\rho Q_\infty\rho^{-1},\mathbf{L}^\pm] = -\frac{i}{2}\mathbf{L}^0\, .
\end{equation}
By using \eqref{def-rho} and the commutation relations between $Q_\infty$ and the $L_\alpha$, one computes
\begin{align}
\rho Q_\infty \rho^{-1}&= e^{\frac{i\pi}{4}\ad\left(L_{1}+L_{-1}\right)} Q_\infty\\
&= Q_\infty+\sum_{n=1}^\infty 4^{n-1}\left[\frac{(i\pi /4)^{2n}}{(2n)!} 2L_0-\frac{(i\pi /4)^{2n-1}}{(2n-1)!}\left(L_{1}-L_{-1}\right) \right]\\
&= Q_\infty - \frac{1}{2}L_0 - \frac{i}{2}L_{1} + \frac{i}{2} L_{-1}\, .
\end{align}
Inserting this result in \eqref{eq:rho_Q_rhoinv} then immediately yields the $Q$-constraint as presented in \eqref{charges_cL}
\begin{align}
\label{eq:Q_constr_coeff}
[2Q_\infty-L_0, \mathbf{L}^0] &= 2i(\mathbf{L}^++\mathbf{L}^-)+i[L_{1},\mathbf{L}^0]-i[L_{-1},\mathbf{L}^0]\, ,\\
[2Q_\infty-L_0, \mathbf{L}^\pm] &= -i\mathbf{L}^0 +i [L_{1}, \mathbf{L}^\pm]-i[L_{-1}, \mathbf{L}^\pm]\, .
\end{align} 

\subsection{Imposing the $Q$-constraint on the CKS input data}

We recall that the input data for the CKS recursion is given by vectors $\Phi^{n,-1}_{n}$ which have $d=n$ and $\epsilon=-1$. In the main text the following expression for the input data $\Phi_n^{n,-1}$ was given in terms of the boundary data
\begin{equation}
\Phi_n^{n,-1}= \sum_{1\leq s,q \leq n-1} a^{n,s}_q \begin{pmatrix}
-\frac{1}{n-s} \left(\mathrm{ad}\;L_{1}\right)^{s+1}\\
2 \left(\mathrm{ad}\;L_{1}\right)^{s} \\
(n-s+1) \left(\mathrm{ad}\;L_{1}\right)^{s-1}
\end{pmatrix}\hat{\eta}^{(-n)}_{-q}\, .
\end{equation}
In this section we will show that indeed this $\Phi_n^{n,-1}$ has eigenvalue $\epsilon=-1$. By the characterization \ref{eq:B_eigenvalue} of the eigenvalue $\epsilon$, it suffices to show that
\begin{equation}
4B_0(\Phi_n^{n,-1})=(n+2)\Phi_n^{n,-1}\, ,
\end{equation}
where 
\begin{equation}
4B_0= \begin{pmatrix}
\ad L_0 & -\ad L_{1} & 0\\
-2\ad L_{-1} & 0 & 2 \ad L_{1}\\
0 & \ad L_{1} & -\ad  L_0
\end{pmatrix}.
\end{equation}
Inserting the expression for $\Phi_n^{n,-1}$ into the above equation, we obtain
\begin{equation}
4B_0(\Phi_n^{n,-1}) = \sum_{s,q} a^{n,s}_q \begin{pmatrix}
-\frac{1}{n-s} \ad L_0 (\ad L_{1})^{s+1} - 2 (\ad L_{1})^{s+1}\\
\frac{2}{n-s} \ad L_{-1} (\ad L_{1})^{s+1} +2(n-s+1)(\ad L_{1})^s\\
2 \ad L_{-1}(\ad L_{1})^s - (n-s+1) \ad L_0 (\ad L_{1})^{s-1}
\end{pmatrix}\hat{\eta}^{(-n)}_{-q}\, .
\end{equation}
Next, we use the Jacobi identity and the $\slt$-algebra to write
\begin{align}
\ad L_0 (\ad L_{1})^{s}&= (\ad L_{1})^s(\ad L_0 +2s)\, ,\\
\ad L_{-1}(\ad L_{1})^s &= s (\ad L_{1})^{s-1}(-\ad L_0-s+1)+(\ad L_{1})^s \ad L_{-1}\, .
\end{align}
Recalling that $\hat{\eta}^{(-n)}_{-q}$ satisfies
\begin{equation}
[L_{-1},\hat{\eta}^{(-n)}_{-q}]=0,\quad [L_0, \hat{\eta}^{(-n)}_{-q}] = -n\; \hat{\eta}^{(-n)}_{-q}\, ,
\end{equation}
it follows that
\begin{align}
\ad L_0 (\ad L_{1})^{s} \hat{\eta}^{(-n)}_{-q}&= (2s-n)(\ad L_{1})^s\hat{\eta}^{(-n)}_{-q}\, ,\\
\ad L_{-1} (\ad L_{1})^s \hat{\eta}^{(-n)}_{-q} &= s (n-s+1)(\ad L_{1})^{s-1}\hat{\eta}^{(-n)}_{-q}\, .
\end{align} 
Using these two results, one may simplify	
\begin{align}
4B_0(\Phi_n^{n,-1}) &=(n+2)\sum_{s,q} a^{n,s}_q \begin{pmatrix}
-\frac{1}{n-s}  (\ad L_{1})^{s+1}\\
2   (\ad L_{1})^{s}\\
(n-s+1)  (\ad L_{1})^{s-1}
\end{pmatrix}\hat{\eta}^{(-n)}_{-q}= (n+2) \Phi_n^{n,-1}\, .
\end{align}
We conclude that the input data indeed has $\epsilon=-1$. Note that this derivation is valid for any choice of $a^{n,s}_q$. To fix the $a^{n,s}_q$, we evaluate the $Q$-constraint at level $n$ and insert the expression for $\Phi_n^{n,-1}$. We follow the computation in \cite{Grimm:2020cda}. Using the fact that
\begin{equation}
[2Q_\infty - L_0, (\ad L_{1})^s\hat{\eta}^{(-n)}_{-q}] = (n-2q)(\ad L_{1})^s\hat{\eta}^{(-n)}_{-q}\, ,
\end{equation}
and as well as the previous relations we find that \eqref{eq:Q_constr_coeff} reduces to
\begin{align}
0&=\sum_{1\leq s,q\leq n-1} \left(i(n-2q)a^{n,s}_q+ \frac{n-s}{n-s+1} a^{n,s-1}_q-s(n-s) a^{n,s+1}_q \right) (\ad L_{1})^{s}\hat{\eta}^{(-n)}_{-q}\, ,
\end{align} 
where in the second line we collected terms according to their power in $\ad L_{1}$. Since this relation must hold for every $s$ separately, we find
\begin{equation}
\label{eq:Q_constraint_coef}
i(n-2q)a^{n,s}_q+ \frac{n-s}{n-s+1} a^{n,s-1}_q-s(n-s) a^{n,s+1}_q=0\, , \quad a^{n,1}_q=\frac{1}{n}\, .
\end{equation}
The particular normalization is due to the fact that one should recover $\hat{\eta}$ from the $L_n^-$ as follows:\footnote{The component of $L_n^-$ which has weight $-n$ corresponds to the $s=1$ part of the sum.}
\begin{equation}
\hat{\eta} = \sum_{n\geq 2} \sum_{q\geq 1} (L_n^-)^{(n,-n)}_{-q}=\sum_{n\geq 2} \sum_{q\geq 1} na^{n,1}_q \hat{\eta}^{(-n)}_{-q}\, .
\end{equation}
In \cite{Grimm:2020cda} it was stated that \eqref{eq:Q_constraint_coef} together with the reality condition $\overline{a^{n,s}_q}=a^{n,s}_{n-q}$ has a unique solution given by
\begin{equation}
a^{n,s}_q = i^{s-1} \frac{(n-s)!}{n!} b^{s-1}_{q-1,n-q-1}\, ,
\end{equation}
where the coefficients $b^k_{p,q}$ are determined by
\begin{equation}
(1-x)^p(1+x)^q = \sum_{k=0}^{p+q} b^k_{p,q} x^k \, .
\end{equation}

\section{Expressions for the bulk reconstruction of $\mathrm{IV}_1$  boundaries}
\label{app:LCS_basis}
In this appendix we collect some relevant expressions for the bulk reconstruction of the $\mathrm{IV}_1$ boundary that complements the derivation in section \ref{ssec:IV1}. 

First we write out the adjoint action of the $\slt$-triple $(L_1,L_0,L_{-1})$ in terms of the basis $T_a$ of the algebra $\mathfrak{g}=\mathfrak{sp}(4)$ introduced in \eqref{eq:basis_LCS}. In this basis the triple $(\ad L_{1}, \ad L_0, \ad L_{-1})$ takes the form
\begin{equation}\label{eq:adjoints}
\begin{aligned}
\ad L_{1} &= \Scale[0.8]{\begin{pmatrix}
0 & 0 & i \sqrt{\frac{3}{2}} & \sqrt{\frac{3}{2}} & 0 & 0 & 0 & 0 & 0 & 0 \\
0 & 0 & -\sqrt{\frac{3}{2}} & i \sqrt{\frac{3}{2}} & 0 & 0 & 0 & 0 & 0 & 0 \\
-i \sqrt{\frac{3}{2}} & \sqrt{\frac{3}{2}} & 0 & 0 & i \sqrt{\frac{5}{2}} & \sqrt{\frac{5}{2}} & 0 & 0 & 0
& 0 \\
-\sqrt{\frac{3}{2}} & -i \sqrt{\frac{3}{2}} & 0 & 0 & -\sqrt{\frac{5}{2}} & i \sqrt{\frac{5}{2}} & 0 & 0 &
0 & 0 \\
0 & 0 & -i \sqrt{\frac{5}{2}} & \sqrt{\frac{5}{2}} & 0 & 0 & i \sqrt{6} & 0 & 0 & 0 \\
0 & 0 & -\sqrt{\frac{5}{2}} & -i \sqrt{\frac{5}{2}} & 0 & 0 & -\sqrt{6} & 0 & 0 & 0 \\
0 & 0 & 0 & 0 & -i \sqrt{6} & \sqrt{6} & 0 & 0 & 0 & 0 \\
0 & 0 & 0 & 0 & 0 & 0 & 0 & 0 & 0 & i \\
0 & 0 & 0 & 0 & 0 & 0 & 0 & 0 & 0 & -1 \\
0 & 0 & 0 & 0 & 0 & 0 & 0 & -i & 1 & 0 \\
\end{pmatrix}}, \\
\ad L_{0} &= \rule[-1.05cm]{.0cm}{2.3cm}\Scale[0.8]{\begin{pmatrix}
0 & -6 i & 0 & 0 & 0 & 0 & 0 & 0 & 0 & 0 \\
6 i & 0 & 0 & 0 & 0 & 0 & 0 & 0 & 0 & 0 \\
0 & 0 & 0 & -4 i & 0 & 0 & 0 & 0 & 0 & 0 \\
0 & 0 & 4 i & 0 & 0 & 0 & 0 & 0 & 0 & 0 \\
0 & 0 & 0 & 0 & 0 & -2 i & 0 & 0 & 0 & 0 \\
0 & 0 & 0 & 0 & 2 i & 0 & 0 & 0 & 0 & 0 \\
0 & 0 & 0 & 0 & 0 & 0 & 0 & 0 & 0 & 0 \\
0 & 0 & 0 & 0 & 0 & 0 & 0 & 0 & -2 i & 0 \\
0 & 0 & 0 & 0 & 0 & 0 & 0 & 2 i & 0 & 0 \\
0 & 0 & 0 & 0 & 0 & 0 & 0 & 0 & 0 & 0 \\
\end{pmatrix}}, \\
\ad  L_{-1} &= \rule[-1.05cm]{.0cm}{2.3cm}\Scale[0.8]{\begin{pmatrix}
0 & 0 & i \sqrt{\frac{3}{2}} & -\sqrt{\frac{3}{2}} & 0 & 0 & 0 & 0 & 0 & 0 \\
0 & 0 & \sqrt{\frac{3}{2}} & i \sqrt{\frac{3}{2}} & 0 & 0 & 0 & 0 & 0 & 0 \\
-i \sqrt{\frac{3}{2}} & -\sqrt{\frac{3}{2}} & 0 & 0 & i \sqrt{\frac{5}{2}} & -\sqrt{\frac{5}{2}} & 0 & 0 &
0 & 0 \\
\sqrt{\frac{3}{2}} & -i \sqrt{\frac{3}{2}} & 0 & 0 & \sqrt{\frac{5}{2}} & i \sqrt{\frac{5}{2}} & 0 & 0 & 0
& 0 \\
0 & 0 & -i \sqrt{\frac{5}{2}} & -\sqrt{\frac{5}{2}} & 0 & 0 & i \sqrt{6} & 0 & 0 & 0 \\
0 & 0 & \sqrt{\frac{5}{2}} & -i \sqrt{\frac{5}{2}} & 0 & 0 & \sqrt{6} & 0 & 0 & 0 \\
0 & 0 & 0 & 0 & -i \sqrt{6} & -\sqrt{6} & 0 & 0 & 0 & 0 \\
0 & 0 & 0 & 0 & 0 & 0 & 0 & 0 & 0 & i \\
0 & 0 & 0 & 0 & 0 & 0 & 0 & 0 & 0 & 1 \\
0 & 0 & 0 & 0 & 0 & 0 & 0 & -i & -1 & 0 \\
\end{pmatrix}}. \\
\end{aligned}
\end{equation}

We next consider the recursion relation \eqref{eq:recursion_CKS} for $\Phi_n=(\mathbf{L}^+, \mathbf{L}^0, \mathbf{L}^-)$ for $\mathrm{IV}_1$ boundaries. By using ansatz \eqref{eq:LCS_Phi_ansatz} we find that the CKS recursion reduces to a system of four coupled recursion relations for the coefficients $a_n$, $b_n$, $c_n$ and $d_n$, given by
\begin{align}\label{eq:recursion}
a_{n}+b_{n} &= \frac{2}{6n+8}\sum_{0<k<n} \Big[a_{n-k} \Big(9 c_{k}+14 i \sqrt{\frac{6}{5}} b_{k}\Big) +3 b_{n-k} \Big( \frac{28 i}{3}
\sqrt{\frac{2}{15}} b_k +4 c_k+d_k \Big) \Big] ,  \nonumber \\
3a_{n}-4b_{n} &= \frac{2}{6n-6}\sum_{0<k<n} \Big[12 b_{n-k} \Big(7 i \sqrt{\frac{2}{15}}
b_{k}+3 c_{k}-d_{k} \Big)-4a_{n-k} \Big(9 c_{k}+14 i \sqrt{\frac{6}{5}} b_{k}\Big)\Big] , \nonumber \\
c_{n}+d_{n} &= \frac{2}{6n+4}\sum_{0<k<n} \frac{1}{30} \Big[90 ( c_{k}+d_{k}) c_{n-k}-1568 b_{k} (3a_{n-k}+b_{n-k}) \Big] , \\
2c_{n}-d_{n} &= \frac{2}{6n-2}\sum_{0<k<n} \Big[\frac{784}{15} b_{n-k}(b_k-6a_k) +3c_{n-k}(2d_k- c_{n-k} )\Big]\, . \nonumber 
\end{align}
Let us remind the reader that the these coupled recursion relations are solved by \eqref{eq:LCS_recursion_solution}.

\bibliographystyle{jhep}
\bibliography{references_2}

\providecommand{\href}[2]{#2}\begingroup\raggedright\begin{thebibliography}{10}

\bibitem{Palti:2019pca}
E.~Palti, \emph{{The Swampland: Introduction and Review}},
  \href{http://dx.doi.org/10.1002/prop.201900037}{\emph{Fortsch. Phys.} {\bf
  67} (2019) 1900037}, [\href{https://arxiv.org/abs/1903.06239}{{\tt
  1903.06239}}].

\bibitem{vanBeest:2021lhn}
M.~van Beest, J.~Calder\'on-Infante, D.~Mirfendereski and I.~Valenzuela,
  \emph{{Lectures on the Swampland Program in String Compactifications}},
  \href{https://arxiv.org/abs/2102.01111}{{\tt 2102.01111}}.

\bibitem{Ooguri2007}
H.~Ooguri and C.~Vafa, \emph{{On the Geometry of the String Landscape and the
  Swampland}},
  \href{http://dx.doi.org/10.1016/j.nuclphysb.2006.10.033}{\emph{Nucl. Phys.}
  {\bf B766} (2007) 21--33}, [\href{https://arxiv.org/abs/hep-th/0605264}{{\tt
  hep-th/0605264}}].

\bibitem{Klaewer:2016kiy}
D.~Klaewer and E.~Palti, \emph{{Super-Planckian Spatial Field Variations and
  Quantum Gravity}},
  \href{http://dx.doi.org/10.1007/JHEP01(2017)088}{\emph{JHEP} {\bf 01} (2017)
  088}, [\href{https://arxiv.org/abs/1610.00010}{{\tt 1610.00010}}].

\bibitem{GPV}
T.~W. Grimm, E.~Palti and I.~Valenzuela, \emph{{Infinite Distances in Field
  Space and Massless Towers of States}},
  \href{http://dx.doi.org/10.1007/JHEP08(2018)143}{\emph{JHEP} {\bf 08} (2018)
  143}, [\href{https://arxiv.org/abs/1802.08264}{{\tt 1802.08264}}].

\bibitem{Heidenreich:2018kpg}
B.~Heidenreich, M.~Reece and T.~Rudelius, \emph{{Emergence of Weak Coupling at
  Large Distance in Quantum Gravity}},
  \href{http://dx.doi.org/10.1103/PhysRevLett.121.051601}{\emph{Phys. Rev.
  Lett.} {\bf 121} (2018) 051601},
  [\href{https://arxiv.org/abs/1802.08698}{{\tt 1802.08698}}].

\bibitem{Lee:2019wij}
S.-J. Lee, W.~Lerche and T.~Weigand, \emph{{Emergent Strings from Infinite
  Distance Limits}},  \href{https://arxiv.org/abs/1910.01135}{{\tt
  1910.01135}}.

\bibitem{Lanza:2020qmt}
S.~Lanza, F.~Marchesano, L.~Martucci and I.~Valenzuela, \emph{{Swampland
  Conjectures for Strings and Membranes}},
  \href{https://arxiv.org/abs/2006.15154}{{\tt 2006.15154}}.

\bibitem{Calderon-Infante:2020dhm}
J.~Calder\'on-Infante, A.~M. Uranga and I.~Valenzuela, \emph{{The Convex Hull
  Swampland Distance Conjecture and Bounds on Non-geodesics}},
  \href{https://arxiv.org/abs/2012.00034}{{\tt 2012.00034}}.

\bibitem{Ashok:2003gk}
S.~Ashok and M.~R. Douglas, \emph{{Counting flux vacua}},
  \href{http://dx.doi.org/10.1088/1126-6708/2004/01/060}{\emph{JHEP} {\bf 01}
  (2004) 060}, [\href{https://arxiv.org/abs/hep-th/0307049}{{\tt
  hep-th/0307049}}].

\bibitem{Acharya:2006zw}
B.~S. Acharya and M.~R. Douglas, \emph{{A Finite landscape?}},
  \href{https://arxiv.org/abs/hep-th/0606212}{{\tt hep-th/0606212}}.

\bibitem{Grimm:2020cda}
T.~W. Grimm, \emph{{Moduli Space Holography and the Finiteness of Flux Vacua}},
   \href{https://arxiv.org/abs/2010.15838}{{\tt 2010.15838}}.

\bibitem{Cecotti:2020rjq}
S.~Cecotti, \emph{{Special Geometry and the Swampland}},
  \href{https://arxiv.org/abs/2004.06929}{{\tt 2004.06929}}.

\bibitem{Cecotti:2020uek}
S.~Cecotti, \emph{{Moduli spaces of Calabi-Yau $d$-folds as
  gravitational-chiral instantons}},
  \href{https://arxiv.org/abs/2007.09992}{{\tt 2007.09992}}.

\bibitem{Grimm:2019ixq}
T.~W. Grimm, C.~Li and I.~Valenzuela, \emph{{Asymptotic Flux Compactifications
  and the Swampland}},
  \href{http://dx.doi.org/10.1007/JHEP06(2020)009}{\emph{JHEP} {\bf 06} (2020)
  009}, [\href{https://arxiv.org/abs/1910.09549}{{\tt 1910.09549}}].

\bibitem{Grimm:2018cpv}
T.~W. Grimm, C.~Li and E.~Palti, \emph{{Infinite Distance Networks in Field
  Space and Charge Orbits}},
  \href{http://dx.doi.org/10.1007/JHEP03(2019)016}{\emph{JHEP} {\bf 03} (2019)
  016}, [\href{https://arxiv.org/abs/1811.02571}{{\tt 1811.02571}}].

\bibitem{Corvilain:2018lgw}
P.~Corvilain, T.~W. Grimm and I.~Valenzuela, \emph{{The Swampland Distance
  Conjecture for K{\"a}hler moduli}},
  \href{http://dx.doi.org/10.1007/JHEP08(2019)075}{\emph{JHEP} {\bf 08} (2019)
  075}, [\href{https://arxiv.org/abs/1812.07548}{{\tt 1812.07548}}].

\bibitem{Font:2019cxq}
A.~Font, A.~Herr{\'a}ez and L.~E. Ib{\'a}{\~n}ez, \emph{{The Swampland Distance
  Conjecture and Towers of Tensionless Branes}},
  \href{http://dx.doi.org/10.1007/JHEP08(2019)044}{\emph{JHEP} {\bf 08} (2019)
  044}, [\href{https://arxiv.org/abs/1904.05379}{{\tt 1904.05379}}].

\bibitem{Grimm:2019wtx}
T.~W. Grimm and D.~Van De~Heisteeg, \emph{{Infinite Distances and the Axion
  Weak Gravity Conjecture}},
  \href{http://dx.doi.org/10.1007/JHEP03(2020)020}{\emph{JHEP} {\bf 03} (2020)
  020}, [\href{https://arxiv.org/abs/1905.00901}{{\tt 1905.00901}}].

\bibitem{Gendler:2020dfp}
N.~Gendler and I.~Valenzuela, \emph{{Merging the Weak Gravity and Distance
  Conjectures Using BPS Extremal Black Holes}},
  \href{https://arxiv.org/abs/2004.10768}{{\tt 2004.10768}}.

\bibitem{BastianGrimmHeisteeg}
B.~Bastian, T.~W. Grimm and D.~van~de Heisteeg, \emph{Weak gravity bounds in
  asymptotic string compactifications},
  \href{https://arxiv.org/abs/2011.08854}{{\tt 2011.08854}}.

\bibitem{Grimm:2020ouv}
T.~W. Grimm and C.~Li, \emph{{Universal Axion Backreaction in Flux
  Compactifications}},  \href{https://arxiv.org/abs/2012.08272}{{\tt
  2012.08272}}.

\bibitem{robles_2016}
C.~Robles, \emph{Classification of horizontal $\text{SL}(2)$s},
  {\emph{Compositio Mathematica} {\bf 152} (2016) 918},
  [\href{https://arxiv.org/abs/arXiv:1405.3163}{{\tt arXiv:1405.3163}}].

\bibitem{Kerr2017}
M.~Kerr, G.~Pearlstein and C.~Robles, \emph{{Polarized relations on horizontal
  SL(2)s}},  \href{https://arxiv.org/abs/1705.03117}{{\tt 1705.03117}}.

\bibitem{Schmid}
W.~Schmid, \emph{{Variation of Hodge structure: the singularities of the period
  mapping}}, \href{http://dx.doi.org/10.1007/BF01389674}{\emph{Invent. Math. ,
  22:211--319, 1973} }.

\bibitem{CKS}
E.~Cattani, A.~Kaplan and W.~Schmid, \emph{{Degeneration of Hodge Structures}},
  \href{http://dx.doi.org/10.2307/1971333}{\emph{Annals of Mathematics} {\bf
  123} (1986) 457--535}.

\bibitem{Harlow:2018fse}
D.~Harlow, \emph{{TASI Lectures on the Emergence of Bulk Physics in AdS/CFT}},
  \href{http://dx.doi.org/10.22323/1.305.0002}{\emph{PoS} {\bf TASI2017} (2018)
  002}, [\href{https://arxiv.org/abs/1802.01040}{{\tt 1802.01040}}].

\bibitem{Hamilton:2006az}
A.~Hamilton, D.~N. Kabat, G.~Lifschytz and D.~A. Lowe, \emph{{Holographic
  representation of local bulk operators}},
  \href{http://dx.doi.org/10.1103/PhysRevD.74.066009}{\emph{Phys. Rev. D} {\bf
  74} (2006) 066009}, [\href{https://arxiv.org/abs/hep-th/0606141}{{\tt
  hep-th/0606141}}].

\bibitem{Nakayama:2015mva}
Y.~Nakayama and H.~Ooguri, \emph{{Bulk Locality and Boundary Creating
  Operators}}, \href{http://dx.doi.org/10.1007/JHEP10(2015)114}{\emph{JHEP}
  {\bf 10} (2015) 114}, [\href{https://arxiv.org/abs/1507.04130}{{\tt
  1507.04130}}].

\bibitem{Beccaria:2020qtk}
M.~Beccaria, H.~Jiang and A.~A. Tseytlin, \emph{{Boundary correlators in WZW
  model on AdS$_{2}$}},
  \href{http://dx.doi.org/10.1007/JHEP05(2020)099}{\emph{JHEP} {\bf 05} (2020)
  099}, [\href{https://arxiv.org/abs/2001.11269}{{\tt 2001.11269}}].

\bibitem{Nakayama:2015hga}
Y.~Nakayama and Y.~Nomura, \emph{{Weak gravity conjecture in the AdS/CFT
  correspondence}},
  \href{http://dx.doi.org/10.1103/PhysRevD.92.126006}{\emph{Phys. Rev. D} {\bf
  92} (2015) 126006}, [\href{https://arxiv.org/abs/1509.01647}{{\tt
  1509.01647}}].

\bibitem{Harlow:2015lma}
D.~Harlow, \emph{{Wormholes, Emergent Gauge Fields, and the Weak Gravity
  Conjecture}}, \href{http://dx.doi.org/10.1007/JHEP01(2016)122}{\emph{JHEP}
  {\bf 01} (2016) 122}, [\href{https://arxiv.org/abs/1510.07911}{{\tt
  1510.07911}}].

\bibitem{Benjamin:2016fhe}
N.~Benjamin, E.~Dyer, A.~L. Fitzpatrick and S.~Kachru, \emph{{Universal Bounds
  on Charged States in 2d CFT and 3d Gravity}},
  \href{http://dx.doi.org/10.1007/JHEP08(2016)041}{\emph{JHEP} {\bf 08} (2016)
  041}, [\href{https://arxiv.org/abs/1603.09745}{{\tt 1603.09745}}].

\bibitem{Montero:2016tif}
M.~Montero, G.~Shiu and P.~Soler, \emph{{The Weak Gravity Conjecture in three
  dimensions}}, \href{http://dx.doi.org/10.1007/JHEP10(2016)159}{\emph{JHEP}
  {\bf 10} (2016) 159}, [\href{https://arxiv.org/abs/1606.08438}{{\tt
  1606.08438}}].

\bibitem{Montero:2017mdq}
M.~Montero, \emph{{Are tiny gauge couplings out of the Swampland?}},
  \href{http://dx.doi.org/10.1007/JHEP10(2017)208}{\emph{JHEP} {\bf 10} (2017)
  208}, [\href{https://arxiv.org/abs/1708.02249}{{\tt 1708.02249}}].

\bibitem{Harlow:2018tng}
D.~Harlow and H.~Ooguri, \emph{{Symmetries in quantum field theory and quantum
  gravity}},  \href{https://arxiv.org/abs/1810.05338}{{\tt 1810.05338}}.

\bibitem{Harlow:2018jwu}
D.~Harlow and H.~Ooguri, \emph{{Constraints on Symmetries from Holography}},
  \href{http://dx.doi.org/10.1103/PhysRevLett.122.191601}{\emph{Phys. Rev.
  Lett.} {\bf 122} (2019) 191601},
  [\href{https://arxiv.org/abs/1810.05337}{{\tt 1810.05337}}].

\bibitem{Bae:2018qym}
J.-B. Bae, S.~Lee and J.~Song, \emph{{Modular Constraints on Superconformal
  Field Theories}},
  \href{http://dx.doi.org/10.1007/JHEP01(2019)209}{\emph{JHEP} {\bf 01} (2019)
  209}, [\href{https://arxiv.org/abs/1811.00976}{{\tt 1811.00976}}].

\bibitem{Conlon:2018vov}
J.~P. Conlon and F.~Quevedo, \emph{{Putting the Boot into the Swampland}},
  \href{http://dx.doi.org/10.1007/JHEP03(2019)005}{\emph{JHEP} {\bf 03} (2019)
  005}, [\href{https://arxiv.org/abs/1811.06276}{{\tt 1811.06276}}].

\bibitem{Montero:2018fns}
M.~Montero, \emph{{A Holographic Derivation of the Weak Gravity Conjecture}},
  \href{http://dx.doi.org/10.1007/JHEP03(2019)157}{\emph{JHEP} {\bf 03} (2019)
  157}, [\href{https://arxiv.org/abs/1812.03978}{{\tt 1812.03978}}].

\bibitem{Lin:2019kpn}
Y.-H. Lin and S.-H. Shao, \emph{{Anomalies and Bounds on Charged Operators}},
  \href{http://dx.doi.org/10.1103/PhysRevD.100.025013}{\emph{Phys. Rev. D} {\bf
  100} (2019) 025013}, [\href{https://arxiv.org/abs/1904.04833}{{\tt
  1904.04833}}].

\bibitem{Conlon:2020wmc}
J.~P. Conlon and F.~Revello, \emph{{Moduli Stabilisation and the Holographic
  Swampland}}, \href{http://dx.doi.org/10.31526/LHEP.2020.171}{\emph{LHEP} {\bf
  2020} (2020) 171}, [\href{https://arxiv.org/abs/2006.01021}{{\tt
  2006.01021}}].

\bibitem{Ooguri:2020sua}
H.~Ooguri and T.~Takayanagi, \emph{{Cobordism Conjecture in AdS}},
  \href{https://arxiv.org/abs/2006.13953}{{\tt 2006.13953}}.

\bibitem{Perlmutter:2020buo}
E.~Perlmutter, L.~Rastelli, C.~Vafa and I.~Valenzuela, \emph{{A CFT Distance
  Conjecture}},  \href{https://arxiv.org/abs/2011.10040}{{\tt 2011.10040}}.

\bibitem{donaldson1984}
S.~K. Donaldson, \emph{Nahm's equations and the classification of monopoles},
  {\emph{Comm. Math. Phys.} {\bf 96} (1984) 387--407}.

\bibitem{Borel1973}
A.~Borel and J.-P. Serre, \emph{{Corners and Arithmetic Groups}},
  {\emph{Commentarii mathematici Helvetici} {\bf 48} (1973) 436--483}.

\bibitem{Grimm:2019bey}
T.~W. Grimm, F.~Ruehle and D.~van~de Heisteeg, \emph{{Classifying Calabi-Yau
  threefolds using infinite distance limits}},
  \href{https://arxiv.org/abs/1910.02963}{{\tt 1910.02963}}.

\bibitem{Pearlstein2006}
G.~Pearlstein, \emph{$\rm{SL}_2$-orbits and degenerations of mixed hodge
  structure}, \href{http://dx.doi.org/10.4310/jdg/1175266181}{\emph{J.
  Differential Geom.} {\bf 74} (09, 2006) 1--67}.

\bibitem{Tyurin:2003}
A.~N. Tyurin, \emph{{Fano versus Calabi--Yau}}, {\emph{The Fano Conference}
  (2004) 701--734}, [\href{https://arxiv.org/abs/math/0302101}{{\tt
  math/0302101}}].

\bibitem{Joshi:2019nzi}
A.~Joshi and A.~Klemm, \emph{{Swampland Distance Conjecture for One-Parameter
  Calabi-Yau Threefolds}},
  \href{http://dx.doi.org/10.1007/JHEP08(2019)086}{\emph{JHEP} {\bf 08} (2019)
  086}, [\href{https://arxiv.org/abs/1903.00596}{{\tt 1903.00596}}].

\bibitem{Candelas:1990rm}
P.~Candelas, X.~C. De~La~Ossa, P.~S. Green and L.~Parkes, \emph{{A Pair of
  Calabi-Yau manifolds as an exactly soluble superconformal theory}},
  \href{http://dx.doi.org/10.1016/0550-3213(91)90292-6}{\emph{Nucl. Phys. B}
  {\bf 359} (1991) 21--74}.

\bibitem{Blumenhagen:2018nts}
R.~Blumenhagen, D.~Klaewer, L.~Schlechter and F.~Wolf, \emph{{The Refined
  Swampland Distance Conjecture in Calabi-Yau Moduli Spaces}},
  \href{http://dx.doi.org/10.1007/JHEP06(2018)052}{\emph{JHEP} {\bf 06} (2018)
  052}, [\href{https://arxiv.org/abs/1803.04989}{{\tt 1803.04989}}].

\bibitem{Demirtas:2019sip}
M.~Demirtas, M.~Kim, L.~Mcallister and J.~Moritz, \emph{{Vacua with Small Flux
  Superpotential}},
  \href{http://dx.doi.org/10.1103/PhysRevLett.124.211603}{\emph{Phys. Rev.
  Lett.} {\bf 124} (2020) 211603},
  [\href{https://arxiv.org/abs/1912.10047}{{\tt 1912.10047}}].

\bibitem{Demirtas:2020ffz}
M.~Demirtas, M.~Kim, L.~Mcallister and J.~Moritz, \emph{{Conifold Vacua with
  Small Flux Superpotential}},  \href{https://arxiv.org/abs/2009.03312}{{\tt
  2009.03312}}.

\bibitem{Blumenhagen:2020ire}
R.~\'Alvarez-Garc\'\i{}a, R.~Blumenhagen, M.~Brinkmann and L.~Schlechter,
  \emph{{Small Flux Superpotentials for Type IIB Flux Vacua Close to a
  Conifold}},  \href{https://arxiv.org/abs/2009.03325}{{\tt 2009.03325}}.

\bibitem{Bastiantoappear}
B.~Bastian, T.~W. Grimm and D.~van~de Heisteeg, \emph{{to appear}},
  {\emph{2021} }.

\bibitem{Teitelboim:1983ux}
C.~Teitelboim, \emph{{Gravitation and Hamiltonian Structure in Two Space-Time
  Dimensions}},
  \href{http://dx.doi.org/10.1016/0370-2693(83)90012-6}{\emph{Phys. Lett. B}
  {\bf 126} (1983) 41--45}.

\bibitem{Jackiw:1984je}
R.~Jackiw, \emph{{Lower Dimensional Gravity}},
  \href{http://dx.doi.org/10.1016/0550-3213(85)90448-1}{\emph{Nucl. Phys. B}
  {\bf 252} (1985) 343--356}.

\bibitem{Lee:2018urn}
S.-J. Lee, W.~Lerche and T.~Weigand, \emph{{Tensionless Strings and the Weak
  Gravity Conjecture}},
  \href{http://dx.doi.org/10.1007/JHEP10(2018)164}{\emph{JHEP} {\bf 10} (2018)
  164}, [\href{https://arxiv.org/abs/1808.05958}{{\tt 1808.05958}}].

\bibitem{Lee:2018spm}
S.-J. Lee, W.~Lerche and T.~Weigand, \emph{{A Stringy Test of the Scalar Weak
  Gravity Conjecture}},
  \href{http://dx.doi.org/10.1016/j.nuclphysb.2018.11.001}{\emph{Nucl. Phys.}
  {\bf B938} (2019) 321--350}, [\href{https://arxiv.org/abs/1810.05169}{{\tt
  1810.05169}}].

\bibitem{Marchesano:2019ifh}
F.~Marchesano and M.~Wiesner, \emph{{Instantons and infinite distances}},
  \href{http://dx.doi.org/10.1007/JHEP08(2019)088}{\emph{JHEP} {\bf 08} (2019)
  088}, [\href{https://arxiv.org/abs/1904.04848}{{\tt 1904.04848}}].

\bibitem{Lee:2019xtm}
S.-J. Lee, W.~Lerche and T.~Weigand, \emph{{Emergent Strings, Duality and Weak
  Coupling Limits for Two-Form Fields}},
  \href{https://arxiv.org/abs/1904.06344}{{\tt 1904.06344}}.

\bibitem{Baume:2019sry}
F.~Baume, F.~Marchesano and M.~Wiesner, \emph{{Instanton Corrections and
  Emergent Strings}},
  \href{http://dx.doi.org/10.1007/JHEP04(2020)174}{\emph{JHEP} {\bf 04} (2020)
  174}, [\href{https://arxiv.org/abs/1912.02218}{{\tt 1912.02218}}].

\bibitem{Klaewer:2020lfg}
D.~Klaewer, S.-J. Lee, T.~Weigand and M.~Wiesner, \emph{{Quantum Corrections in
  4d N=1 Infinite Distance Limits and the Weak Gravity Conjecture}},
  \href{https://arxiv.org/abs/2011.00024}{{\tt 2011.00024}}.

\bibitem{Lanzatoappear}
S.~Lanza, F.~Marchesano, L.~Martucci and I.~Valenzuela, \emph{{to appear}},
  {\emph{2021} }.

\bibitem{Hain:2003}
R.~Hain, \emph{{Periods of Limit Mixed Hodge Structures}}, {\emph{Current
  Developments in Mathematics.} {\bf 2002} (2003) 113--133},
  [\href{https://arxiv.org/abs/math/0305090}{{\tt math/0305090}}].

\end{thebibliography}\endgroup

\end{document}